\documentclass[usegraphicx,usenatbib,useAMS]{mn2e}

\usepackage{multirow}
\usepackage{xcolor}
\usepackage{aas_macros}
\usepackage{amssymb}
\usepackage{xspace}
\usepackage[normalem]{ulem}

\newcommand{\gsim}{\gtrsim} % alias for symbol available through amssymb 
\newcommand{\lsim}{\lesssim} % alias for symbol available through amssymb

\newcommand{\kms}{\,{\rm km}\,{\rm s}^{-1}}
\newcommand{\Msol}{\,{\rm M}_{\odot}}

\newcommand{\GALFORM}{\textsc{galform}\xspace}
\newcommand{\lgalaxy}{\textsc{l-galaxies}\xspace}
\newcommand{\MORGANA}{\textsc{morgana}\xspace}

\title[New cooling model]
{A new gas cooling model for semi-analytic galaxy formation models}

\author[Hou et al.]  {Jun Hou\thanks{E-mail:
    jun.hou@durham.ac.uk},$^1$ Cedric G. Lacey,$^1$
  Carlos. S. Frenk$^1$
  \\
  $^{1}$Institute for Computational Cosmology, Department of Physics,
  University of Durham, South Road, Durham, DH1 3LE, UK}

\begin{document}

\maketitle

\begin{abstract}
  Semi-analytic galaxy formation models are widely used to gain insight
  into the astrophysics of galaxy formation and in model testing,
  parameter space searching and mock catalogue building. In this work we
  present a new model for gas cooling in halos in semi-analytic models,
  which improves over previous cooling models in several ways. Our new
  treatment explicitly includes the evolution of the density profile of
  the hot gas driven by the growth of the dark matter halo and by the
  dynamical adjustment of the gaseous corona as gas cools down. The
  effect of the past cooling history on the current mass cooling rate is
  calculated more accurately, by doing an integral over the past
  history. The evolution of the hot gas angular momentum profile is
  explicitly followed, leading to a self-consistent and more detailed
  calculation of the angular momentum of the cooled down gas. This
  model predicts higher cooled down masses than the cooling models
  previously used in \GALFORM, closer to the predictions of the cooling
  models in \lgalaxy and \MORGANA, even though those models are
  formulated differently.  It also predicts cooled down angular momenta
  that are higher than in previous \GALFORM cooling models, but generally
  lower than the predictions of \lgalaxy and \MORGANA.  When used in
  a full galaxy formation model, this cooling model improves the
  predictions for early-type galaxy sizes in \GALFORM.
\end{abstract}

\begin{keywords}
methods: analytical -- galaxies: evolution -- galaxies:formation
\end{keywords}

%%%%%%%%%%%%%%%%%%%%%%%%%%%%%%%%%%%%%%%%%%%%%%%%%%%%%%%%%%%%%%%%%%%%%%%%%%%%%%
\section{Introduction}
Understanding galaxy formation is a central aim of
astrophysics. Galaxies are interesting objects in their own right. In addition, they are 
a tracer of the large-scale matter distribution, which is important
for the study of cosmology, and also provide the background
environment for astrophysical processes happening on small scales,
such as star formation and black hole growth. Despite its importance, 
many aspects of galaxy formation remain poorly understood, 
because of the complexities of the physical processes involved.

Currently there are two major theoretical approaches to studying
galaxy formation: hydrodynamical simulations and
semi-analytic (SA) models, both of which have advantages and
disadvantages. Hydrodynamical simulations provide a more detailed
picture of galaxy formation by numerically solving the equations
governing this process, but at large computational expense. This
limits their ability to generate large galaxy samples. To derive a
representative sample of galaxies, hydrodynamical simulations have to
be performed in cosmological volumes. Such simulations necessarily employ
parametrized sub-grid models for many physical processes happening on
small scales, due to limited numerical resolution; their large
computational expense makes it difficult to explore the entire
parameter space. In contrast, semi-analytic models \citep[e.g.][]{WF_1991,Baugh_2006_SA_review} develop a coarse-grained picture of galaxy formation by focusing on global properties of a galaxy, such as total stellar mass, total cold gas mass, etc. SA models view many such quantities as reservoirs, and the physical processes driving the evolution of them, such as gas cooling, star formation, feedback and galaxy mergers, are viewed as channels connecting the corresponding reservoirs. Simplified analytic descriptions are used to model these channels, and to evolve the global properties from the initial time to the output time. Many SA models also contain simplified recipes for calculating galaxy sizes. SA models calculate the evolution in less detail than hydrodynamical simulations, but are much less computationally expensive.
SA models make it easy to
generate large mock catalogues and to search parameter space, so
semi-analytic models can be very complementary to hydrodynamical
simulations. Moreover, semi-analytic models are more flexible, and
one can easily apply different models for a given physical process,
which makes these models an ideal tool for testing different
modeling approaches and different ideas about which physical processes
are important.

Although the prescriptions in semi-analytic models are generally 
simplified, it is still important to make them as physically
consistent as possible. This lays the foundation for the realism and
reliability of the resulting mock catalogues, and also reduces the extent of 
false degrees of freedom generated by the model parametrization, so
that parameter space searches produce more physically useful
information. In this work we focus on the modelling of gas cooling and
accretion in haloes. In hieararchical
structure formation models, dark matter haloes grow in mass through
both accretion and mergers. Baryons in the form of gas are accreted
into haloes along with the dark matter. However, only some fraction of
this gas is accreted onto the central galaxy in the halo, this being
determined by the combined effects of gravity, pressure, shock heating
and radiative cooling. This whole process of gas accretion onto
galaxies in haloes is what we mean by ``halo gas cooling''.  This is a
crucial process in galaxy formation, for, along with galaxy mergers,
it determines the amount of mass and angular momentum delivered to a
galaxy, and thus is a primary determinant of the properties and
evolution of galaxies.

Currently, most semi-analytic models use treatments of halo gas
cooling that are more or less based on the gas cooling picture set out
in \citet{WF_1991} [also see \citet{classical_cooling_binney,
classical_cooling_rees, classical_cooling_silk} and
\citet{classical_cooling_white78}], in which the gas in a dark matter
halo initially settles in a spherical pressure-supported hot gas halo,
and this gas gradually cools down and contracts under gravity as it
loses pressure support, while new gas joins the halo due to structure
growth or to the reincorporation of the gas ejected by feedback from
supernovae (SN) and AGN.

The above picture has been challenged by the so-called ``cold accretion
scenario'' \citep[e.g.][]{cold_accretion_dekel, cold_accretion_keres},
in which the accreted gas in low mass haloes
($M_{\rm halo} \lsim 3\times 10^{11}\Msol$) does not build a hot
gaseous halo, but rather stays cold and falls freely onto the central
galaxy. However, in these small haloes, the cooling time scale of the
assumed hot gas halo in SA models is very short, and the gas accretion
onto central galaxies is in practice limited by the free-fall time
scale, both in the original \citet{WF_1991} model and in most current
SA models. Therefore the use of the \citeauthor {WF_1991} cooling
picture for these haloes should not introduce large errors in the
accreted gas masses \citep{Benson11}. In the cold accretion picture,
cold gas flows through the halo along filaments
\citep{cold_accretion_keres}, and it has been argued that even in more
massive haloes some gas from the filaments can penetrate the hot gas
halo and deliver cold gas directly to the central galaxy
\citep[e.g.][]{Keres2009}, or to a shock close to the central galaxy
\citep[e.g.][]{nelson2016_filaments}. However, this only happens when
the temperature of the hot gas halo is not very high and the filaments are
still narrow, and so only in a limited range of redshift and halo mass
\citep[e.g.][]{Keres2009}. Furthermore, the effects of accretion along
filaments within haloes are expected to be reduced when the effects of
gas heating by SN and AGN are included
\citep[e.g.][]{Benson11}. Therefore the cooling picture of
\citet{WF_1991} should remain a reasonable approximation for the cold
gas accretion rate.

There are three main gas cooling models used in SA models, namely
those in the Durham model \GALFORM
\citep[e.g.][]{cole2000,galform_baugh2005, galform_bower2006,
galform_lacey2015}, in the Munich model \lgalaxy \citep{munich_model1,
munich_model2, DeLucia07, munich_model_Guo11, munich_model3} and in the \MORGANA
model \citep{morgana1, morgana2}. Most other SA models
\citep[e.g.][]{Somerville08} use a variant of one of these. We outline the key
differences between the three cooling models here, and give more
details in \S\ref{sec:other_models}.

The \GALFORM cooling model calculates the evolution of a cooling
front (i.e. the boundary separating the hot gas and the cooled down gas), integrating outwards from the centre. However, it introduces
artificial `halo formation' events, when the halo mass doubles; at this time the halo gas density profile
is reset, and the radius of the cooling front is reset to zero. Between
these formation events, there is no contraction in the profile of the
gas that is yet to cool. An improved version of this model, in which
the artificial halo formation events are removed, was introduced in
\citet{benson_bower_2010_cooling}, but the treatment of the cooling
history and contraction of the hot gas halo is still fairly
approximate.

The \lgalaxy cooling model is simpler to calculate than that in
\GALFORM. It is motivated by the \citet{Bertschinger_cooling_solution} self-similar solution for gas cooling. However, the original solution is derived for a static gravitational potential, while in cosmological structure formation, the halo grows and its potential evolves with time, so this self-similar solution is not directly applicable.

The \MORGANA cooling model incorporates a more detailed calculation of
the contraction of the hot gas halo due to cooling compared to the
above models, but instead of letting the gas at small radius cool
first, it assumes that hot gas at different radii contributes to the
mass cooling rate simultaneously. However in a perfectly spherical system, as assumed in \MORGANA, the gas cooling timescale is a unique function of radius, and the gas should cool shell by shell.

Furthermore, while the \GALFORM cooling model accounts for an angular
momentum profile in the halo gas when calculating the angular momentum
of the cooled down gas, the \lgalaxy and \MORGANA models are much more
simplified in this respect.

In summary, all of the main cooling models used in current
semi-analytic models have important limitations. In this paper, we
introduce a new cooling model. This new
model treats the evolution of the hot gas density profile and of the
gas cooling more self-consistently compared to the models mentioned
above, while also incorporating a detailed treatment of the angular
momentum of the cooled down gas.  This new cooling model is still
based on the cooling picture in \citet{WF_1991}. In particular, it
still assumes a spherical hot gas halo. As argued above, this picture
may be a good approximation, but it needs to be further checked by
comparing with hydrodynamical simulations in which shock heating and
filamentary accretion are considered in detail. We leave this
comparison for a future work. Note that even if accretion of cold gas
along filaments within haloes is significant, this does not exclude the
existence of a diffuse, roughly spherical hot gas halo, and our new
model should provide a better modeling of this component than the
previous models mentioned above, and thus constitutes a step
towards an even more accurate and complete model of halo gas cooling.

This paper is organized as follows. Section~\ref{sec:models} first describes
our new cooling model, and then the other main cooling
models used in semi-analytic modelling. Then Section~\ref{sec:results}
compares predictions from the new cooling model with those from other
models, first in static haloes and then in hierarchically growing
haloes. The effects of the new cooling model on a full galaxy formation
model are also shown and briefly discussed in this section. Finally a
summary is given in Section~\ref{sec:summary}.

%%%%%%%%%%%%%%%%%%%%%%%%%%%%%%%%%%%%%%%%%%%%%%%%%%%%%%%%%%%%%%%%%%%%%%%%%%%%

\section{Models} \label{sec:models}

\subsection{The new cooling model} \label{sec:new_model}
\subsubsection{Overview of the new cooling model}
The hot gas inside a dark matter halo is assumed to form a spherical
pressure-supported halo in hydrostatic equlibrium. The gas
accreted during halo growth and also the reincorporated gas that was
previously ejected by SN feedback are shock heated and join this hot
gas halo. The hot gas halo itself can cool down due to radiation, and
this cooling removes gas from the halo. The cooled down gas,
which lacks pressure support, falls into the central region of the
dark matter halo and delivers mass and angular momentum to the central
galaxy. We call this compoment of cold infalling gas the cold gas
halo. Typically, the gas at smaller radii cools faster, and this kind
of cooling leads to the reduction of pressure support from the centre
outwards. The hot gas halo then contracts under gravity.

\begin{figure*}
 \centering
\includegraphics[width=0.7\textwidth]{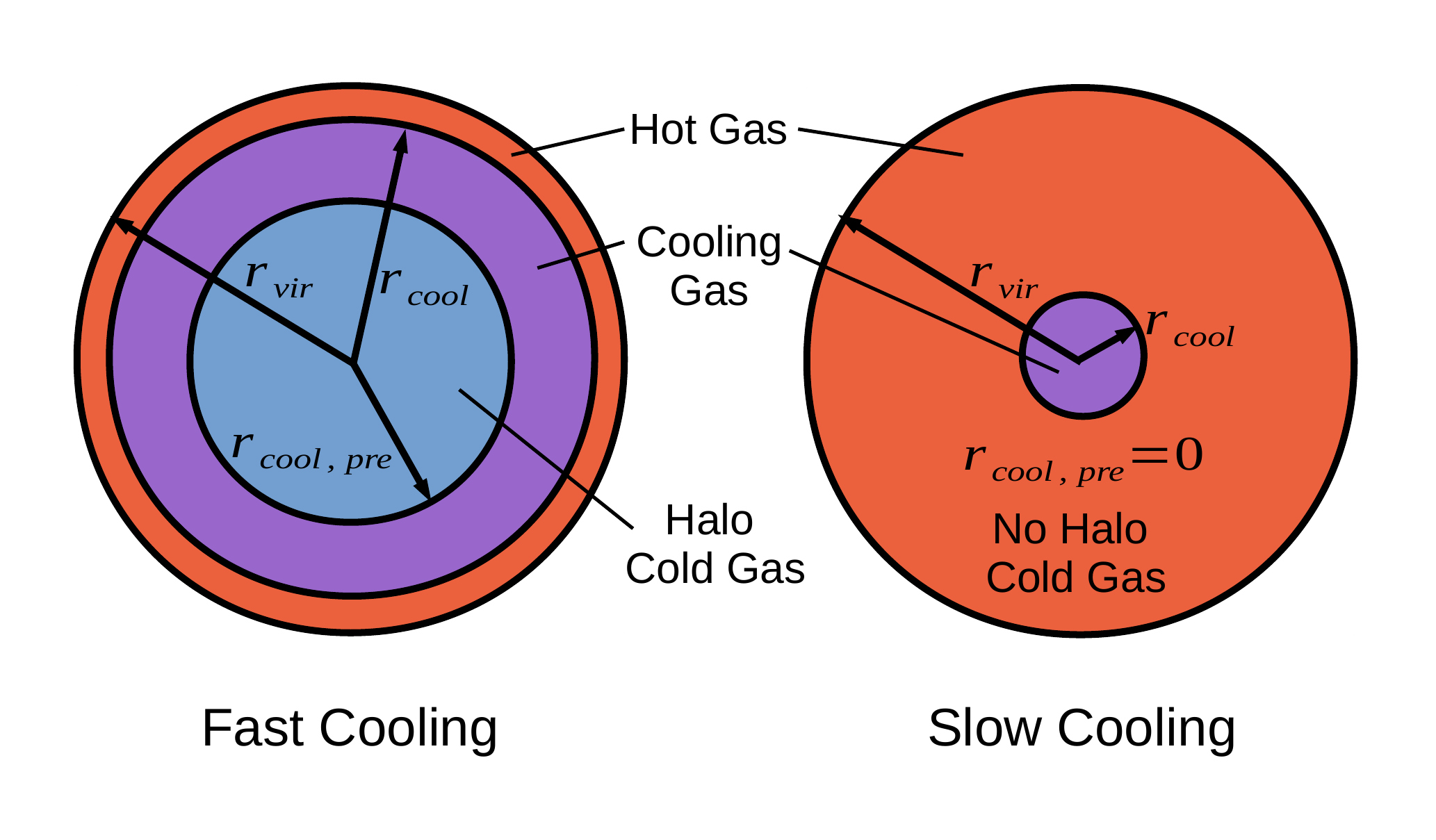}
\caption{Sketch of the new cooling model.}

\label{fig:new_cooling_picture}
\end{figure*}

The boundary between the cold gas halo and the hot gas halo is the
so-called cooling radius, $r_{\rm cool}$, at which the gas just has
enough time to cool down [the mathematical definition of $r_{\rm cool}$ is given in equation~(\ref{eq:r_cool_eq})]. When discrete timesteps are used, we
introduce another quantity, $r_{\rm cool,pre}$, which is the boundary
at the beginning of a timestep. The hot gaseous halo is treated as
fixed during a timestep, $r_{\rm cool}$ is calculated based on this
fixed halo, and the gas between $r_{\rm cool,pre}$ and $r_{\rm cool}$
cools down in this timestep, and is called the cooling gas. Note that
$r_{\rm cool,pre}$ is identical to $r_{\rm cool}$ calculated in the
previous timestep only if there is no contraction of the hot gas
halo. This picture is sketched in Fig.~\ref{fig:new_cooling_picture}.

The above scheme is similar to that in \citet{WF_1991} and to those in
many other semi-analytic models, but most of these other models
(apart from \MORGANA) do not explicitly introduce the cold gas halo
component or the contraction of the hot gas halo. Unlike the \MORGANA
model, in which the whole hot gas halo contributes to the cooled down
gas in any timestep, here the hot gas cools gradually from halo center
outwards. A more detailed discussion of the relation of the new
cooling model to those in other semi-analytic models is given in
\S\ref{sec:other_models}.

%---------------------------------------------------------------------------------------------------

\subsubsection{Basic assumptions of the new cooling model} \label{sec:new_cool_assumption}
Based on the above picture, we impose our basic assumptions about the
cooling as follows:
\begin{enumerate}
\item The hot gas in a dark matter halo is in a spherical hot gas
halo, with a density distribution described by the so-called
$\beta$-distribution:
\begin{equation}
\rho_{\rm hot}(r)\propto \frac{1}{r^2+r_{\rm core}^2},\ r_{\rm cool,pre}\leq r\leq r_{\rm vir},
\end{equation}
where $r_{\rm core}$ is called the core radius and is a parameter of
this density distribution, while $r_{\rm vir}$ is the virial radius of
the dark matter halo, defined as
\begin{equation}
r_{\rm vir} = \left( \frac{3 M_{\rm halo}}{4\pi \Delta_{\rm vir}
    \overline\rho} \right)^{1/3} ,
\label{eq:rvir}
\end{equation}
where $\overline\rho$ is the mean density of the universe at that
redshift, and the overdensity, 
$\Delta_{\rm vir}(\Omega_{\rm m},\Omega_{\rm v})$, is calculated from
the spherical collapse model \citep[e.g.][]{Eke1996}. In \GALFORM,
typically $r_{\rm core}$ is set to be a fixed fraction of
$r_{\rm vir}$ or of the NFW scale radius $r_{\rm NFW}$
\citep{Navarro1997}.

\item The hot gas has only one temperature at any time, and it is set
to be the dark matter halo virial temperature $T_{\rm vir}$, where
\begin{equation}
T_{\rm vir} = \frac{\mu_{\rm m} V_{\rm vir}^2}{2 k_{\rm B}} ,
\label{eq:Tvir}
\end{equation}
where $k_{\rm B}$ is the Boltzmann constant, $\mu_{\rm m}$ is the mean
mass per particle, and
$V_{\rm vir} = (G M_{\rm halo}/r_{\rm vir})^{1/2}$ is the circular
velocity at $r_{\rm vir}$.

\item When new gas is added to the hot gas halo, it is assumed to mix homogeneously
with the existing hot gas halo. This also means that the
hot gas halo has a single metallicity, $Z_{\rm hot}$, at any given
time.

\item In the absence of cooling, the specific angular momentum
distribution of the hot gas, $j_{\rm hot}(r) \propto r$, corresponding
to a mean rotation velocity in spherical shells that is constant with
radius. This applies to the initial time when no cooling has
happened and also to the gas newly added to the hot gas halo, which is
newly heated up. When cooling induces contraction of the hot gas halo,
the angular momentum of each Lagrangian hot gas shell is conserved
during the contraction, and after this, the rotation velocity is no longer a constant with radius.
\end{enumerate}
Our choices of $\rho_{\rm hot}(r)$ and of the initial $j_{\rm hot}(r)$
follow those of \citet{cole2000}, which are based on hydrodynamical
simulations without cooling. This is reasonable because here they
only apply to the hot gas.

%---------------------------------------------------------------------------------------------------

\subsubsection{Cooling calculation}
We describe the calculation for a single timestep, starting at time
$t$ and ending at time $t+\Delta t$.  The timestep, $\Delta t$, should
generally be chosen to be small compared to the halo dynamical
timescale, so that the evolution in the halo mass and the contraction
of the hot gas halo over a timestep are small. At the beginning of
each step, $M_{\rm halo}$ is updated according to the halo merger tree,
and $r_{\rm vir}$ and $T_{\rm vir}$ are then updated according to the
current values of $\Delta_{\rm vir}$ and $\overline\rho$. Next, the
hot gas density profile, $\rho_{\rm hot}(r,t)$, is updated, which
involves two quantities, namely $r_{\rm core}$ and the density
normalization. As mentioned above, $r_{\rm core}$ is calculated from
the halo radius $r_{\rm vir}$ or $r_{\rm NFW}$. The normalization is
fixed by the integral
\begin{equation}
4\pi\int_{r_{\rm cool,pre}(t)}^{r_{\rm vir}(t)}\rho_{\rm
  hot}(r,t) \, r^2dr=M_{\rm hot}(t) ,
\end{equation}
where $M_{\rm hot}$ is the total hot gas mass, and $r_{\rm cool,pre}$
the inner boundary of the hot gas halo at time $t$. Initially $r_{\rm
cool,pre}=0$ and is updated (see below) in each timestep for the
calculation of the next timestep. For a static halo, $r_{\rm
  cool,pre}(t) = r_{\rm cool}(t)$, but this no longer applies if the
halo grows or the hot gas distribution contracts.

With the density profile determined, the cooling radius $r_{\rm
cool}(t+\Delta t)$ at the end of the timestep can be
calculated. $r_{\rm cool}$ is defined by
\begin{equation}
t_{\rm cool}(r_{\rm cool},t+\Delta t)=\tilde{t}_{\rm cool,avail}(r_{\rm
  cool},t+\Delta t) ,
\label{eq:r_cool_eq}
\end{equation}
where $t_{\rm cool}(r,t)$ is the cooling timescale of a shell at
radius $r$ at time $t$, and $\tilde{t}_{\rm cool,avail}(r,t)$ is the time
available for cooling for that shell. $t_{\rm cool}(r,t)$ is defined
as
\begin{equation}
t_{\rm cool}=\frac{\delta U}{\delta L_{\rm cool}} 
=\frac{3k_{\rm B}}{2\mu_{\rm m}}\frac{T_{\rm vir}}{\tilde\Lambda(T_{\rm
    vir},Z)\rho_{\rm hot}} ,
\label{eq:t_cool}
\end{equation}
where $\delta U$ is the total thermal energy of this shell, while
$\delta L_{\rm cool}$ is its current cooling luminosity. For gas with
temperature $T_{\rm vir}$ and metallicity $Z_{\rm hot}$, we express
the thermal energy density as $(3/2) (\rho_{\rm hot}/\mu_{\rm m})
k_{\rm B} T_{\rm vir}$, and the radiative cooling rate per unit volume
as $\tilde\Lambda(T_{\rm vir},Z_{\rm hot})\rho_{\rm hot}^2$, assuming
collisional ionization equilibrium. This then
leads to the final expression on the RHS above.

The calculation of the time available for cooling, $\tilde{t}_{\rm
cool,avail}(r,t)$, is more complicated. For a halo in which the hot
gas density distribution, temperature and metallicity are static, and
in which the gas started cooling at a halo formation time $t_{\rm form}$, we would
define $\tilde{t}_{\rm cool,avail} = t - t_{\rm form}$, as in
\citet{cole2000}. However, this definition is not applicable to an
evolving halo. Instead, we would like to define a gas shell as having
cooled when $\delta U = \delta E_{\rm cool}$, where $\delta U$ is
defined as above, and $\delta E_{\rm cool}$ is the total energy
that would have been radiated away by this hot gas shell over its past history when we
track the shell in a Lagrangian sense. When we calculate $\delta U$
and $\delta E_{\rm cool}$ for a gas shell, we include the effects of
evolution in $\rho_{\rm hot}$, $T_{\rm vir}$ and $Z_{\rm hot}$ due to
halo growth, reaccretion of ejected gas and contraction of the hot
gas. However, in our approach, $\rho_{\rm hot}$ and $T$ in a gas shell
are assumed to be unaffected by radiative cooling within that shell,
up until the time when the cooling condition is met, when the hot
gas shell is assumed to lose all of its thermal energy in a single
instant, and be converted to cold gas. Combining the condition $\delta
U = \delta E_{\rm cool}$ with equation~(\ref{eq:t_cool}) then leads to a
cooling condition of the form $t_{\rm cool}(r,t)=\tilde{t}_{\rm
cool,avail}(r,t)$ if $\tilde{t}_{\rm cool,avail}$ for a shell is defined as
\begin{equation}
\tilde{t}_{\rm cool,avail}=\frac{\delta E_{\rm cool}}{\delta L_{\rm cool}}. 
\label{eq:t_cool_avail_def}
\end{equation}
This is just the time that it would take for the gas shell to radiate
the energy actually radiated over its past history, if it were
radiating at its current rate.  Note that for a static halo cooling
since time $t_{\rm form}$, $L_{\rm cool}$ is constant over the past
history of a hot gas shell, so
$\delta E_{\rm cool} = \delta L_{\rm cool} \, (t-t_{\rm form})$, and
the above definition reduces to
$\tilde{t}_{\rm cool,avail}= t - t_{\rm form}$.

The quantity $t_{\rm cool}$ is easy to calculate for each hot gas
shell because it only involves quantities at time $t$. In contrast,
the calculation of $\tilde{t}_{\rm cool,avail}$ is more difficult, because
$\delta E_{\rm cool}$ involves the previous cooling history. To
calculate $\tilde{t}_{\rm cool,avail}$ exactly, the cooling history of each
Lagrangian hot gas shell would have to be stored. However, this is too
computationally expensive for a semi-analytic model, and some
further approximations are needed.  We first note that for a discrete
timestep of length $\Delta t$ and starting at $t$,
\begin{eqnarray}
\tilde{t}_{\rm cool,avail}(r_{\rm cool},t+\Delta t) &=& \tilde{t}_{\rm cool,avail}(r_{\rm cool},t)+\Delta t 
\label{eq:t_cool_avail_a} \\
&\approx& \tilde{t}_{\rm cool,avail}(r_{\rm cool,pre},t)+\Delta t .
\label{eq:t_cool_avail_b}
\end{eqnarray}
The first line above comes from the assumption that the hot gas halo
is fixed within a given timestep, and thus the increase of
$\tilde{t}_{\rm cool,avail}$ over the step is just the increase of the
physical time. To justify the approximation in the second line,
we consider two cases: (a) $r_{\rm cool} \sim r_{\rm cool,pre}$. In
this case, which typically happens when the gas cools slowly compared
to the halo dynamical timescale,
$\tilde{t}_{\rm cool,avail}(r_{\rm cool},t) \approx \tilde{t}_{\rm cool,avail}(r_{\rm
  cool,pre},t)$. (b) $r_{\rm cool} \gg r_{\rm cool,pre}$. This
typically happens when the gas cools fast compared to the halo
dynamical timescale, but in that case, halo growth and hot gas halo contraction play only a
weak role in cooling, which means that $\tilde{t}_{\rm cool,avail}$ is nearly
the same for all gas shells (as in a completely static halo), so again
$\tilde{t}_{\rm cool,avail}(r_{\rm cool},t) \approx \tilde{t}_{\rm cool,avail}(r_{\rm
  cool,pre},t)$.

Finally, we make the approximation 
\begin{equation}
\tilde{t}_{\rm cool,avail}(r_{\rm cool,pre},t) 
= \frac{\delta E_{\rm cool}(r_{\rm cool,pre},t)}{\delta L_{\rm cool}(r_{\rm cool,pre},t)} 
\approx \frac{E_{\rm cool}(t)}{L_{\rm cool}(t)}
\label{eq:t_cool_avail_c}
\end{equation}
Here, $L_{\rm cool}$ is the cooling luminosity of the whole hot gas
halo at time $t$,
\begin{equation}
L_{\rm cool}(t) = 4\pi 
\int_{r_{\rm cool,pre}}^{r_{\rm vir}} 
\tilde\Lambda(T_{\rm vir},Z_{\rm hot}) \,
\rho_{\rm hot}^2(r,t)\, r^2 \, dr ,
\label{eq:new_cool_L_cool1}
\end{equation}
and $E_{\rm cool}(t)$ is the total energy radiated away over its past
history by all of the hot gas that is within the halo at time $t$, 
\begin{equation}
E_{\rm cool}(t) = 4\pi\int_{t_{\rm init}}^{t}\int_{r_{\rm p}(\tau)}^{r_{\rm
    vir}(\tau)} 
\tilde \Lambda(T_{\rm vir},Z_{\rm hot}) \, \rho_{\rm hot}^2(r,\tau) \, r^2
    dr d\tau .
\label{eq:new_cool_E_cool1}
\end{equation}
In the above integral, $t_{\rm init}$ is the starting time for the
cooling calculation, and $r_{\rm p}(\tau)$ is the radius at time
$\tau$ of the shell that has radius $r_{\rm cool,pre}$ at time $t$.

To justify the approximation made in equation~(\ref{eq:t_cool_avail_c}), we
first note that, due to the integrals in
equations~(\ref{eq:new_cool_L_cool1}) and (\ref{eq:new_cool_E_cool1})
involving $\rho_{\rm hot}^2$, both are dominated by the densest
regions in the hot gas halo. We now need to consider two cases. (a)
$r_{\rm cool,pre} \gsim r_{\rm core}$. In this case, the gas density
decreases monotonically for $r\gsim r_{\rm core,pre}$, so that both
integrals are dominated by the contributions from the gas shells near
the lower limit of the integral, i.e.\ near $r_{\rm cool,pre}$. It
follows that
$E_{\rm cool}(t)/L_{\rm cool}(t) \approx \delta E_{\rm cool}(r_{\rm
  cool,pre},t)/\delta L_{\rm cool}(r_{\rm cool,pre},t)$. (b)
$r_{\rm cool,pre} \lsim r_{\rm core}$. In this case,
$\delta E_{\rm cool}(r,t)/\delta L_{\rm cool}(r,t)$ is approximately
independent of radius for $r \lsim r_{\rm core}$ due to the
approximately constant density, while the integrals for
$E_{\rm cool}(t)$ and $L_{\rm cool}(t)$ are dominated by the region
$r \lsim r_{\rm core}$, so that we again have
$E_{\rm cool}(t)/L_{\rm cool}(t) \approx \delta E_{\rm cool}(r_{\rm
  cool,pre},t)/\delta L_{\rm cool}(r_{\rm cool,pre},t)$.

By combining equations~(\ref{eq:t_cool_avail_b}) and
(\ref{eq:t_cool_avail_c}), we obtain the expression for
$t_{\rm cool,avail}$ that we actually use:
\begin{eqnarray}
t_{\rm cool,avail}(t+\Delta t) & = & \frac{E_{\rm cool}(t)}{L_{\rm cool}(t)}+\Delta t \nonumber \\
                               & \approx & \tilde{t}_{\rm cool,avail}(r_{\rm cool},t+\Delta t) ,
\label{eq:t_cool_avail}
\end{eqnarray}
In the above, the term $E_{\rm cool}(t)/L_{\rm cool}(t)$
represents the available time at the start of the step, calculated
from the previous cooling history.

The calculation of $E_{\rm cool}$ from equation~(\ref{eq:new_cool_E_cool1})
appears to require storing the histories of all of the shells of hot
gas in order to evaluate the integral. However, from its definition,
it is easy to derive an approximate recursive equation for it (see Appendix~\ref{app:app_E_cool})
\begin{eqnarray}
E_{\rm cool}(t+\Delta t) & \approx & E_{\rm cool}(t)+L_{\rm cool}(t)\times \Delta t \nonumber \\
                         & - & L'_{\rm cool}(t)\times t_{\rm
                         cool,avail}(t+\Delta t) ,
\label{eq:recursive_E_cool}
\end{eqnarray}
where
\begin{equation}
L'_{\rm cool}(t)
=4\pi\int_{r_{\rm cool,pre}}^{r_{\rm cool}}
\tilde\Lambda(T_{\rm vir},Z_{\rm hot}) \, \rho_{\rm hot}^2 \, r^2 \,
dr .
\label{eq:L_cool_p}
\end{equation}
The second term in equation~(\ref{eq:recursive_E_cool}) adds the energy
radiated away in the current timestep, while the third term removes
the contribution from gas between $r_{\rm cool,pre}$ and $r_{\rm
cool}$, because it cools down in the current timestep and therefore
is not part of the hot gas halo at the next timestep. Starting from
the initial value $E_{\rm cool}=0$, equation~(\ref{eq:recursive_E_cool}) can
be used to derive $E_{\rm cool}$ for the subsequent timesteps, and
then equations~(\ref{eq:r_cool_eq}), (\ref{eq:t_cool}) and
(\ref{eq:t_cool_avail}) can be used to calculate $r_{\rm cool}$. For
a static halo, in which there is no accretion and no contraction of
the hot gas, it can be shown that
equations~(\ref{eq:t_cool_avail})-(\ref{eq:L_cool_p}) lead to $t_{\rm
avail}(t + \Delta t) = t + \Delta t - t_{\rm init}$, the same as in
\citet{cole2000}.

With $r_{\rm cool,pre}$ and $r_{\rm cool}$ determined, the mass and angular
momentum of the gas cooled down over the time interval $(t,t+\Delta
t)$ are calculated from
\begin{eqnarray}
\Delta M_{\rm cool} & = & 4\pi\int_{r_{\rm cool,pre}}^{r_{\rm cool}}
\rho_{\rm hot}r^2dr \\
\Delta J_{\rm cool} & = & 4\pi\int_{r_{\rm cool,pre}}^{r_{\rm cool}}
j_{\rm hot}\rho_{\rm hot}r^2dr ,
\end{eqnarray}
where $j_{\rm hot}(r)$ is the specific angular momentum distribution
of the hot gas, which is calculated as described in
\S\ref{sec:j_calculation}. $\Delta M_{\rm cool}$ and $\Delta J_{\rm
cool}$ are used to update the mass, $M_{\rm halo,cold}$, and angular
momentum, $J_{\rm halo,cold}$, of the cold halo gas component.

Gas in the cold halo gas component is not pressure supported, and so
is assumed to fall to the central galaxy in the halo on the freefall
timescale. We therefore calculate the mass, $\Delta M_{\rm acc,gal}$,
and angular momentum, $\Delta J_{\rm acc,gal}$, accreted onto the
central galaxy over a timestep as
\begin{eqnarray}
\Delta M_{\rm acc,gal} & = & M_{\rm halo,cold}\times 
\min [1,\Delta t/t_{\rm ff}(r_{\rm cool})] 
\label{eq:new_cool_m_acc_gal} \\
\Delta J_{\rm acc,gal} & = & J_{\rm halo,cold}\times 
\min [1,\Delta t/t_{\rm ff}(r_{\rm cool})]
\end{eqnarray}
where $t_{\rm ff}(r_{\rm cool})$ is the free-fall time scale at the 
cooling radius. Note that in the slow cooling regime, where $t_{\rm
ff}(r_{\rm cool}) < t_{\rm cool}(r_{\rm cool})$, the mass of the
cold halo gas component remains relatively small, since the timescale
for draining it ($t_{\rm ff}$) is short compared to the timescale for
feeding it ($t_{\rm cool}$).

Note that here we treat the angular momentum of the cooled down gas as a scalar. This means that the axis of the galaxy spin is assumed to be always aligned with the axis of the hot gas halo spin. We adopt this assumption mainly because the halo spin parameter, which is the basis of the calculation of hot gas angular momenta, only contains information on the magnitude of the angular momentum. This is an important limitation, and a calculation of the angular momentum of the hot gas considering both its magnitude and direction should be developed. However, this is beyond the scope of this paper, and we leave it for future work.

Finally, we consider the contraction of the hot gas halo. The gas
between the cooling radius and the virial radius is assumed to remain
in approximate hydrostatic equilibrium, so for simplicity we assume
that it always follows the $\beta$-profile. The hot gas at the cooling
radius is not pressure-supported by the cold gas at smaller radii, so
we assume that this gas contracts towards the halo centre on a
timescale $t_{\rm ff}(r_{\rm cool})$.  The new $r_{\rm cool,pre}$ at
the next timestep starting at $t+\Delta t$ is therefore estimated as
\begin{equation}
r_{\rm cool,pre}(t+\Delta t)=r_{\rm cool}(t+\Delta t) \times 
\max [0,1-\Delta t/t_{\rm ff}(r_{\rm cool})].
\end{equation}
The above equation only applies if the gravitational potential of the
halo is fixed. When the halo grows in mass, and when the mean halo
density within $r_{\rm vir}$ adjusts with the mean density of the
universe, the gravitational potential also changes, and this affects the
contraction of the hot gas halo. We estimate the effect of this on the
inner boundary of the hot halo gas by requiring that
the mass of dark matter contained inside $r_{\rm cool,pre}$ remains
the same before and after the change in the halo potential, i.e.\
\begin{equation}
M'_{\rm halo}[r'_{\rm cool,pre}(t+\Delta t)]=
M_{\rm halo}[r_{\rm cool,pre}(t+\Delta t)] ,
\label{eq:new_cool_shrink_halo_growth}
\end{equation}
where the quantities with apostrophes are after halo growth, while
those without apostrophes are before halo growth. The reason for using
the dark matter to trace this contraction is that the gas within
$r_{\rm cool,pre}$ is cold with negligible pressure effects, so its
dynamics should be similar to those of the collisionless dark matter.

%---------------------------------------------------------------------------------------------------

\subsubsection{Calculating $j_{\rm hot}(r)$} 
\label{sec:j_calculation}
The specific angular momentum of the hot gas averaged over spherical
shells is assumed to follow $j_{\rm hot}(r) \propto r$ at the initial
time, as stated in \S\ref{sec:new_cool_assumption}, with the
normalization set by the assumption that the mean specific angular
momentum of the hot gas in the whole halo, $\overline{j}_{\rm hot}$,
is initially equal to that of the dark matter, $J_{\rm halo}/M_{\rm halo}$ (see
\S\ref{sec:halo_spin}). Later on, the dark matter halo growth, the contraction 
of the hot gas halo and the addition of new gas all can change the angular momentum
profile. In this new cooling model, at the beginning of each timestep, we first 
consider the angular momentum profile change of the existing hot gas due to the hot gas halo contraction and the 
dark matter halo growth that took place during the last 
timestep, and then add the contribution from the newly added hot gas to this 
adjusted profile.

In deriving the change of angular momentum profile of the existing hot gas, 
we assume mass and angular momentum conservation for each Lagrangian shell. 
Consider a shell with mass $dm$, original radius $r$ and specific angular 
momentum $j_{\rm hot}(r)$, which, after the dark matter growth and hot gas 
halo contraction, moves to radius $r'$ with specific angular momentum $j'_{\rm hot}(r')$. 
The shell mass is unchanged because of mass conservation. Then angular 
momentum conservation implies $j'_{\rm hot}(r')=j_{\rm hot}(r)$. In other 
words, the angular momentum profile after these changes is $j_{\rm hot}[r(r')]$. 
Given $j_{\rm hot}(r)$ from the last timestep, the major task for 
deriving $j'_{\rm hot}(r')$ is to derive $r(r')$. This can be done by considering 
shell mass conservation and the density profiles of the hot gas. Specifically, 
assuming $\rho_{\rm hot}(r)$ and $\rho'_{\rm hot}(r')$ are respectively the 
density profiles of the existing hot gas before and after the dark matter 
halo growth and hot gas halo contraction, then one has
\begin{equation}
4\pi\rho_{\rm hot}(r)r^2dr = dm = 4\pi\rho'_{\rm hot}(r')r'^2dr'.
\label{eq:m_conservation}
\end{equation}
This, together with the assumption that $\rho_{\rm hot}(r)$ and $\rho'_{\rm hot}(r')$ 
follow the $\beta$-distribution, can then be solved for $r(r')$. Unfortunately, this
equation can only provide an implicit form for $r'(r)$, and does not
lead to an explicit analytical expression for $j'_{\rm hot}(r')$. A
straightforward way to deal with this is to evaluate $j'_{\rm
hot}(r')$ numerically for a grid of radii and then store this
information, however, this is computationally expensive. Instead, we
apply further approximations to reduce the computational cost of
solving for $j'_{\rm hot}(r')$, as described in detail in
Appendix~\ref{app:app_j}.

To derive the final angular momentum distribution, $j''_{\rm hot}(r')$, one still 
needs to consider the contribution from the newly added hot gas. Assuming the gas 
newly added to a given shell with radius $r'$ has mass $dm_{\rm new}$ and specific 
angular momentum $j_{\rm new}(r')$, then one has
\begin{equation}
j''_{\rm hot}(r')(dm+dm_{\rm new})=j'_{\rm hot}(r')dm+j_{\rm new}(r')dm_{\rm new}.
\end{equation}

Since the newly added gas is assumed to be mixed
homogeneously with the hot gas halo, so all $dm_{\rm new}/(dm+dm_{\rm new})$ should be
the same for all shells, and hence
\begin{equation}
\frac{dm_{\rm new}}{dm_{\rm new}+dm}=\frac{M_{\rm new}}{M_{\rm new}+M_{\rm old}} ,
\label{eq:mass_ratio1}
\end{equation} 
where $M_{\rm new}$ is the total mass added to the hot gas halo during
the timestep, while $M_{\rm old}$ is the previous
mass.

Further, according to the assumption in \S\ref{sec:new_cool_assumption}, $j_{\rm new}(r') \propto r'$. 
In general, there are two components to the newly added hot gas: (a)
gas brought in through growth of the dark matter halo; and (b) gas
that has been ejected from the galaxy by SN feedback, has joined the
ejected gas reservoir, and then has been reaccreted into the hot gas
halo. Their contributions to the total angular momentum of the newly added 
gas are described in \S\ref{sec:merger_and_ejected_gas_calculation}. With this, 
the normalization of $j_{\rm new}(r')$ can be determined.

Finally, with $j'_{\rm hot}(r')$ and $j_{\rm new}(r')$ known, 
the specific angular momentum distribution at the current timestep, 
$j''_{\rm hot}(r')$, is determined as
\begin{equation}
j''_{\rm hot}(r')(M_{\rm new}+M_{\rm old})=j'_{\rm hot}(r')M_{\rm old}+j_{\rm new}(r')M_{\rm new}.
\label{eq:j_recursive}
\end{equation}
In this way, the specific angular momentum distribution for any 
given timestep can be derived recursively from the initial distribution.

%---------------------------------------------------------------------------------------------------

\subsubsection{Treatments of gas ejected by feedback and halo mergers}
\label{sec:merger_and_ejected_gas_calculation}

The SN feedback can heat and eject gas in galaxies, and the ejected gas is added to 
the so-called ejected gas reservoir. This transfers mass and angular momentum from 
galaxies to that reservoir. The gas ejected from both the central galaxy and its 
satellites is added to the ejected gas reservoir of the central galaxy. The ejected 
mass is determined by the SN feedback prescription, and is typically proportional to the 
instantaneous star formation rate. The angular momentum of this ejected gas is 
calculated as follows. 

The total angular momentum of the ejected gas can be expressed 
as the product of its mass and its specific angular momentum. For the gas ejected 
from the central galaxy, its specific angular momentum is estimated as that of the 
central galaxy, while for the gas ejected from satellites, its specific angular 
momentum is estimated as the mean specific angular momentum of the central galaxy's 
host dark matter halo, i.e.\ $J_{\rm halo}/M_{\rm halo}$, in order roughly to include 
the contribution to the ejected angular momentum from the satellite orbital motion. This is only a rough estimate. A better estimate would be obtained by following the satellite orbit, but we leave this for future work.

This ejected gas can later be reaccreted onto the hot gas halo, thus delivering mass 
and angular momentum to it. The reaccretion rates of mass and angular momentum are 
respectively estimated as
\begin{eqnarray}
\dot{M}_{\rm return} & = & \alpha_{\rm return}\times M_{\rm eject}/t_{\rm dyn} \\
\dot{J}_{\rm return} & = & \alpha_{\rm return}\times J_{\rm eject}/t_{\rm dyn},
\end{eqnarray}
where $\dot{M}_{\rm return}$ and $\dot{J}_{\rm return}$ are respectively the mass 
and angular momentum reaccretion rates, $M_{\rm eject}$ and $J_{\rm eject}$ are 
respectively the total mass and angular momentum of the ejected gas reservoir, 
$t_{\rm dyn}=r_{\rm vir}/v_{\rm vir}$ is the halo dynamical timescale 
and $\alpha_{\rm return}\sim 1$ a free parameter. For a timestep of finite 
length $\Delta t$, the mass and angular momentum reaccreted within it is then 
calculated as the products of the corresponding rates and $\Delta t$.

When a halo falls into a larger halo, it becomes a subhalo, while the larger one 
becomes the host halo of this subhalo. The halo gas in the subhalo could be ram-pressure or tidally stripped. This process can be calculated within the semi-analytic framework [see e.g.\ \citet{font2008_ram_pressure} or \citet{munich_model_Guo11}], but here we assume for simplicity that the relevant gas is instantaneously removed on infall. The new cooling model assumes that the hot 
gas and ejected gas reservoir associated with this subhalo are instantaneously 
transferred to the corresponding gas components of the host halo at infall. The masses of 
these transferred components can be simply added to the corresponding components 
of the host halo. However, the angular momentum cannot be directly added, because 
it is calculated before infall, when the subhalo was still an isolated halo, 
and the reference point for this angular momentum is the centre of the subhalo, 
while after the transition, the reference point becomes the centre of the host 
halo. 

Here the angular momentum transferred is estimated as follows. The total 
angular momentum transferred is expressed as a product of the total transferred 
mass and the specific angular momentum. The latter one is estimated 
as $j_{\rm new,halo}=\Delta J_{\rm halo}/\Delta M_{\rm halo}$, 
where $\Delta J_{\rm halo}$ and $\Delta M_{\rm halo}$ are the angular momentum and 
mass changes in dark matter halo during the halo merger, and they can be determined 
when the mass and spin, $\lambda_{\rm halo}$, of each halo in a merger tree are 
given (see \S\ref{sec:halo_spin}). The reason for this estimation is that the dark 
matter and baryon matter accreted by the host halo have roughly the same motion, and 
thus should gain similar specific angular momentum through the torque exerted by the surrounding 
large-scale structures. The mass and angular momentum transferred during the 
halo merger can be summarized as:
\begin{eqnarray}
\Delta M_{\rm hot,mrg} & = & \sum_{i=1}^{N_{\rm mrg}}M_{\rm hot,i}, \\
\Delta J_{\rm hot,mrg} & = & j_{\rm new,halo}\times \Delta M_{\rm hot,mrg}, \\
\Delta M_{\rm eject,mrg} & = & \sum_{i=1}^{N_{\rm mrg}}M_{\rm eject,i}, \\
\Delta J_{\rm eject,mrg} & = & j_{\rm new,halo}\times \Delta M_{\rm eject,mrg},
\end{eqnarray}
where $\Delta M_{\rm hot,mrg}$ and $\Delta J_{\rm hot,mrg}$ are respectively the 
total mass and angular momentum transferred to the hot gas halo of the host halo 
during the halo merger, while $\Delta M_{\rm eject,mrg}$ and $\Delta J_{\rm eject,mrg}$ 
are the mass and angular momentum transferred to the ejected gas reservoir; 
$N_{\rm mrg}$ is the total number of infalling haloes over the timestep, 
$M_{\rm hot,i}$ is the total mass of the hot gas halo of the $i$th infalling 
halo, and $M_{\rm eject,i}$ is the mass of its ejected gas reservoir.

In this cooling model, by default, the halo cold gas is not transferred during 
halo mergers, because it is cold and in the central region of the 
infalling halo, and thus is less affected by ram pressure and tidal stripping. 
After infall, this cold gas halo can still deliver cold gas to the satellite for 
a while. There are also options in the code to transfer the halo cold gas to 
the hot gas halo or halo cold gas of the host halo. In this work, we always 
adopt the default setting.

A dark matter halo may also accrete smoothly. The accreted gas is assumed to 
be shock heated and join the hot gas halo. In each timestep, the mass of this 
gas, $\Delta M_{\rm hot,smooth}$, is given as 
$\Delta M_{\rm hot,smooth}=[\Omega_{\rm b}/\Omega_{\rm m}]\Delta M_{\rm halo, smooth}$, 
with $\Delta M_{\rm halo, smooth}$ the mass of smoothly accreted dark matter, 
which is provided by the merger tree, while the associated angular momentum is 
estimated as $\Delta J_{\rm hot,smooth}=j_{\rm new,halo}\times \Delta M_{\rm hot,smooth}$.

In each timestep, $\Delta M_{\rm return}$, $\Delta M_{\rm hot,mrg}$ 
and $\Delta M_{\rm hot,smooth}$ increase the mass of the hot gas halo, 
but do not increase $E_{\rm cool}$. This means the newly added gas has 
no previous cooling history, consistently with the assumption that this 
gas is newly heated up by shocks. The total angular momentum of this 
newly added gas is $\Delta J_{\rm return}+\Delta J_{\rm hot,mrg}+\Delta J_{\rm hot,smooth}$, 
and, together with the assumption that $j_{\rm new}(r)\propto r$, it 
completely determines the specific angular momentum distribution of 
the newly added gas.

%====================================================================================================================================================

\subsection{Previous cooling models} 
\label{sec:other_models}

\subsubsection{\GALFORM cooling model GFC1} 
\label{sec:other_models_GFC1}

The GFC1 (GalForm Cooling 1) cooling model is used in all recent versions of \GALFORM
\citep[e.g.][]{galform_gonzalez2014, galform_lacey2015}, and is based
on the cooling model introduced in \citet{cole2000}, and modified in
\citet{galform_bower2006}. The \citet{cole2000} cooling model
introduced so-called halo formation events. These are defined such
that the appearance of a halo with no progenitor in a merger tree is a
halo formation event, and the time when a halo first becomes at least
twice as massive as at the last halo formation event is a new
halo formation event. The \citeauthor{cole2000} model then assumes that the
hot gas halo is set between two adjacent halo formation events, and
is reset at each formation event. Under this assumption, $\tilde{t}_{\rm
cool,avail}(r_{\rm cool},t)$ is always the time elapsed since the latest halo
formation event, which is straightforward to calculate. As in the 
new cooling model, we denote the actual $\tilde{t}_{\rm cool,avail}(r_{\rm cool},t)$ 
used in this model as $t_{\rm cool,avail}(t)$. With $t_{\rm cool,avail}$ 
given, $r_{\rm cool}$ can be then calculated from
equation~(\ref{eq:r_cool_eq}), and the mass and angular momentum cooled
down can be calculated as described below. The assumption of a fixed
hot gas halo between two halo formation events means that changes in
$r_{\rm vir}$ and $T_{\rm vir}$ induced by halo growth, and by the
addition of new hot gas either by halo growth or by the
reincorporation of gas ejected by feedback between halo formation
events, are not considered until the coming of a halo formation
event. While this may be reasonable for halo formation events induced
by halo major mergers, in which the hot gas halo properties change
fairly abruptly, it is not physical if the halo formation event is
triggered through smooth halo growth, in which case the changes in the
hot gas halo should also happen smoothly, instead of happening in a
sudden jump at the halo formation event.

The GFC1 model \citep{galform_bower2006} improves the \citeauthor{cole2000} model by updating
some hot gas halo properties at each timestep instead of only at halo
formation events. Specifically, the hot gas is still
assumed to settle in a density profile described by the
$\beta$-distribution, with temperature equal to the current halo
virial temperature, $T_{\rm vir}$, and $r_{\rm core}$ set to be a fixed
fraction of the current $r_{\rm vir}$. The halo mass is updated at
each timestep, and the total hot gas mass and metallicity include the
contributions from the hot gas newly added at each timestep. However,
$V_{\rm vir}$ and $T_{\rm vir}$ are fixed at the values calculated at the
last halo formation event. Unlike in the new cooling model, the
normalization of the density profile is determined by requiring that
\begin{equation}
4\pi\int_{0}^{r_{\rm vir}}\rho_{\rm hot}r^2dr=M_{\rm hot}+M_{\rm
  cooled} ,
\label{eq:rho_norm_deter_GFC1}
\end{equation}
where $M_{\rm hot}$ is the total mass of the hot gas, while
$M_{\rm cooled}$ is the total mass of the gas that has cooled down from 
this halo since the last halo formation event, and is either in the central 
galaxy or ejected by SN feedback but not yet reaccreted by the hot gas 
halo. Accordingly, $M_{\rm cooled}$ is reset to $0$ at each halo formation 
event, while the ejected gas reservoir mass, $M_{\rm eject}$, evolves 
smoothly and is not affected by halo formation events.

This is not very physical 
because the cooled down gas might have collapsed onto the central
galaxy long ago, while the ejected gas is outside the halo. This also
means that the contraction of the hot gas halo due to cooling is
largely ignored in the determination of its density profile. This
point is most obvious in the case of a static halo, when the dark
matter halo does not grow. In this case, if there is no feedback 
and subsequent reaccretion, then the amount of hot gas gradually 
reduces due to cooling, and the hot gas halo should gradually contract 
in response to the reduction of pressure support caused by this 
cooling. However, in the GFC1 model, in this situation, the hot 
gas profile remains fixed, because $M_{\rm hot}+M_{\rm cooled}$ 
always equals the initial total hot gas mass. For a dynamical 
halo, $M_{\rm cooled}$ is reset to zero at each halo formation 
event, and thus the hot gas contracts to halo center at these 
events. In this way, the halo contraction due to cooling is included 
to some extent.

In the GFC1 model $r_{\rm cool}$ is calculated in the same way as in
\citet{cole2000}. For the estimation of $\tilde{t}_{\rm cool,avail}(r_{\rm cool},t)$, the GFC1
model retains the artificial halo formation events. This means that in
both the GFC1 and \citet{cole2000} cooling models, the hot gas cooling
history is effectively reset at each halo formation event. While this
might be physical when the halo grows through major
mergers\footnote{Although, \citet{monaco_2014_comp} suggests that halo
major mergers do not strongly affect cooling.}, it is artificial when a
halo grows smoothly, in which case the cooling history is expected to
evolve smoothly as well. Moreover, in principle $\tilde{t}_{\rm cool,avail}(r_{\rm cool},t)$
should change when the hot gas halo changes, which happens between
halo formation events in the GFC1 model, so estimating $\tilde{t}_{\rm
cool,avail}(r_{\rm cool},t)$ in the GFC1 model in the same way as in \citet{cole2000}
is not self-consistent.

Unlike the new cooling model that explicitly introduces a cold halo gas
component that drains onto the central galaxy on the free-fall
timescale, the GFC1 and \citet{cole2000} cooling models introduce a 
free-fall radius, $r_{\rm ff}$, to allow for the fact that gas
cannot accrete onto the central galaxy more rapidly than on a
free-fall timescale, no matter how rapidly it cools. $r_{\rm ff}$ is
calculated as
\begin{equation}
t_{\rm ff}(r_{\rm ff})=t_{\rm ff,avail} ,
\label{eq:rff}
\end{equation}
where $t_{\rm ff}(r)$ is the free-fall timescale at radius $r$,
defined as the time for a particle to fall to $r=0$ starting at rest
at radius $r$, and $t_{\rm ff,avail}$ is the time available for
free-fall, which is set to be the same as $t_{\rm cool,avail}$ in
these two cooling models. Then, the mass accreted onto the cental
galaxy over a timestep is given by
\begin{equation}
\Delta M_{\rm acc,gal}=4\pi\int_{r_{\rm infall,pre}}^{r_{\rm
    infall}}\rho_{\rm hot} \, r^2dr ,
\end{equation}
where $\rho_{\rm hot}$ is the current halo gas density distribution,
while $r_{\rm infall}=\min(r_{\rm cool},r_{\rm ff})$, and
$r_{\rm infall,pre}$ is determined by
$4\pi\int_{0}^{r_{\rm infall,pre}}\rho_{\rm hot}r^2dr=M_{\rm
  cooled}$.

The introduction of $r_{\rm ff}$ and $r_{\rm infall}$ leaves part of
the cooled down gas in the nominal hot gas halo when
$r_{\rm cool}>r_{\rm ff}$, which is the case in the fast cooling
regime. This gas is treated as hot gas in subsequent timesteps. While
in the fast cooling regime this should not strongly affect the final
results for the amount of gas that cools, due to the cooling and
accretion being rapid, this misclassification of cold gas as hot is
still an unwanted physical feature of a cooling model.

The calculation of the angular momentum of the gas accreted onto the
central galaxy is the same in the cooling model in \citet{cole2000}
and GFC1 model. The angular momentum is calculated as
\begin{equation}
\Delta J_{\rm acc,gal}=4\pi\int_{r_{\rm infall,pre}}^{r_{\rm
    infall}}j_{\rm hot}\rho_{\rm hot}r^2dr ,
\end{equation}
where $j_{\rm hot}$ is the specific angular momentum distribution of
the hot gas halo, which is assumed to vary as $j_{\rm hot} \propto
r$. As mentioned in \S\ref{sec:new_cool_assumption}, this assumption
is based on hydrodynamical simulations without cooling. Assuming
it applies unchanged in the presence of cooling means that the effect of
contraction of the hot gas halo due to cooling is ignored.

This model adopts treatments for the gas ejected by feedback and 
for halo mergers similar to those of the new cooling model. Since 
the GFC1 model assumes that $t_{\rm cool,avail}$ is always the physical 
time since the last halo formation event, here the gas newly added 
through halo growth and reaccretion of the feedback ejected gas would 
share this $t_{\rm cool,avail}$ and thus implicitly gain some previous 
cooling history. As a result, the newly added gas is effectively not 
actually newly heated up.

%---------------------------------------------------------------------------------------------------

\subsubsection{\GALFORM cooling model GFC2} 
\label{sec:gfc2_model}

The GFC2 (GalForm Cooling 2) model was introduced by \citet{benson_bower_2010_cooling}. It makes
several improvements over the GFC1 model. The
assumptions about the density profile\footnote{
  \citet{benson_bower_2010_cooling} actually adopt a different
  density profile for the hot gas halo; however, here for a fair
  comparison with other \GALFORM cooling models, the $\beta$-profile
  is adopted instead for this model.}, temperature and metallicity of
the hot gas halo are the same as in GFC1, but the influence of halo
formation events is mostly removed. The density profile of the hot gas
is normalized by requiring
\begin{equation}
4\pi\int_{0}^{r_{\rm vir}}\rho_{\rm hot}r^2dr=M_{\rm hot}+M_{\rm
  cooled}+M_{\rm eject} ,
\label{eq:rho_norm_deter_GFC2}
\end{equation}
where $M_{\rm eject}$ is the mass of gas ejected by SN feedback and not yet reaccreted, while the
definition of $M_{\rm cooled}$ is modified: (a) It is incremented by
the mass cooled and accreted onto the central galaxy, and reduced
by the mass ejected by SN feedback. (b) A 
gradual reduction of $M_{\rm cooled}$ as
\begin{equation}
\dot{M}_{\rm cooled}=-\alpha_{\rm remove}\times M_{\rm cooled}/t_{\rm ff}(r_{\rm vir}) ,
\label{eq:gfc2_mcooled_remove}
\end{equation}
with $\alpha_{\rm remove}\sim 1$ being a free parameter. (c) When a
halo merger occurs, the value of $M_{\rm cooled}$ is propagated to the
current halo from its most massive progenitor (rather than being reset
to 0 at each halo formation event as in the GFC1 model).  Since the
density profile normalization for the hot gas is determined by
equation~(\ref{eq:rho_norm_deter_GFC2}), for a given $M_{\rm hot}$ and
$M_{\rm eject}$, the gradual reduction of $M_{\rm cooled}$ due to
equation~(\ref{eq:gfc2_mcooled_remove}) lowers the normalization, and so to
include the same mass, $M_{\rm hot}$, in the density profile, the hot
gas must be distributed to smaller radii. This gradual reduction of
$M_{\rm cooled}$ thus effectively leads to a contraction of the hot
gas halo in response to the removal of hot gas by cooling, which is
more physical than the treatment in the GFC1 model. However, here the
timescale for this contraction is $t_{\rm ff}(r_{\rm vir})$, while the
region where the contraction happens has a radius $\sim r_{\rm cool}$,
so there is still a physical mismatch in this scale. This is
improved in the new cooling model introduced in \S\ref{sec:new_model},
where the timescale $t_{\rm ff}(r_{\rm cool})$ is adopted instead.

In the GFC2 model, as in the new cooling model, $r_{\rm cool}$ is
calculated using equation~(\ref{eq:r_cool_eq}), with $\tilde{t}_{\rm cool,avail}(r_{\rm cool},t)$
being estimated from the energy radiated away. By doing this, the
effect of artificial halo formation events on the gas cooling is
largely removed.  However, instead of directly accumulating this
radiated energy as in the new cooling model, the GFC2 model further
approximates the integrals involving $\rho_{\rm hot}^2$ in
equations~(\ref{eq:new_cool_L_cool1}) and (\ref{eq:new_cool_E_cool1}) as
\begin{eqnarray}
4\pi\int_{0}^{r_{\rm vir}}\rho^2_{\rm hot}r^2dr 
& \approx & \bar{\rho}_{\rm hot}\times 4\pi\int_{0}^{r_{\rm vir}}\rho_{\rm hot}r^2dr \nonumber \\
& = & \bar{\rho}_{\rm hot}(M_{\rm hot}+M_{\rm cooled}+M_{\rm eject}) ,
\label{eq:gfc2_rough_approximation}
\end{eqnarray}
where $\bar{\rho}_{\rm hot}$ is the mean density given by the density
profile. This approximation is very rough, and while in the new
cooling model the integral is limited to the gas that is hot,
i.e. between $r_{\rm cool,pre}$ and $r_{\rm vir}$, in the GFC2 model
the integration range is extended to $r=0$, which, according to equation~(\ref{eq:rho_norm_deter_GFC2}), includes the part of
the density profile where the gas has already cooled down. These
approximations make the calculation of $\tilde{t}_{\rm cool,avail}(r_{\rm cool},t)$ faster but less
accurate and physical than in the new cooling model.

With these approximations, for any time $t$, the GFC2 model adopts the
following equations in place of equations~(\ref{eq:new_cool_L_cool1}) and
(\ref{eq:new_cool_E_cool1}) in the new cooling model:
\begin{eqnarray}
L_{\rm cool}(t) & = &\tilde\Lambda(T_{\rm vir},Z)\bar{\rho}_{\rm hot}(M_{\rm hot}+M_{\rm cooled}+M_{\rm eject}) \\
E_{\rm cool}(t) & = & \int_{t_{\rm init}}^t \tilde\Lambda(T_{\rm vir},Z)\bar{\rho}_{\rm hot}\times \nonumber \\
                &   & [M_{\rm hot}(\tau)+M_{\rm cooled}(\tau)+M_{\rm eject}(\tau)]d\tau \nonumber \\
                & + & \int_{t_{\rm init}}^t \frac{3k_{\rm B}}{2\mu_{\rm m}}T_{\rm vir}\dot{M}_{\rm
                      cooled}d\tau .
\label{eq:gfc2_E_cool}
\end{eqnarray}
The second term in equation~(\ref{eq:gfc2_E_cool}), which is negative, is
equal in absolute value to the total thermal energy of the cooled mass
removed according to equation~(\ref{eq:gfc2_mcooled_remove}), and is
designed to remove the contribution to $E_{\rm cool}$ from this cooled
mass. Given $E_{\rm cool}$ and $L_{\rm cool}$, $\tilde{t}_{\rm cool,avail}(r_{\rm cool},t)$
for a given timestep is calculated from equation~(\ref{eq:t_cool_avail}),
as in the new model. Again, the actual $\tilde{t}_{\rm cool,avail}(r_{\rm cool},t)$ 
used in this model is denoted as $t_{\rm cool,avail}(t)$. Note that the approximation made in
equation~(\ref{eq:gfc2_rough_approximation}) leads to the derived
$t_{\rm cool,avail}$ being closer to the average cooling history of
all shells instead of the cooling history of gas near $r_{\rm cool}$,
and so leads to less accurate results than in the new cooling model.

The GFC2 model allows for the effect of the free-fall timescale on the
gas mass accreted onto the central galaxy in a similar way to the GFC1
model, by introducing the radius, $r_{\rm ff}$, calculated from
equation~(\ref{eq:rff}), but with $t_{\rm ff,avail}$ calculated in a way similar 
to that for $t_{\rm cool,avail}$. Specifically, a quantity with dimensions of energy similar to $E_{\rm cool}$ is accumulated for $t_{\rm ff,avail}$, but 
this quantity has an upper limit, $t_{\rm ff}(r_{\rm vir})\times L_{\rm cool}$, and 
once it exceeds this limit, it is then reset to this limit value. This 
limit ensures $t_{\rm ff,avail}\leq t_{\rm ff}(r_{\rm vir})$. Note that 
the effect of imposing this limit is usually to lead to a $t_{\rm ff,avail}$ 
different from both $t_{\rm cool,avail}$ and $t_{\rm ff}(r_{\rm vir})$. This 
calculation of $t_{\rm ff,avail}$ is not very physical because the calculation of
$t_{\rm cool,avail}$ here is based approximately on the
total energy released by the cooling radiation, while the accretion of
the cooled gas onto the central galaxy is driven by gravity, which
does not depend on the energy lost by radiation. In addition, by
introducing $r_{\rm ff}$, the GFC2 model inherits the associated
problems already identified for the GFC1 model.

The GFC2 model also adopts a specific angular momentum distribution
for the hot gas to calculate the angular momentum of the gas that
cools down and accretes onto the central galaxy. The simpler method to
specify this angular momentum distribution is as a function of radius,
namely $j_{\rm hot}(r)$. But, in principle, this requires calculating
the subsequent evolution of $j_{\rm hot}(r)$ as the hot gas
halo contracts, which is considered in the new cooling model but not
in the GFC1 or GFC2 models. A more complex method is to specify
$j_{\rm hot}$ as a function of the gas mass enclosed by a given
radius, i.e.\ $j_{\rm hot}(<M)$. This implicitly includes the effect
of contraction of the hot gas halo in the case of a static halo, where
no new gas joins the hot gas halo, because while the radius of each
gas shell changes during contraction , the enclosed mass is kept
constant and can be used to track each Lagrangian shell. However, when
there is new gas being added to the hot gas halo, this method also
fails, because the newly joining gas mixes with the hot gas halo
after contraction, and, in this case, the contraction has to be
considered explicitly. Since even the more complex method is not fully
self-consistent, for the sake of simplicity, in this work we adopt the
simpler method to calculate the angular momentum, without allowing for
contraction of the hot gas halo.

This model also adopts the treatments for the gas ejected by feedback 
and for halo mergers similar to those of the new cooling model, but 
unlike in the latter, here $E_{\rm cool}$ of the hot gas in the infalling 
haloes is also transferred. This again gives the newly added gas some 
previous cooling history, so it is not newly heated up.

%---------------------------------------------------------------------------------------------------

\subsubsection{Cooling model in \lgalaxy}

The cooling model used in \lgalaxy (see e.g.\ \citealt{munich_model2,
  munich_model_Guo11,munich_model3}) assumes that the hot gas is
always distributed from $r=0$ to $r=r_{\rm vir}$, and that its density
profile is singular isothermal, namely
$\rho_{\rm hot}(r)\propto r^{-2}$, with a single metallicity and a
single temperature equaling $T_{\rm vir}$.
The total mass inside this profile is $M_{\rm hot}$.

Then, a cooling radius, $r_{\rm cool}$, is calculated from
$t_{\rm cool}(r_{\rm cool}) = t_{\rm cool,avail}$, with
$t_{\rm cool,avail}=t_{\rm dyn} = r_{\rm vir}/V_{\rm vir}$.  If $r_{\rm cool}\leq r_{\rm vir}$, then the mass accreted
onto the central galaxy in a timestep, $\Delta t$, is
%--------------------
\footnote{Here, we adopted the equation for $\Delta M_{\rm acc,gal}$ from
  recent versions of the \lgalaxy model
  \citep[e.g.][]{munich_model_Guo11,munich_model3}.}
% In earlier versions
%   \citep[e.g.][]{munich_model1}, an extra factor $0.5$ is introduced
%   in front of the second line of equation~(\ref{eq:munich_slow_cooling}). See
%   the footnote in \citet{munich_model_Guo11} for more information.}
%--------------------
\begin{eqnarray}
\Delta M_{\rm acc,gal} & = & 4\pi\rho_{\rm hot}(r_{\rm cool})\times r_{\rm cool}^2
\frac{dr_{\rm cool}}{dt}\Delta t \nonumber \\
                & = & \frac{M_{\rm hot}}{r_{\rm vir}}
\frac{r_{\rm cool}} {t_{\rm dyn}}\Delta t ,
\label{eq:munich_slow_cooling}
\end{eqnarray}
with $dr_{\rm cool}/dt$ being estimated as
$dr_{\rm cool}/dt=r_{\rm cool}/t_{\rm cool,avail}=r_{\rm cool}/t_{\rm
  dyn}$. If instead $r_{\rm cool}> r_{\rm vir}$, then
\begin{equation}
\Delta M_{\rm acc,gal}=\frac{M_{\rm hot}}{t_{\rm dyn}}\Delta t .
\label{eq:munich_fast_cooling}
\end{equation}

Note that earlier predecessors of the \lgalaxy model
made slightly different assumptions. \citet{Kauffmann1993} and
subsequent papers in that series followed the approach of
\citet{WF_1991}, assuming that $t_{\rm cool,avail}=t$, with $t$
being the age of the Universe, and also that $dr_{\rm
  cool}/dt=r_{\rm cool}/(2 t_{\rm cool,avail})$, where the latter
follows mathematically from the result that $r_{\rm cool} \propto
t^{1/2}$ for a static halo with $\rho_{\rm hot}(r)\propto r^{-2}$ and $T_{\rm
  hot}(r)={\rm const}$. \citet{munich_model1} modified the first of these
assumptions by instead assuming $t_{\rm cool,avail}=t_{\rm
  dyn}$. This change in $t_{\rm cool,avail}$ was
effectively justified by the work of \citet{yoshida2002_sph}, who
compared the \lgalaxy cooling model with results from the ``stripped-down''
cosmological gasdynamical simulation of galaxy formation described below. As described in
\cite{munich_model_Guo11}, versions of \lgalaxy
from \citet{munich_model2} onwards then changed to using $dr_{\rm
  cool}/dt=r_{\rm cool}/t_{\rm cool,avail}$. This originates from an erroneous omission of the factor $0.5$ in the \lgalaxy code (see the footnote to equation~(5) in \citeauthor{munich_model_Guo11} for more details). Note that the SAGE model \citep[e.g.][]{SAGE_model} uses the same cooling model, but keeps the factor $1/2$, adopting $dr_{\rm cool}/dt=r_{\rm cool}/(2t_{\rm dyn})$.

The \lgalaxy cooling model is motivated by the self-similar cooling solution for a static halo derived in \citet{Bertschinger_cooling_solution}, in which the evolution of the hot gas profile driven by cooling is expressed in terms of a characteristic scale length $r_{\rm cool}(t)$. \citeauthor{Bertschinger_cooling_solution} defines $r_{\rm cool}$ by $t_{\rm cool}(r_{\rm cool})=t$, where $t_{\rm cool}(r)$ is the cooling timescale profile of the hot gas profile at the initial time, i.e.\ before the start of cooling, while $t$ is the physical time elapsed since then. \citeauthor{Bertschinger_cooling_solution} found that the mass accretion rate onto the centre is approximately the same as the mass cooling rate at $r_{\rm cool}$, leading to an expression similar to the first line of equation~(\ref{eq:munich_slow_cooling}). Note that the $r_{\rm cool}$ introduced in \citet{Bertschinger_cooling_solution} is a scale radius in the hot gas profile, while the $r_{\rm cool}$ in other cooling models considered in this paper is the inner boundary of the hot gas halo, which separates hot and cooled down gas, and thus they have different physical meanings.

However, the \citet{Bertschinger_cooling_solution} solution does not provide a complete justification for the \lgalaxy cooling model. The \lgalaxy cooling model does not follow the original definition of $r_{\rm cool}$ in \citet{Bertschinger_cooling_solution}. It instead defines $r_{\rm cool}$ as $t_{\rm cool}(r_{\rm cool})=t_{\rm dyn}$, where $t_{\rm cool}(r)$ is the cooling timescale profile of the current hot gas halo (including the evolution of the density of the hot gas halo driven by cooling) rather than that at the initial time, and the halo dynamical timescale $t_{\rm dyn}$ is adopted instead of the time elapsed since the initial time. Moreover, the solution in \citet{Bertschinger_cooling_solution} is for a static gravitational potential, while in the cosmological structure formation context, the halo grows and its potential evolves with time.

Mass accretion rates onto central galaxies calculated using equations~(\ref{eq:munich_slow_cooling}) and (\ref{eq:munich_fast_cooling}) have been shown to be in good agreement with stripped-down
	SPH hydrodynamical simulations, in which cooling is included but other
	processes, such as star formation and feedback, are ignored
	\citep{yoshida2002_sph,monaco_2014_comp}, but because of the inconsistencies in its
	physical formulation, this agreement is more in the nature of a fit to the results of these simplified simulations, and does not imply the physical validity of this calculation
	in the full galaxy formation context.

The angular momentum of the cooled down gas that accretes onto the
central galaxy is calculated as
\begin{equation}
\Delta J_{\rm acc,gal}= \Delta M_{\rm acc,gal}\times \bar{j}_{\rm
  halo} ,
\end{equation}
where $\bar{j}_{\rm halo}=J_{\rm halo}/M_{\rm halo}$ is the specific
angular momentum of the entire dark matter halo, with $J_{\rm halo}$
and $M_{\rm halo}$ being the total angular momentum and mass of the
dark matter halo respectively. This correponds to a specific angular
momentum distribution for the hot halo gas very different from the
$j_{\rm hot}(r)$ adopted in \GALFORM cooling models.

When a halo falls into a larger halo and becomes a subhalo, the \lgalaxy 
model assumes that its hot gas halo is instantaneously stripped and added 
to the hot gas halo of its host halo [see e.g.\ equation~(1) in \citet{de_lucia_2010_cmp}, 
but note that a more complex gradual stripping model also exists in the \lgalaxy model, 
see e.g.\ \citet{munich_model_Guo11}]. In this work we only use the \lgalaxy 
cooling model in the stripped down model (without other physical processes 
such as galaxy mergers, star formation and feedback), so we do not consider 
the treatment of gas ejected by SN feedback.
%---------------------------------------------------------------------------------------------------

\subsubsection{Cooling model in \MORGANA}
The full details of this cooling model are given in \citet{morgana1}
and \citet{morgana2}. The hot gas in a dark matter halo is assumed to
be in hydrostatic equilibrium, and a cold halo gas component similar
to that in the new cooling model is also introduced. As in the new
cooling model, in the continuous time limit, the boundary between the
hot gas halo and the cold halo gas is the cooling radius
$r_{\rm cool}$. The hot gas halo density and temperature profiles are
determined by the assumptions of hydrostatic equlibrium and that the
hot gas between $r_{\rm cool}$ and $r_{\rm vir}$ follows a polytropic
equation of state. This generally gives more complex profiles than
those used in \GALFORM and \lgalaxy, but typically the derived
density profile is close to the cored $\beta$-distribution used in
\GALFORM, while the temperature profile is very flat and close to
$T_{\rm vir}$. Therefore in this work, when calculating predictions
for this cooling model, for simplicity we will adopt the
$\beta$-distribution as the hot gas density profile and a constant
temperature, $T_{\rm vir}$, as the temperature profile. Just as
in the new cooling model, the density profile and temperature of the
hot gas halo are updated at every timestep.

The \MORGANA cooling model then calculates the cooling rate
$\dot{M}_{\rm cool}$. However, unlike the cooling models described
previously, this does not explicitly depend on the cooling history of
the hot gas, as expressed in $t_{\rm cool,avail}$, but instead it
assumes that at any given time, each shell of hot gas contributes to
$\dot{M}_{\rm cool}$ according to its own cooling time scale~\footnote{\citet{morgana2} introduced a modification of this for a
  static halo, in which the onset of cooling is delayed by a time
  interval equaling $t_{\rm cool}(r=0)$. But this modification is not
  applied in the full \MORGANA model, so here we ignore it and use the
  cooling model described in \citet{morgana1}.}.
Specifically, this is
\begin{equation}
\dot{M}_{\rm cool}=4\pi\int_{r_{\rm cool}}^{r_{\rm vir}}
\frac{\rho_{\rm hot}(r)}{t_{\rm cool}(r)}r^2dr ,
\label{eq:morgana_cooling_rate}
\end{equation}
where $\rho_{\rm hot}(r)$ is the hot gas density at radius $r$, while
$t_{\rm cool}(r)$ is the cooling time scale corresponding to gas
density $\rho_{\rm hot}(r)$ and temperature $T_{\rm vir}$, and is
given by equation~(\ref{eq:t_cool}). This equation is supplemented by another
equation, 
\begin{equation}
\dot{r}_{\rm cool}=\frac{\dot{M}_{\rm cool}}{4\pi\rho_{\rm hot}(r_{\rm
    cool})r_{\rm cool}^2}-c_{\rm s}(r_{\rm cool}) ,
\label{eq:morgana_r_cool}
\end{equation}
where $c_{\rm s}(r_{\rm cool})$ is the local sound speed at radius
$r_{\rm cool}$. The first term in equation~(\ref{eq:morgana_r_cool})
describes the increase of $r_{\rm cool}$ due to cooling. The form of
this term is derived from the picture that the cooled down gas all
comes from the region near $r_{\rm cool}$, and then mass conservation
for a spherical shell gives
$\dot{M}_{\rm cool}dt=4\pi\rho_{\rm hot}(r_{\rm cool})r_{\rm
  cool}^2dr_{\rm cool}$. The second term describes the contraction of
the hot gas halo due to the reduction of pressure support induced by
cooling. Since the hot gas halo is in hydrostatic equilibrium in the
gravitational potential well of the dark matter halo,
$c_{\rm s}(r_{\rm cool})$ is close to the circular velocity at
$r_{\rm cool}$, so the contraction time scale is comparable to
$t_{\rm ff}(r_{\rm cool})$. Thus, the contraction here is similar to
that introduced in the new cooling model, but in the
\MORGANA cooling model the contraction does not include the effect
of halo growth, which is included explicitly in the new cooling
model using equation~(\ref{eq:new_cool_shrink_halo_growth}). Together,
equations~(\ref{eq:morgana_cooling_rate}) and (\ref{eq:morgana_r_cool})
enable the calculation of $r_{\rm cool}$ and $\dot{M}_{\rm cool}$ for
each timestep.

There are some physical inconsistencies between
equations~(\ref{eq:morgana_cooling_rate}) and (\ref{eq:morgana_r_cool}). In
equation~(\ref{eq:morgana_cooling_rate}), it is assumed that the cooled down gas
comes from the whole region between $r_{\rm cool}$ and $r_{\rm vir}$,
but in equation~(\ref{eq:morgana_r_cool}) the cooled down gas is assumed to
only come from a shell around $r=r_{\rm cool}$. Unless $r_{\rm cool}$
is very close to $r_{\rm vir}$, these two assumptions about the
spatial origin of the cooled down gas conflict with each
other. Furthermore, equation~(\ref{eq:morgana_cooling_rate}) implies that
there is differential cooling within a single hot gas shell, with a
fraction of the gas cooling completely and the remainder not cooling
at all. However, since in a perfectly spherical system the gas inside one shell all has the same
density and temperature, the whole shell should cool down
simultaneously, namely all gas in it cools down after a time
$t_{\rm cool}$, but no gas cools down before that time. Of course, in reality deviations from spherical symmetry will make the cooling process more complex.

The mass of gas cooled down in one timestep is then
$\Delta M_{\rm cool}=\dot{M}_{\rm cool}\Delta t$. This mass is used to
update the mass of the cold halo gas component, $M_{\rm halo,cold}$,
and then the mass accreted onto the central galaxy,
$\Delta M_{\rm acc,gal}$, is derived assuming gravitational infall of
the cold halo gas component, which is calculated in the same way as our
new cooling model, using equation~(\ref{eq:new_cool_m_acc_gal}).

The \MORGANA cooling model does not explicitly follow the flow of
angular momentum. Instead, it assumes that the central galaxy always
has a specific angular momentum equal to that of its host dark
matter halo, $\bar{j}_{\rm halo}$, with
$\bar{j}_{\rm halo}=J_{\rm halo}/M_{\rm halo}$, and $J_{\rm halo}$ and
$M_{\rm halo}$ the total angular momentum and mass of the dark matter
halo respectively. This assumption is even cruder than
the treatment in \lgalaxy. \citet{Stevens2017} compare $\bar{j}_{\rm halo}$ and the specific angular momentum of central galaxies in the EAGLE simulation, and find that this assumption is indeed very crude.

The \MORGANA model adopts a relatively complex treatment of halo gas 
components during halo mergers \citep[e.g.][]{morgana1}. One important 
feature of the original \MORGANA treatment is that gas cooling is forced 
to pause for several halo dynamical timescales after halo major mergers. 
However, \citet{monaco_2014_comp} argued that this suppression of cooling 
seems to be too strong when compared with SPH simulations and suggested 
turning it off. Here, for simplicity, and in order to concentrate on the 
cooling calculation, we adopt the same treatment for the \MORGANA cooling 
model as in the new cooling model, and the suppression of cooling during 
halo major mergers is not included.

In this paper, the \MORGANA cooling model is only used in the stripped down 
model, therefore we do not consider here the treatment of the gas ejected 
by SN feedback in the \MORGANA model.

%====================================================================================================================================================
\subsection{Halo spin and concentration}
\label{sec:halo_spin}
All of the cooling models described above require knowledge of the
density profile and angular momentum of the dark matter halo. The
former is needed for calculating the free-fall time scale from a given
radius, while the latter is required for the calculations of the
angular momentum of the gas. Assuming the NFW profile for the dark
matter halo, the remaining major task for characterizing the profile is to determine
the halo concentration, $c_{\rm NFW}$; other parameters of the
profile, such as halo mass and virial radius, are relatively
straightforward to derive given the merger tree. The angular momentum
of a halo is usually expressed in terms of the halo spin parameter, 
$\lambda_{\rm halo}$, which is defined as,
\begin{equation}
\lambda_{\rm halo}=\frac{J_{\rm halo}\left| E_{\rm halo}\right|
  ^{1/2}}{GM_{\rm halo}^{5/2}} ,
\label{eq:lambda_def}
\end{equation}
where $J_{\rm halo}$, $E_{\rm halo}$ and $M_{\rm halo}$ are the total
angular momentum, energy and mass of a dark matter halo respectively,
and $G$ is the gravitational constant. Thus, the major task of
determining halo angular momentum is to determine $\lambda_{\rm halo}$
for a given halo.

Different semi-analytic models use different methods to assign these
two parameters to each halo in a merger tree. The main \GALFORM models
\citep[e.g.][]{galform_baugh2005,galform_bower2006,
  galform_gonzalez2014, galform_lacey2015} follow the method
introduced in \citet{cole2000}, in which a halo inherits the
$c_{\rm NFW}$ and $\lambda_{\rm halo}$ of its most massive progenitor
until it undergoes a halo formation event.  At a halo formation event,
a new $c_{\rm NFW}$ is assigned according to the mass and redshift of
this halo through the $M_{\rm halo}$-$c_{\rm NFW}$ correlation
\citep{Navarro1997}, and a new
$\lambda_{\rm halo}$ is randomly selected according to a lognormal
distribution derived from N-body simulations [e.g.\
\citet{Cole1996,spin_distribution_warren, spin_distribution_gardner},
but see \citet{spin_distribution_bett} for a different fitting
form]. This method introduces sudden jumps in $c_{\rm NFW}$ and
$\lambda_{\rm halo}$ at halo formation events even if the halo growth
is smooth, which is unphysical. Also, the possible evolution of
$c_{\rm NFW}$ and $\lambda_{\rm halo}$ between two adjacent halo
formation events is ignored.

\lgalaxy models use halo merger trees from N-body simulations, and
adopt $c_{\rm NFW}$ and $\lambda_{\rm halo}$ measured directly for
the haloes in these simulations. In principle, this provides the most
accurate way to assign $c_{\rm NFW}$ and $\lambda_{\rm halo}$ to a
given halo; however, it also has some limitations. Firstly, resolving
the halo mass accretion history and thus building merger trees only
requires marginal resolution, i.e. a halo should be resolved by at
least several tens of particles, but robust measurement of
$c_{\rm NFW}$ and $\lambda_{\rm halo}$ requires higher resolution,
i.e. a halo should be resolved by at least several hundred particles 
\citep{concentration_measure_neto2007,spin_distribution_bett}. Therefore
$c_{\rm NFW}$ and $\lambda_{\rm halo}$ values measured for the smaller
haloes from an N-body simulation are not reliable. Secondly, a
semi-analytic model employing this method cannot use Monte Carlo
merger trees, which limits its applicability, particularly in building
large statistical samples.

The \MORGANA model also assigns $c_{\rm NFW}$ according to the
$M_{\rm halo}$-$c_{\rm NFW}$ correlation, but it does this at each
timestep instead of at each halo formation event. By doing so, the
artificial sudden jumps in $c_{\rm NFW}$ at halo formation events is
removed. In the \MORGANA model each halo inherits the
$\lambda_{\rm halo}$ of its most massive progenitor, while for each
halo without progenitor, a value of $\lambda_{\rm halo}$ is assigned
randomly according to the lognormal distribution. In this way,
$\lambda_{\rm halo}$ is constant in each branch of a merger tree, and
there is no artificial jump in its value as in \GALFORM models, but
the evolution of $\lambda_{\rm halo}$ due to halo growth is completely
ignored.

\citet{benson_bower_2010_cooling} and \citet{vitvitska_2002_spin} [see
also \citet{maller_2002_spin}] proposed another way to assign a value
of $\lambda_{\rm halo}$ to each halo. In their method, haloes with no
progenitor are assigned $\lambda_{\rm halo}$ values randomly according
to the $\lambda_{\rm halo}$ distribution derived from N-body
simulations, but then the evolution of $\lambda_{\rm halo}$ is
calculated based on the orbital angular momenta of accreted
haloes. With the halo accretion history given by the merger tree and
distributions of orbital parameters derived from N-body simulations,
the evolution of $\lambda_{\rm halo}$ can be calculated. One potential
problem with this method is that it assumes that smoothly
accreted mass makes no contribution to the evolution of
$\lambda_{\rm halo}$. This may not be true, and also whether the
accretion is smooth or clumpy is resolution dependent, so this
approach omits the effect of unresolved accreted haloes, which may
affect the long term evolution of $\lambda_{\rm halo}$.

In the present paper, we follow \citet{cole2000} to set $c_{\rm NFW}$
and $\lambda_{\rm halo}$ for the GFC1 model, to remain consistent with
its original assumptions. For other models, we adopt the method used
in the \MORGANA model for setting $c_{\rm NFW}$ (i.e. setting it
according to the adopted $c_{\rm NFW}$-$M_{\rm halo}$ relation at each
timestep), while for the assignment of $\lambda_{\rm halo}$, we
introduce a new and simple method. Specifically, a lognormal
distribution is adopted to randomly generate spins for haloes at the
tips of merger trees. The subsequent evolution of $\lambda_{\rm halo}$
is then modelled by a Markov random walk, in which the spins of a halo
and its progenitor become approximately uncorrelated when this halo
reaches twice its progenitor's mass. In each timestep, a conditional
probability distribution for the new spin can be constructed for each
halo given the mass increase and progenitor $\lambda_{\rm halo}$, and
then a value of $\lambda_{\rm halo}$ is assigned randomly according to
this conditional distribution. This method allows large spin changes
when the halo mass increases by a large factor, i.e.\ in major
mergers, and small, but usually nonzero, changes for small mass
increases, which are typical in smooth halo growth. More details of
this random walk method are provided in
Appendix~\ref{app:lambda_random_walk}, together with some comparisons
of the predictions of this method with results from N-body
simulations.

We have checked that all the results presented in this paper are not
sensitive to the method used for assigning $c_{\rm NFW}$ and
$\lambda_{\rm halo}$.

%%%%%%%%%%%%%%%%%%%%%%%%%%%%%%%%%%%%%%%%%%%%%%%%%%%%%%%%%%%%%%%%%%%%%%%%%%%%%%
\section{Results} \label{sec:results}

This section presents predictions from the new cooling model, and
compares them with the corresponding results from the earlier cooling
models described in the previous section. We start, in
\S\ref{sec:results_static_halo}, by considering the cooling histories
for the simplest case of a static haloes, and then, in
\S\ref{sec:results_dynamic_halo}, consider the more realistic case of
evolving haloes with full merger histories. Finally, in
\S\ref{sec:results_full_model}, we show the effects of using the new
cooling model within a full galaxy formation model. All the calculations adopt the cooling function tabulated in \citet{Sutherland_1993_cooling_function}.

\subsection{Static halo} 
\label{sec:results_static_halo}
For the static halo case, we consider dark matter haloes of fixed
mass, $M_{\rm halo}$, and also a fixed density profile, corresponding
to a halo that forms at redshift $z$. We present 4 cases that illustrate the
range of behaviours: $M_{\rm halo}=10^{11}\,{\rm M}_{\odot}$ (low mass
and fast cooling halo), $M_{\rm halo}=10^{12}\,{\rm M}_{\odot}$ (Milky
Way like halo), $M_{\rm halo}=10^{13}\,{\rm M}_{\odot}$ (group halo)
and $M_{\rm halo}=10^{14}\,{\rm M}_{\odot}$ (cluster halo). For
$M_{\rm halo}=10^{11}\,{\rm M}_{\odot}$ we choose $z=3$, while for the
other cases we choose $z=0$. The core radius of the
$\beta$-distribution for hot gas is set to be
$r_{\rm core}=0.07r_{\rm vir}$. The redshift is introduced here to determine $r_{\rm vir}$, which then enters the calculation of the virial temperature $T_{\rm vir}$, free-fall timescale $t_{\rm ff}(r)$ and core radius $r_{\rm core}$ of the hot gas density profile. To isolate the effects of the
different cooling models, star formation and feedback processes are
turned off.

\begin{figure*}
\centering
\includegraphics[width=0.8\textwidth]{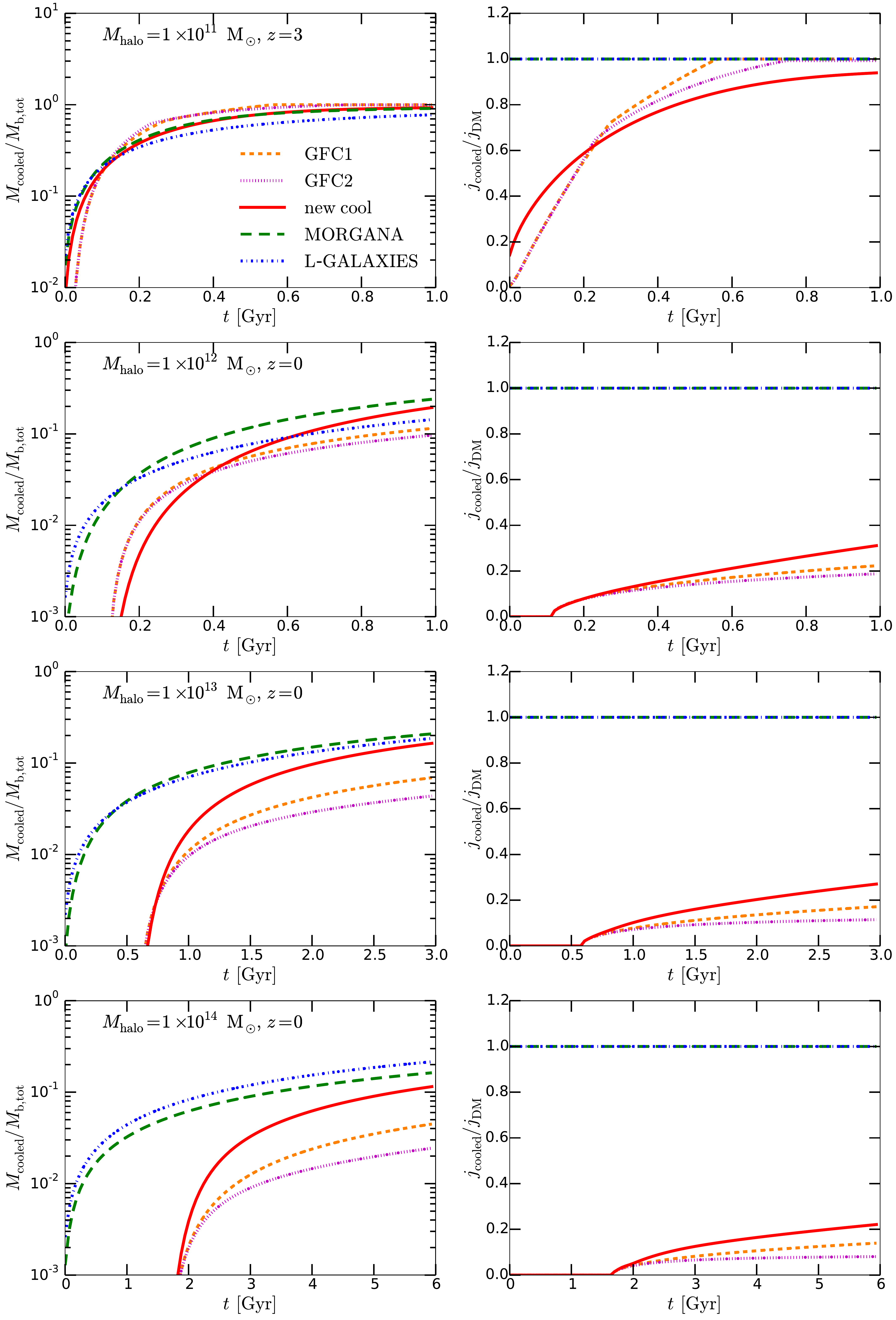}
\caption{Cooling histories for static haloes, measured from the time when
  radiative cooling is turned on. The different line styles show the
  predictions for different cooling models, as labelled in the line
  key. Each row corresponds to a different halo mass and assembly
  redshift, as labelled. Left column: the
  ratio of cooled mass to total baryon mass. Right
  column: the ratio of the specific angular momentum of the cooled gas
  to the specific angular momentum of the halo. Note the different time spans in the different panels.}
\label{fig:cool_result_static_halo}
\end{figure*}

Fig.~\ref{fig:cool_result_static_halo} shows the total mass 
and the specific angular momentum of the gas that has cooled
down and accreted onto the central galaxy, as predicted by the
different cooling models. Results are plotted against the time, $t$,
since radiative cooling is turned on in the halo. For the fast cooling
halo ($M_{\rm halo}=10^{11}\,{\rm M}_{\odot}$), all cooling models
predict very similar results for the two quantities. This is because
in the fast cooling regime, the accretion of the cooled down gas is
mainly limited by the timescale for free-fall rather than that for
radiative cooling, and all of the cooling models calculate the free-fall
accretion rate in a similar way. For the more massive haloes
($M_{\rm halo}=10^{12} - 10^{14}\,{\rm M}_{\odot}$), the results
for the \lgalaxy and \MORGANA cooling models remain very similar, but the
results for the GFC1, GFC2 and new cooling models diverge from those
models and from each other.

For haloes of all masses, gas starts to accrete onto the central galaxy
from $t=0$ in the \lgalaxy and \MORGANA cooling models; for the
GFC1, GFC2 and new cooling models there is a time delay that varies
with halo mass. This time delay is equal to the central radiative
cooling timescale, $t_{\rm cool}(r=0)$. It is a consequence of the
assumption that the hot gas density decreases monotonically with
radius, so that $t_{\rm cool}\propto \rho_{\rm hot}(r)^{-1}$ increases
with radius. In the \GALFORM cooling models, the hot gas density at
$r=0$ is finite, and gas cools shell by shell, so no gas can cool and
accrete before the gas at the centre cools. In contrast, in the
\lgalaxy cooling model, the hot gas density at $r=0$ is infinite,
while in \MORGANA, the gas does not cool shell by shell, so there is
no time delay.

\begin{figure*}
\includegraphics[width=1.0\textwidth]{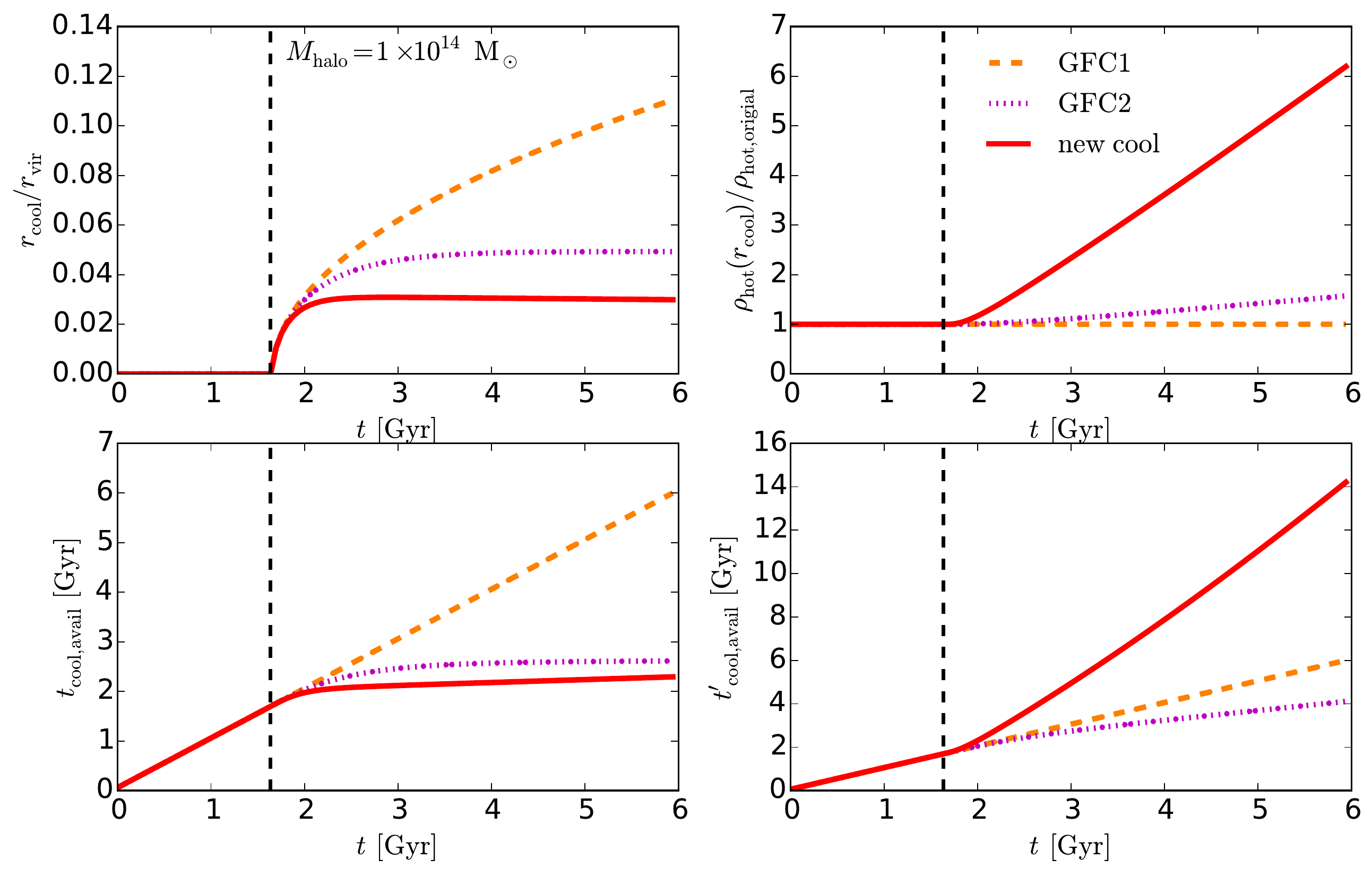}
\caption{More detailed information on the cooling in static haloes in
  the three \GALFORM cooling models for $M_{\rm halo}=10^{14}\,{\rm M}_{\odot}$. From 
  upper left to lower right, these four panels respectively show the time evolution of 
  the cooling radius, $r_{\rm cool}$, ratio of the density of the shell at $r_{\rm cool}$ 
  to the density of the same Lagrangian shell at $t=0$, the time available for cooling, 
  $t_{\rm cool,avail}$, and the scaled time available for cooling, $t'_{\rm cool,avail}$, 
  predicted by the three models. Each line style corresponds to a different model, with 
  the model name given in the key in the upper right panel. The vertical dashed line in 
  each panel indicates the moment at which cooling starts.}
\label{fig:static_halo_cooling_detail}
\end{figure*}

For the Milky Way like halo, the GFC1 and GFC2 models generally
predict lower accreted masses than the new cooling model, and this
difference grows with halo mass. For the $10^{14}\,{\rm M}_{\odot}$
halo, the difference can be a factor $\gsim 4$. The origin of this
difference can be understood by looking at the cooling in more detail,
as is done in Fig.~\ref{fig:static_halo_cooling_detail}.  For
conciseness, we only show the most massive halo, where the
abovementioned difference is largest. The results for less massive 
haloes are similar.

The upper left panel of Fig.~\ref{fig:static_halo_cooling_detail} shows the
time evolution of the cooling radius, $r_{\rm cool}$. The GFC1 model
predicts that $r_{\rm cool}$ increases monotonically with time. This
is expected for a fixed hot gas halo, in which the hot gas cools down
at larger and larger radii with increasing time. For the GFC2 and new
cooling models, however, $r_{\rm cool}$ tends to reach a stable value
instead of increasing with time. This is caused by the contraction of
the hot gas halo included in these two models. Although radiative
cooling leads to an increase of $r_{\rm cool}$, just as in the GFC1
model, the contraction moves the hot gas shells to smaller radii, and
the competition of these two factors results in $r_{\rm cool}$
approaching a nearly constant value. The GFC2 model predicts larger values of
$r_{\rm cool}$ than the new cooling model, because, as mentioned in
\S\ref{sec:gfc2_model}, these two models adopt different contraction
time scales, and the GFC2 model tends to overestimate the contraction
timescale, leading to slower contraction, and resulting in values of
$r_{\rm cool}$ intermediate between the GFC1 and new cooling models.

When a hot gas shell moves to smaller radius, it is compressed to
a higher density. This effect is shown in the upper right panel of
Fig.~\ref{fig:static_halo_cooling_detail}. This panel gives the ratio
of the density of the gas at $r_{\rm cool}$ to the density, $\rho_{\rm hot,original}$, in the same
Lagrangian gas shell at $t=0$. This ratio is always $1$ for the GFC1
model, because it assumes a static hot gas halo, while for the GFC2
and new cooling models it is larger than $1$, due to the compression
induced by the hot halo contraction.

The lower left panel of Fig.~\ref{fig:static_halo_cooling_detail} shows 
the $t_{\rm cool,avail}$ predicted by the three models. The prediction 
of the GFC1 model is just the physical time, while those of the GFC2 and 
new cooling models tend to level off at constant values. $t_{\rm cool,avail}$ 
encodes the previous cooling history of the hot gas. The advance of cooling 
tends to increase $t_{\rm cool,avail}$ by increasing $E_{\rm cool}$ in 
equation~(\ref{eq:t_cool_avail}), while the hot gas halo contraction in the GFC2 
and new cooling models increases the shell density, which leads to an increase of the cooling rate, and 
so tends to reduce $t_{\rm cool,avail}$ by increasing $L_{\rm cool}$ in 
equation~(\ref{eq:t_cool_avail}). The combination of these two effects causes 
$t_{\rm cool,avail}$ to approach a roughly stable value.

In the GFC2 and the new cooling models $t_{\rm cool,avail}$ is used to calculate 
the cooled mass for the hot gas halo after contraction. As shown in the upper right 
panel of Fig.~\ref{fig:static_halo_cooling_detail}, the extent of contraction is 
different in these two models, while the GFC1 model does not have this contraction. 
Thus, the $t_{\rm cool,avail}$ in these three models are for different hot gas haloes. 
This makes it complicated to analyze the origin of the differences in predicted 
cool mass based on $t_{\rm cool,avail}$. Therefore, we introduce another quantity, 
$t'_{\rm cool,avail}$, which is defined as
\begin{equation}
t'_{\rm cool,avail}=t_{\rm cool,avail}\frac{\rho_{\rm hot}(r_{\rm cool})}{\rho_{\rm hot,original}} , 
\label{eq:t_cool_avail_prime}
\end{equation}
where $\rho_{\rm hot}(r_{\rm cool})$ is the density of the shell that has just cooled 
down, while $\rho_{\rm hot,original}$ is the density at $t=0$ of the same Lagrangian 
shell, and this density ratio is that shown in the upper right panel of Fig.~\ref{fig:static_halo_cooling_detail}. 
Since for the shell just cooled down one has $t_{\rm cool,avail}=t_{\rm cool}(r_{\rm cool})$, 
and from equation~(\ref{eq:t_cool}), $t_{\rm cool}\propto \rho_{\rm hot}^{-1}$, 
equation~(\ref{eq:t_cool_avail_prime}) implies that
\begin{eqnarray}
t'_{\rm cool,avail} & = & t_{\rm cool}(r_{\rm cool})\frac{\rho_{\rm hot}(r_{\rm cool})}{\rho_{\rm hot,original}} \nonumber \\
	& = & t_{\rm cool,original} ,
\end{eqnarray}
where $t_{\rm cool,original}$ is the cooling timescale of this Lagrangian shell at $t=0$. 
Then it is clear that $t'_{\rm cool,avail}$ is linked to the cooling timescale at the 
initial moment, at which the hot gas halo is the same in all three models, and so is easier 
to compare between models. 

The lower right panel of Fig.~\ref{fig:static_halo_cooling_detail}
shows the $t'_{\rm cool,avail}$ predicted by the three models. The new
cooling model predicts the highest $t'_{\rm cool,avail}$, which means
that at any given time, the shell at $r_{\rm cool}$ in this model has
the largest initial radius among the three models, and since at $t=0$
the hot gas halo density profile is the same for these three models,
the largest radius implies the highest cooled mass. In contrast, the
GFC2 model predicts the smallest $t'_{\rm cool,avail}$, so it predicts
the lowest cool mass (see Fig.~\ref{fig:cool_result_static_halo}).

The density enhancement ($\rho_{\rm hot}(r_{\rm cool})/\rho_{\rm
  hot,original}>1$) seen in the GFC2 and new cooling models implies a
higher cooling luminosity than for the case of a fixed hot gas halo as
in the GFC1 model. This higher cooling luminosity means more thermal
energy is radiated away by a given time, and since the hot gas haloes
in these three models all have the same temperature, this higher
thermal energy loss should mean higher cooled mass. Therefore, it
would be expected that for a cooling model with density enhancement,
its predicted cooled mass should be higher than for a model assuming a
fixed hot gas halo. Also, a higher cooled mass means the shell cooled
down was initially at larger radius, and since the density decreases
with increasing radius for the assumed initial density profile, this
larger radius implies lower initial density and longer original
cooling timescale, $t_{\rm cool,original}$. Therefore, for a given
$\rho_{\rm hot}(r_{\rm cool})/\rho_{\rm hot,original}$, insofar as
this ratio is greater than one, the expected $t'_{\rm cool,avail}$
should be larger than in a model with a fixed hot gas halo, i.e.\ the
GFC1 model.

The new cooling model does predict cooled mass and $t'_{\rm
  cool,avail}$ larger than those in the GFC1 model, but the GFC2 model
predicts these to be lower than in the GFC1 model, which contradicts
the physical expectation above. Thus, the GFC2 model appears to be
physically inconsistent, and the $t'_{\rm cool,avail}$ in it tends to
be too small.  Furthermore, because $t'_{\rm cool,avail}$ and $t_{\rm
  cool,avail}$ are related by the density ratio through
equation~(\ref{eq:t_cool_avail_prime}), for a given density ratio, the
underestimation of $t'_{\rm cool,avail}$ also implies an
underestimation of $t_{\rm cool,avail}$.

To understand why $t_{\rm cool,avail}$ is underestimated in the GFC2
model, consider the following.  As described in
\S\ref{sec:gfc2_model}, the GFC2 model accumulates the total energy
radiated away for the current hot gas halo (equation~\ref{eq:gfc2_E_cool})
and then divides it by the current halo cooling luminosity to estimate
$t_{\rm cool,avail}$. When some gas cools down from the hot gas halo,
its contribution to the total energy radiated away should be removed,
because this gas is no longer part of the hot gas halo, and this is
the motivation for the second term in equation~(\ref{eq:gfc2_E_cool}). This
term basically removes the total thermal energy corresponding to the
mass removed from the hot gas halo. This would be correct if the GFC2
model exactly accumulated the total energy radiated away by
cooling. However, the GFC2 model adopts a very rough approximation
(equation~\ref{eq:gfc2_rough_approximation}), whereby the cooling
luminosity of a gas shell is approximated as
$\delta L_{\rm cool}=4\pi \tilde\Lambda\rho^2_{\rm hot}(r)r^2dr\approx
4\pi \tilde\Lambda\bar{\rho}_{\rm hot}\rho_{\rm hot}(r)r^2dr$, with
$\tilde\Lambda$ being the cooling function and $\bar{\rho}_{\rm hot}$
the mean density of the hot gas. For the $\beta$-distribution used for
the static halo comparison,
$\bar{\rho}_{\rm hot}\sim\rho_{\rm hot}(r=0.5r_{\rm vir})$, and for
the group and cluster haloes, cooling happens in the region where
$\rho_{\rm hot}(r)>\bar{\rho}_{\rm hot}$. Thus, the approximation
underestimates the energy radiated away, and so the second term in
equation~(\ref{eq:gfc2_E_cool}) removes more energy than necessary, leading
to an underestimation of $t_{\rm cool,avail}$. The final cooling
depends on the relative strength of this underestimation and the
density enhancement. For the static halo, this
underestimation of $t_{\rm cool,avail}$ exceeds the effects of the
density enhancement and leads to even less gas cooling down than in
the GFC1 model, but for other cases, the results could be different.

Overall, the introduction of the contraction of the hot gas halo in
the new cooling model results in more efficient cooling than in the
more traditional \GALFORM cooling model GFC1. Some previous works
\citep{de_lucia_2010_cmp, monaco_2014_comp} also noticed that the GFC1
model tends to predict less gas cooling than other cooling models such
as \MORGANA and \lgalaxy, and also less than SPH hydrodynamical
simulations. These works suggested using more centrally concentrated
hot gas density profiles such as the singular isothermal profile to
bring semi-analytic predictions into better agreement with SPH
simulations. However, the results here suggest that at least part of
the reason for the GFC1 model giving low cooling rates is that it does
not include contraction of the hot gas halo as cooling proceeds. Note
that the enhancement of hot gas density and hence cooling induced
by contraction is also mentioned in \MORGANA papers
\citep[e.g.][]{morgana2}, but taking the average over all hot gas shells
to calculate the mass cooling rate (as is done in the \MORGANA cooling
model) may not be the best way to model this effect.

Fig.~\ref{fig:cool_result_static_halo} also shows that for the haloes
other than the fast cooling halo, the different cooling models predict
different specific angular momenta for the gas in the central galaxies. The
\lgalaxy and \MORGANA cooling models give higher 
specific angular momentum than the GFC1, GFC2 and new
cooling models. They predict higher specific angular momentum 
mainly because they (implicitly) assume specific angular
momentum distributions of the hot gas, $j_{\rm hot}(r)$, that are very
different from the three \GALFORM models. The \lgalaxy cooling model
assumes that the gas accreting in the current timestep has specific
angular momentum equal to the mean specific angular momentum of the
dark matter halo. This corresponds to $j_{\rm hot}(r)={\rm constant}$,
i.e. no dependence on the radius from which the gas is cooling. The
\MORGANA cooling model instead assumes that the mean specific angular
momentum of all the gas that has cooled down and accreted onto the
central galaxy over its past history is equal to the mean specific
angular momentum of the current dark matter halo. In the static halo
case, in which the mean specific angular momentum of the halo does not
change with time, the assumption in the \MORGANA model is equivalent
to that in \lgalaxy cooling model. As shown in the right column of
Fig.~\ref{fig:cool_result_static_halo}, this results in the mean specific angular
momentum of the cold gas in central galaxies being equal to the mean halo
specific angular momentum at all times for these two models, in the
case of a static halo.

In contrast, the GFC1, GFC2 and new cooling models assume that
$j_{\rm hot}(r)$ increases with radius, and that the mean specific
angular momentum of all the baryons in a halo is equal to the mean
specific angular momentum of the halo. Under this assumption, the hot
gas in the central region has lower specific angular momentum than the
mean for the halo. For the haloes other than the fast cooling halo,
typically only part of the hot gas cools down, and since the cooling
proceeds from halo center outwards, the hot gas having low specific
angular momentum cools first, so the predicted mean specific angular
momentum of the cold gas in central galaxies is lower than that of the
dark matter halo. The new cooling model predicts higher specific
angular momentum for the cooled gas in central galaxies compared to
the GFC1 and GFC2 models, because it cools more effectively, and so
can cool gas shells that were originally at larger radii, which,
according to the assumed $j_{\rm hot}(r)$, have higher specific
angular momentum.

%%%%%%%%%%%%%%%%%%%%%%%%%%%%%%%%%%%%%%%%%%%%%%%%%%%%%%%%%%%%%%%%%%%%%%%%%%%%%%
\subsection{Cosmologically evolving haloes} 
\label{sec:results_dynamic_halo}
Having understood the behaviours of the different cooling models in the
simplified case of static haloes, the next step is to compare them
in the context of cosmic structure
formation. To achieve this, we run the cooling models in
cosmologically evolving haloes, whose formation histories are described
by merger trees. As before, we choose 4 different halo masses at $z=0$, namely
$M_{\rm halo}=10^{11},\ 10^{12},\ 10^{13}$ and $10^{14}\Msol$. For
each of these masses, we generate $100$ independent merger trees
to sample the range of formation histories, using the Monte Carlo
method of \citet{Parkinson2008} that is based on the Extended
Press-Schechter approach \citep[e.g.][]{LC93}. ( We use Monte Carlo
rather than N-body merger trees for this comparison because it is then
easier to build equal size samples for different $z=0$ halo masses.)
The merger trees are built with halo mass resolution,
$M_{\rm res}=5\times 10^9\Msol$. We choose this relatively high $M_{\rm res}$ 
mainly to avoid too much cooling in small haloes, which would leave 
little gas for the slow cooling regime in high mass haloes. Star 
formation, SN and AGN feedback processes and galaxy
mergers are all turned off in order to isolate the effects of the
different cooling models. For each merger tree, the mass and angular
momentum of the gas cooled and accreted onto the central galaxy in the
haloes in the major branch of this merger tree are recorded.

\begin{figure*}
\centering
\includegraphics[width=0.8\textwidth]{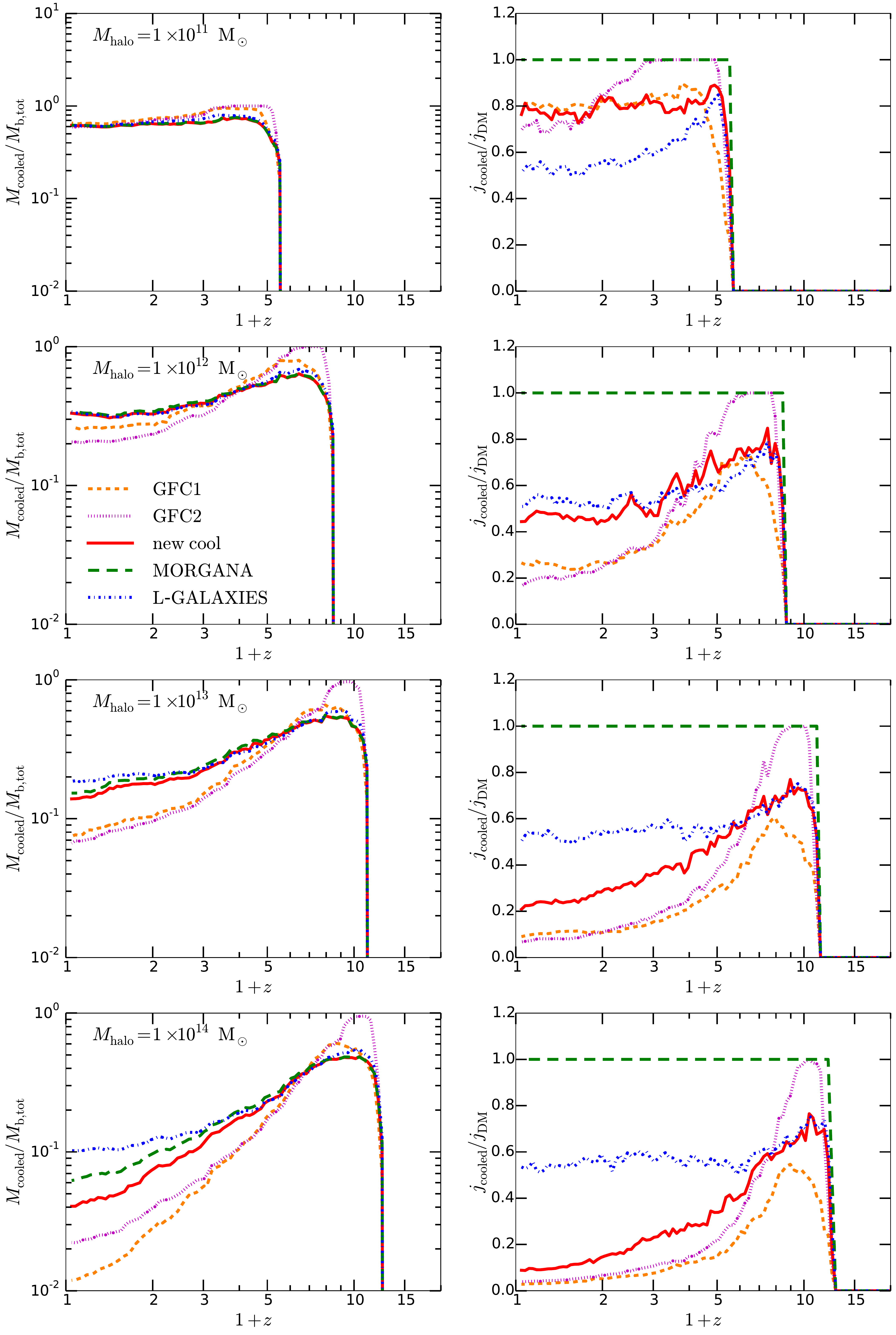}
\caption{The same as Fig.~\ref{fig:cool_result_static_halo}, but for
  cosmologically evolving haloes. For each $z=0$ halo mass shown in the
  figure, $100$ merger trees are constructed, and the cooling models
  run on all branches of each merger tree. Star formation, SN and AGN
  feedback, and galaxy mergers are turned off. The results are
  recorded for the main branch of each merger tree, and the medians
  of each halo sample are plotted.}
\label{fig:cool_result_dynamic_halo}
\end{figure*}

Fig.~\ref{fig:cool_result_dynamic_halo} shows the medians of $100$
realizations for each halo mass of the mass and
specific angular momentum of gas accreted onto the central galaxy in
the main branch of the merger tree.  Many features seen in the static
halo case also appear here. For the fast cooling haloes
($M_{\rm halo}=10^{11}\Msol$ at $z=0$), the predictions of the
different cooling models are similar, again because in the fast
cooling regime the accretion of gas onto galaxies is limited by the
free-fall timescale and largely insensitive to the details of the
cooling calculation. For the slower-cooling haloes
($M_{\rm halo}\ge 10^{12}\Msol$), the new cooling model predicts
larger cooling masses than the GFC1 and GFC2 models, because of the
contraction of the hot gas halo. For haloes less massive than
$10^{14}\Msol$, the predictions of the new cooling model for the mass
cooled down are close to those of the \lgalaxy and \MORGANA cooling
models, but for $10^{14}\Msol$ haloes, the predictions of the new
cooling model at $z=0$ are about a factor of $2$ lower than those of
\MORGANA, and a factor of $3$ lower than those of \lgalaxy.

In the static halo case, the cooled down mass predicted by the GFC2
model is always lower than that of the GFC1 model, but here the
relation of their predictions is more complex. For some halo masses,
the GFC2 model gives higher cooled down masses, but for other halo
masses, its predictions are lower. This is because the diverse halo
merger histories affect the comparative strengths of the
underestimation of $t_{\rm cool,avail}$ and the enhancement of the hot
gas density in the GFC2 model, and the competition of these two
factors determines the final cooling efficiency of this model, as
described in \S\ref{sec:results_static_halo}.

The \MORGANA cooling model forces the specific angular momentum of the
cooled down gas to always equal the mean specific angular momentum
of the halo by construction. Although the \lgalaxy cooling model makes
the same prediction in the static halo case, for dynamically evolving
haloes, the \lgalaxy cooling model predicts lower specific angular
momenta. This is because \lgalaxy assumes that the gas currently
cooling and accreting onto the central galaxy has specific angular
momentum equal to that of the current halo. For cosmologically
evolving haloes, the halo specific angular momentum typically increases
as the halo grows, so the gas cooled at earlier times tends to have
lower specific angular momentum, and so the total mean specific
angular momentum of all of the gas that has cooled up to that time is
lower than the mean value of the current halo.

The new cooling model tends to give higher specific angular momentum
than the GFC1 and GFC2 models, mainly because the new cooling model
can cool gas that was originally at larger radii, which according to
the assumed $j_{\rm hot}(r)$ has higher specific angular momentum.

For the dynamical halo case, a new phenomenon is that for haloes with
$M_{\rm halo}\gsim 10^{12}\Msol$ at $z=0$, the GFC1, GFC2 and new
cooling models predict lower specific angular momentum for the cooled
down gas at $z=0$ than the \lgalaxy cooling model, while for haloes
with $M_{\rm halo}\lsim 10^{12}\Msol$ at
$z=0$, the reverse is true. This can be understood as follows:

In the absence of cooling, all four models would predict that the mean
specific angular momentum of the hot gas always equals that of the
dark matter halo. Typically the specific angular momentum of the dark
matter halo increases as it grows, which means that the specific
angular momentum of gas accreting later is higher than that of gas
accreting earlier. In the presence of cooling, the gas that accreted
earlier is more likely to cool, so the mean specific angular momentum
of the remaining gas is higher than that of the dark matter halo.

For slower-cooling haloes (those with $M_{\rm halo}\gsim 10^{12}\Msol$
at $z=0$), typically only a small fraction of the hot gas halo cools
down, so the mean specific angular momentum of the hot gas cannot be
much different from that of the dark matter halo. Moreover, the
cooling in this case typically happens at small radii, and because the
GFC1, GFC2 and new cooling models all assume $j_{\rm hot}(r)$
increases with $r$, they predict that the gas that is currently
cooling has lower specific angular momentum than the dark matter halo,
and so also lower than the predictions of the \lgalaxy cooling model.

For the faster cooling haloes (those with
$M_{\rm halo}\lsim 10^{12}\Msol$ at $z=0$), most of the gas cools, so
the specific angular momentum of the remaining hot gas can end up being
significantly larger than that of the halo. Since the gas ends up
cooling from large radii, the specific angular momentum of the gas
that cools in a single timestep may be larger than the mean for the
dark halo. This effect is more or less captured in the GFC1, GFC2 and
new cooling models, but not in the \lgalaxy cooling model, which is
why for this case \lgalaxy predicts lower specific angular momentum
for the cooled down gas as a whole compared to the \GALFORM cooling
models. 

%%%%%%%%%%%%%%%%%%%%%%%%%%%%%%%%%%%%%%%%%%%%%%%%%%%%%%%%%%%%%%%%%%%%%%%%%%%%%%
\subsection{Full galaxy formation model} 
\label{sec:results_full_model}
In this section we show the effects of implementing the new cooling
model in a full galaxy formation model. The \GALFORM, \lgalaxy and
\MORGANA semi-analytic models have very different modeling of galaxy
sizes, star formation, black hole growth and SN and AGN feedback. A
full comparison of these models is not the aim of this paper, so here
we restrict our scope to the \GALFORM model, and investigate the
effects of the new cooling model on a recent version of \GALFORM,
namely `Lacey16' \citep{galform_lacey2015}. The `Lacey16' model is calibrated primarily on eight observational constraints: at $z=0$, the $b_{\rm J}$ and $K$ band galaxy luminosity functions; the HI mass function; the morphological fractions; the black hole \-- bulge mass relation; in the range $z=0-3$, the evolution of the $K$ band galaxy luminosity function; the sub-mm galaxy number counts and redshift distributions; the far-IR number counts; and at higher redshift still, the far-UV luminosity functions of Lyman-break galaxies. 
As previously mentioned, the
`Lacey16' model adopts the GFC1 model for gas cooling in haloes.

In our comparison, we focus on three important galaxy properties. The
first is the galaxy luminosity function (LF)
at $z=0$. This gives the abundance of galaxies of different masses,
and reproducing the observed LFs is typically a basic
requirement for any successful galaxy formation model. The second 
is the halo mass \-- stellar mass/total galaxy mass (stars + cold gas) 
correlations at $z=0$, which are also basic propertities. The third 
is the galaxy size-luminosity relation. This is of special interest
because the new cooling model predicts specific angular momenta for
galaxies that are significantly different from previous cooling models.

We first compare the original `Lacey16' model to variants using the new
cooling model, while keeping the other parameters fixed at their
original values. In the original `Lacey16' model, as in earlier
published \GALFORM models using the GFC1 cooling model, the halo
virial velocity, $V_{\rm vir}$, is updated only at halo formation
events, while in the new cooling model $V_{\rm vir}$ is normally
updated at every timestep. Changing how $V_{\rm vir}$ is updated by
itself results in significant changes in some \GALFORM predictions. To
separate more clearly the effects of the new cooling model from the
effects of how $V_{\rm vir}$ is updated, we define several variants
which we then compare: `Lacey16+cv', which is identical to the original
`Lacey16' model except that $V_{\rm vir}$ is updated at every timestep;
`Lacey16+new\ cool', which is the `Lacey16' model with the new cooling
model except with $V_{\rm vir}$ updated at formation events; and
`Lacey16+cv+new\ cool' model, which is the `Lacey16' model with the new
cooling model and with $V_{\rm vir}$ updated at every timestep (the
default case for the new cooling model). These variants are discussed
in \S\ref{sec:full_model_original}. 

As shown below, the
`Lacey16+cv+new\ cool' model without retuning does not provide a good
match to the observed galaxy luminosity functions at $z=0$, so
we then introduce a retuned model, `Lacey16+cv+new\ cool + retuned', in
which some of the other \GALFORM parameters are adjusted to provide a
better fit to these data. This retuned model is discussed in
\S\ref{sec:full_model_retuned}. For simplicity, the retuning here is limited, and only considers the $z=0$ $b_{\rm J}$ and $K$ band luminosity functions and the $z=0$ morphological fractions as constraints. This retuning also tries to maintain the improvement in early-type galaxy sizes achieved by the `Lacey16+cv+new\ cool' model.

All of these models are run on merger trees extracted from Millennium-WMAP7 N-body simulation. More details of these merger trees are given in \citet{galform_lacey2015}.

\begin{figure*}
\centering
\includegraphics[width=1.0\textwidth]{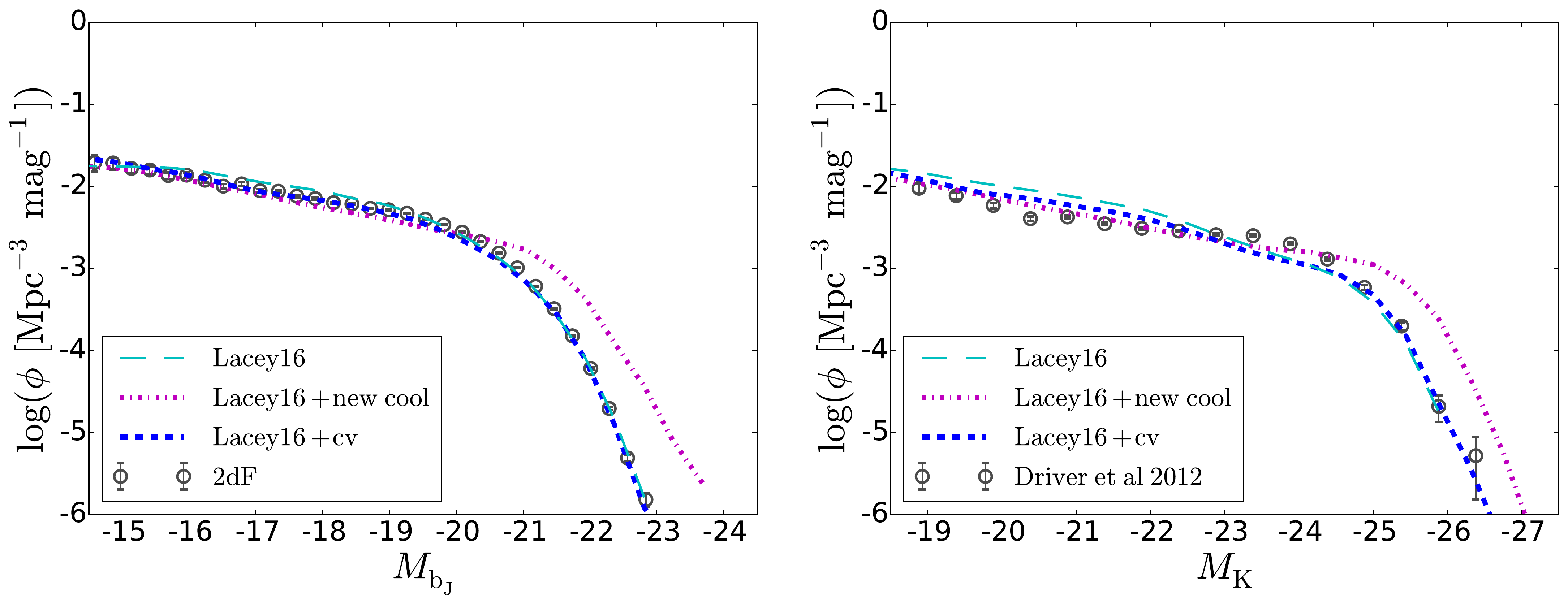}
\includegraphics[width=1.0\textwidth]{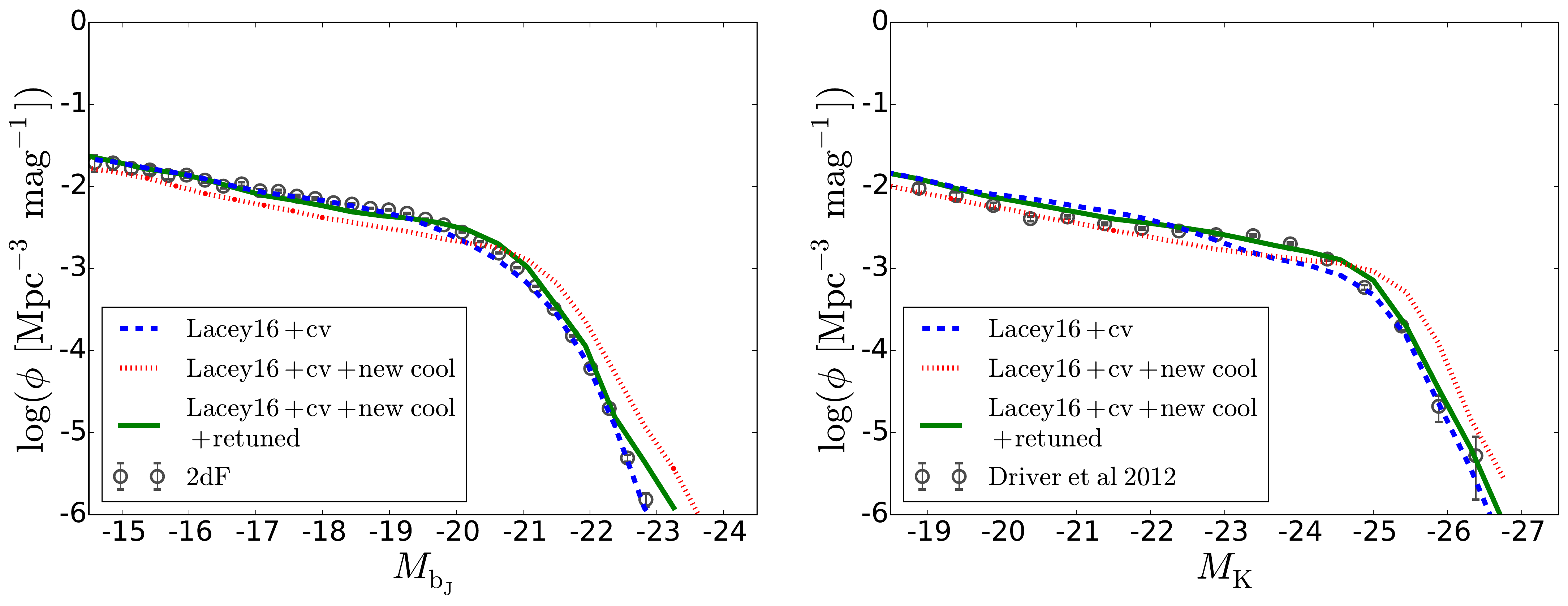}
\caption{The galaxy luminosity functions at $z=0$ in the
  $b_{\rm J}$ and $K$ bands for different variants of the `Lacey16'
  model. Each line shows the prediction for a different model, with
  the corresponding model name given in the key. The different models
  are described in the text. The top panels show the original `Lacey16'
  model and the separate effects of changing to the new cooling model
  or of updating the halo virial velocity at every timestep. In the
  bottom panels all models have the halo virial velocity updated at
  every timestep.  These panels include the retuned `Lacey16' model
  incorporating the new treatment of gas cooling. The gray open
  circles with error bars are observational data, from
  \citet{b_J_lf_obs} for the $b_{\rm J}$-band, and from
  \citet{K_lf_obs} for the $K$-band.}

\label{fig:field_lf}
\end{figure*}

\subsubsection{Original Lacey16 model and variations} 
\label{sec:full_model_original}
We first consider galaxy luminosity functions (LFs). The intrinsic luminosity of a given galaxy is calculated self-consistently by convolving its star formation history with the luminosities of single stellar populations, while the extinction is calculated self-consistently based on the galaxy's cold gas mass and metallicity and its radius. More details are given in \cite{galform_lacey2015}. Fig.~\ref{fig:field_lf}
shows the present-day $b_{\rm J}$- and $K$-band luminosity
functions predicted by the different model variants described above,
compared with observational data. The original `Lacey16' model was
calibrated to provide a good fit to the observed LFs. Updating the
halo virial velocity, $V_{\rm vir}$, at every timestep, as for the
variant `Lacey16+cv', is seen by itself to produce only small changes in
the LFs, reducing them slightly at the faint end. However, replacing
the original cooling model (GFC1) with the new cooling model is seen
to produce a large increase in the number of bright galaxies, although
this effect is smaller in the model `Lacey16+cv+new\ cool' where
$V_{\rm vir}$ is updated at every timestep (lower panels), compared to
the model `Lacey16+new\ cool' where it is only updated at formation
events (upper panels).  In the `Lacey16' model, the bright ends of the
LFs at $z=0$ are controlled mainly by AGN feedback. The excesses seen
at the bright ends show that the AGN feedback is too weak when the new
cooling model is introduced without adjusting any other
parameters. There are two reasons for this. Firstly, as shown in
\S\ref{sec:results_static_halo} and \ref{sec:results_dynamic_halo}, by
more carefully modeling the contraction of the hot gaseous halo, the
new cooling model predicts higher cooling luminosity and more
efficient cooling, which requires stronger AGN feedback to balance
it. Secondly, the efficiency of the AGN feedback that is available in the model is tightly correlated
with the growth of supermassive black holes (SMBH) at the centres of
galaxies. As discussed in \citet{galform_lacey2015}, in the `Lacey16'
model, the black hole accretion triggered by bar instabilities in galaxy disks
is a major contributor to black hole growth.  The new cooling model
generally predicts higher angular momentum for the cooled down gas,
resulting in larger disk sizes, and delaying the onset of disk
instabilities (typically by $\sim 5\,{\rm Gyr}$). This then delays the
onset of efficient AGN feedback, leading to ineffective AGN feedback
over most of the history of a galaxy.

A further effect of using the new cooling model that is apparent in
Fig.~\ref{fig:field_lf} is to lower the faint ends of the LFs
relative to the corresponding models using the GFC1 cooling
model. However, this change is fairly modest, less than a factor of
$2$. This difference indicates that the new cooling model predicts
less gas cooling in the haloes forming these faint galaxies, which are
typically low mass ($M_{\rm halo} \lsim 10^{12}\Msol$) and close to
the fast cooling regime. At first sight, this seems to contradict the
conclusions in \S\ref{sec:results_static_halo} and
\ref{sec:results_dynamic_halo}, which claim that the cooling in low
mass haloes predicted by the different cooling models is
similar. However, the models used in \S\ref{sec:results_static_halo}
and \ref{sec:results_dynamic_halo} do not include SN feedback and so
there is no reincorporation of the gas ejected out of the halo by SN
feedback. In the full model here, this ejected gas plays a central
role in gas cooling because faint galaxies have very strong
SN feedback, and so a large fraction of their gas is ejected and later
reaccreted.

Both the new cooling model and the GFC1 model assume that the ejected
gas is gradually reincorporated into the hot gas halo, and when it
joins the hot gas halo, it is shock heated to $T_{\rm vir}$, so that
it joins as hot gas without any previous cooling history.  However, as
mentioned in \S\ref{sec:other_models_GFC1}, the GFC1 model always
calculates $t_{\rm cool,avail}$ as the time since the last halo
formation event, which means that ejected gas that is reincorporated
between two halo formation events is treated as having been cooling
for longer than it has been part of the hot halo.  In contrast, the
new cooling model estimates the cooling history by accumulating the
energy previously radiated away, $E_{\rm cool}$, and the
reincorporation of the ejected gas does not change $E_{\rm
  cool}$. This difference in the treatment of the reincorporated gas
causes the new cooling model to predict less cooling in these low mass
haloes. The strength of this effect depends on the amount of gas
ejected, so only the galaxies experiencing strong SN feedback are
strongly affected.

\begin{figure*}
\centering
\includegraphics[width=1.0\textwidth]{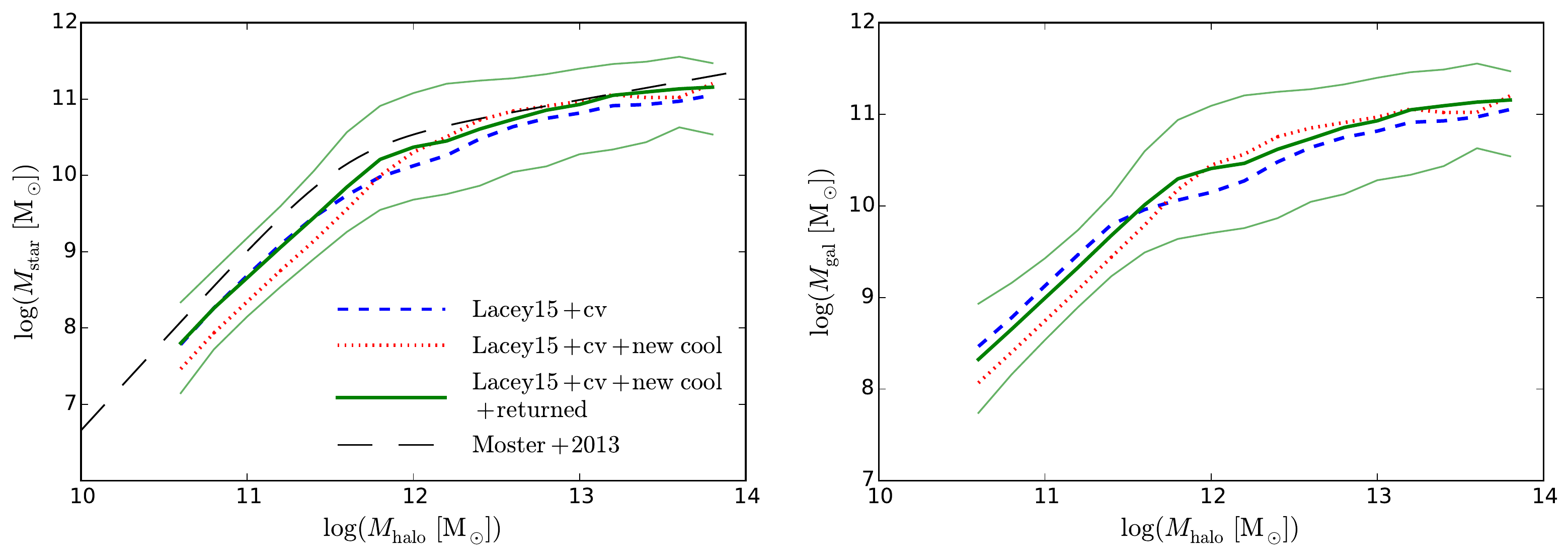}
\includegraphics[width=1.0\textwidth]{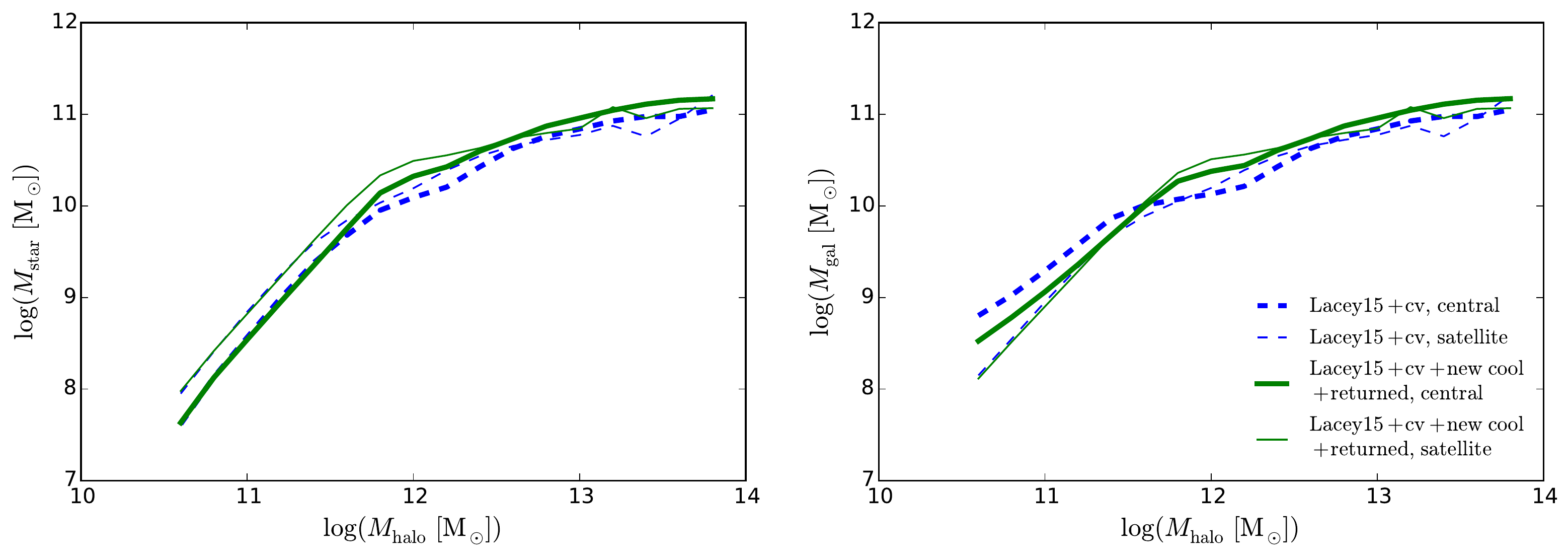}
\caption{Upper left panel: The halo mass \-- stellar mass correlations at 
  $z=0$ predicted by different models. For central galaxies, the halo masses are their host halo masses at $z=0$, while for satellites, the halo masses are the masses of haloes at infall. The thick lines of different styles 
  are median stellar masses at a given halo mass predicted by different 
  models, with the corresponding model names given in the key. The thin 
  solid lines show the $5{\rm th}-95{\rm th}$ percentiles of stellar mass in the retuned 
  model, indicating the scatter around the median correlation. The scatters 
  in other models' predictions are similar to the retuned model, and they 
  are omitted here for clarity. The long dashed line indicates the correlation 
  in \citet{Moster2013_mhalo_mstar_relation} for reference. It is derived by 
  using an abundance matching method. Upper right panel: Similar to the upper 
  left panel, but for the halo mass \-- total galaxy mass correlations at $z=0$ 
  predicted by different models. Here the total galaxy mass is the sum of the 
  stellar mass and the cold gas mass. Lower left panel: The halo mass \-- stellar 
  mass correlations at $z=0$ predicted by different models, shown separately for 
  central and satellite galaxies. The dashed lines show the median stellar masses 
  for a given halo mass predicted by the `Lacey16+cv' model, while the solid lines 
  are the predictions of the retuned model. The thick lines are for the central 
  galaxies, and the thin lines are for the satellites. The scatters around the 
  median correlations are similar in each model's predictions and similar to 
  those shown in the upper two panels. They are omitted here for clarity. Lower 
  right panel: similar to the lower left panel, but for the halo mass \-- total 
  galaxy mass correlations at $z=0$. 
}

\label{fig:mhalo_mgal_correlation}
\end{figure*}

The top row of Fig.~\ref{fig:mhalo_mgal_correlation} shows the halo 
mass \-- stellar mass/total galaxy mass (stars + cold gas) correlations 
predicted by different models. Here, for conciseness, we only show the 
results of models in which $V_{\rm vir}$ is updated at every timestep. 
With the new cooling model, i.e.\ in the `Lacey16+cv+new{\,}cool' model, 
the galaxies in haloes with $M_{\rm halo}\lsim 10^{12}\Msol$ tend to 
have lower stellar masses and total galaxy masses, which is again 
caused by the reduction of cooling in the new cooling model when 
including the reincorporated gas.

\begin{figure*}
\centering
\includegraphics[width=1.0\textwidth]{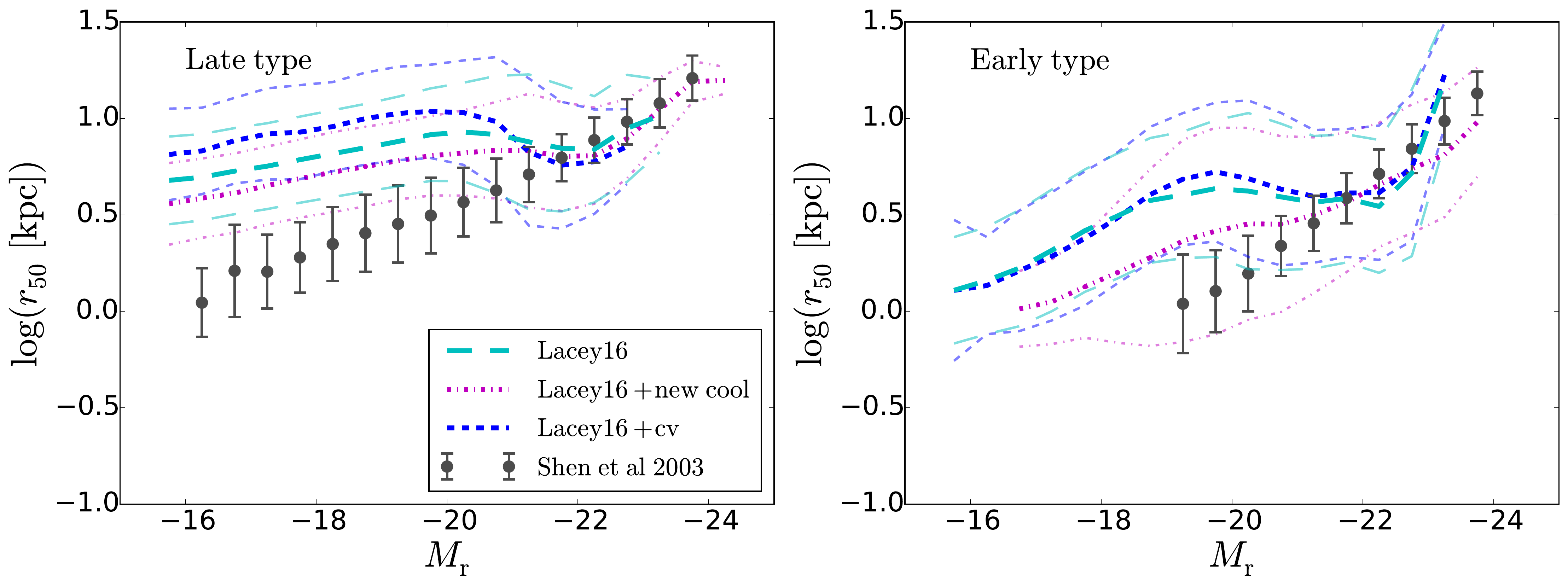}
\includegraphics[width=1.0\textwidth]{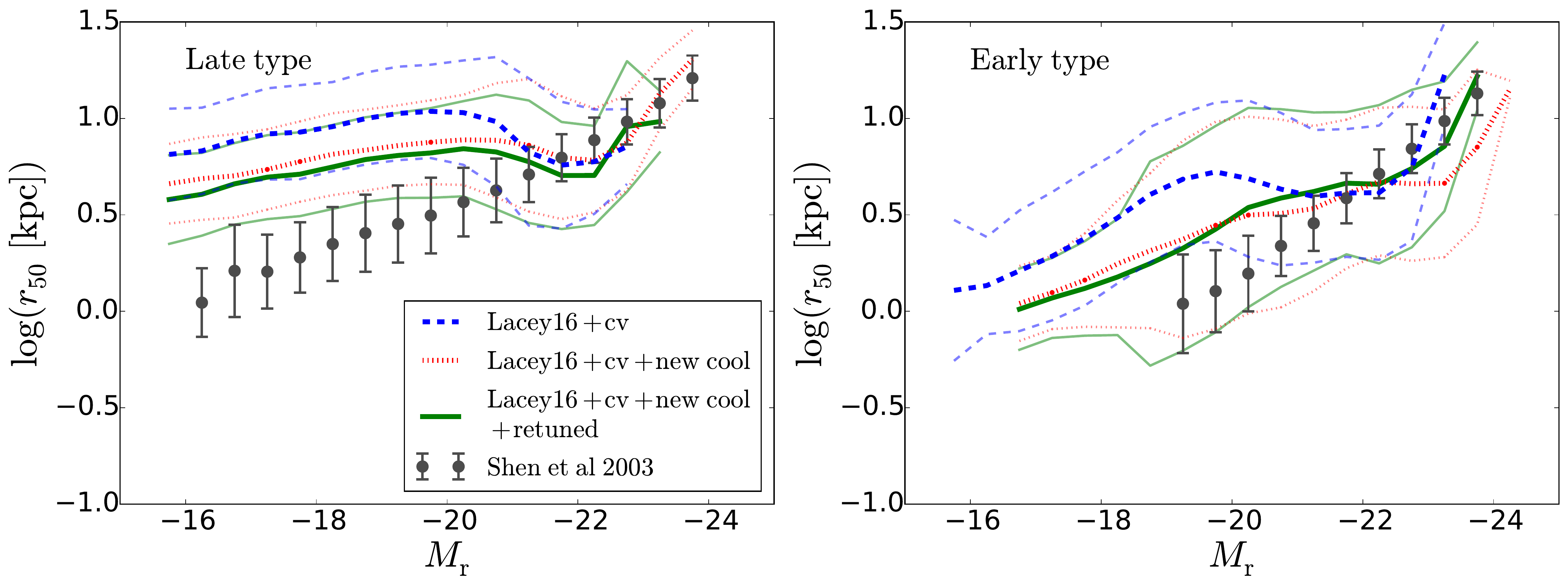}
\caption{Half-light radii of late-type (left column) and early-type
  (right column) galaxies vs. luminosity at $z=0$ . Both the
  half-light radius and luminosity are in the $r$-band. The models
  plotted and their arrangement between top and bottom rows are the
  same as in Fig.~\ref{fig:field_lf}.  The thick lines show the median
  relations, while the corresponding thin lines indicate the 10-90\%
  ranges around this. In the models, galaxies are defined as late- or
  early-type according to their $r$-band bulge to total ratio,
  $(B/T)_r$, with $(B/T)_r<0.5$ for late-type and $(B/T)_r>0.5$ for
  early-type. The gray dots with errorbars show medians and 10-90\%
  ranges based on observational data from
  \citet{size_obs_shen2003}. \citet{size_obs_shen2003} measured the
  half-light radii by fitting Sersic profiles to galaxy images and
  defined the late-type and early-type galaxies by Sersic index
  $n<2.5$ and $n>2,5$ respectively. The observed late-type galaxy sizes have
  been multiplied by $1.34$ to make an average correction to face-on
  values (see \S4.3.2 of \citet{galform_lacey2015} for more details).}

\label{fig:size}
\end{figure*}

We now consider galaxy sizes. In \GALFORM, the disk size is related to the disk specific angular momentum, while the bulge size at formation is estimated based on energy conservation and the virial theorem, and these sizes are then adjusted adiabatically until the disk and bulge reach equilibrium under the gravity of each other and the halo. More details of the size calculation are given in \citet{cole2000}. Fig.~\ref{fig:size} shows the $r$-band
half-light radius vs.\ $r$-band absolute magnitude relations for both
late-type and early-type galaxies at $z=0$. The original `Lacey16' model
predicts too large sizes for faint late-type galaxies
($M_{\rm r}\gsim -20$) and for faint early-type galaxies $V_{\rm vir}$
($M_{\rm r} \gsim -21$). The `Lacey16+cv' model, in which $V_{\rm vir}$
is updated at every timestep, gives similar results, with the
predicted sizes of faint late-type galaxies being even larger. Using
the new cooling model, as in `Lacey16+cv+new\ cool', then reduces the
sizes of faint late-type galaxies compared to the `Lacey16+cv' model,
due to the reduction of gas cooling when including the reincorporated
gas. Lower cooled down mass implies that gas has cooled from smaller 
radii in the hot gas halo, because the cooling proceeds from the halo 
center outwards. Then, since in \GALFORM the assumed hot gas specific 
angular momentum distribution predicts lower specific angular momentum 
at smaller radii, the reduction of cooling leads to the reduction of 
galaxy specific angular momenta and thus galaxy sizes.
However, the sizes of late type galaxies in the `Lacey16+cv+new\
cool' model are almost the same when compared to the original `Lacey16'
model. This indicates that some physical effect other than gas cooling
in haloes must be responsible for the deviation of the model prediction
from observations for late-type galaxies. One possibility is that the current \GALFORM model assumes the same radius for both stellar and gas disks in a galaxy. In reality, the gas disk could be more extended than the stellar disk, because star formation happens mainly in the central region of the gas disk, where the gas density is higher.

Using the new cooling model results in a larger improvement in the
size-luminosity correlation of the early-type galaxies at $z=0$. The
predicted relation is now in better agreement with observations,
much better than both the original `Lacey16' and `Lacey16+cv' models,
although the scatter around the median is still much larger than
observed. This improvement is mainly due to the reduction in the
sizes of the faint early-type galaxies. This can again be understood 
as a consequence of the reduction of cooling in relatively low mass haloes 
when including the reincorporated gas.

\subsubsection{Retuned Lacey16 model}
\label{sec:full_model_retuned}
As already discussed, we retune some of the parameters in the version
of the `Lacey16' model incorporating the new cooling model, in order to
match better the $z=0$ $b_{\rm J}$ and $K$-band LFs at $z=0$,
using the early-type galaxy fraction at different luminosities as a
secondary constraint (see \S{4.2.3} in \citet{galform_lacey2015}). 
At the same time, we ensure that the improvement in the size-luminosity 
correlation of the early-type galaxies at $z=0$ is not spoilt. The retuned parameters 
are summarized in Table~\ref{table:retuned_parameter}.

\begin{table}
\centering

\caption{Retuned parameters and their original values in the `Lacey16' model.}

\begin{tabular}{c|ccl}
\hline
parameter & Lacey16 & retuned & description \\
\hline
$\alpha_{\rm cool}$ & $0.8$ & $1.4$ & threshold of the \\
& & & ratio of the free-fall/cooling \\
& & & time scale\\
\hline
$\gamma_{\rm SN}$ & $3.2$ & $2.8$ & slope of the SN feedback\\
& & & power-law scaling\\
\hline
$f_{\rm df}$ & $1.0$ & $0.7$ & normalization of the \\
& & & dynamical friction sinking \\
& & & time scale\\
\hline
\end{tabular}

\label{table:retuned_parameter}
\end{table}

To match the LF measurements, the major problem needing to be
solved is the excess at the bright end. As discussed in
\S\ref{sec:full_model_original}, this is due to the ineffectiveness of
AGN feedback, which is a combined effect of enhanced cooling and the
less efficient black hole growth induced by the suppression of the
disk instabilities. One direct solution would be to increase the
number of disk instabilities by raising the stability
threshold, which is somewhat uncertain. However, the faint early-type galaxies are mainly produced 
by disk instabilities, and raising the stability threshold would let 
disks with higher specific angular momentum, and thus larger sizes, 
be turned into spheroids. This would increase the median size of the 
faint early-type galaxies, and thus spoil the improvement achieved by using the new cooling model.
Therefore other ways of enhancing the AGN feedback effect should
be considered first.  The effect of the AGN feedback can also be
increased by turning on AGN feedback earlier.  This can be done by
increasing the parameter $\alpha_{\rm cool}$, which sets the threshold of
the ratio of the free-fall timescale over the cooling timescale (both
evaluated at $r=r_{\rm cool}$) at which AGN feedback turns on 
(for more details see Appendix \ref{app:agn_feedback_model}). Here, we
increase $\alpha_{\rm cool}$ from 0.8 to 1.4.

We also slightly reduce the uncertain SN feedback strength in low-mass galaxies
to improve the predictions for the faint ends of the LFs. In
\GALFORM, the strength of the SN feedback scales with galaxy circular
velocity, $V_{\rm c}$, as a power-law, $V_{\rm c})^{-\gamma_{\rm SN}}$. 
We reduce $\gamma_{\rm SN}$ from $3.2$ to $2.8$.

We also slightly reduce the also uncertain galaxy merger timescale to improve the
predicted early-type fraction for bright galaxies. The original
`Lacey16' model and all the variations considered here adopt the fitting
formula from \citet{Jiang_dynamicl_friction} to calculate the galaxy
merger timescale due to dynamical friction. We modify this by
introducing an extra factor $f_{\rm df}$ in the formula for the galaxy
merger timescale (equation~(14) in \citet{galform_lacey2015}). The original
fit in \citet{Jiang_dynamicl_friction} implies $f_{\rm df}=1$, and
this value was effectively assumed in \citet{galform_lacey2015}. Here
we reduce $f_{\rm df}$ to $0.7$, which is still roughly consistent
with the simulation data in \citet{Jiang_dynamicl_friction} (see their
Fig.~10). The most important effect of this is to increase the number
of mergers, particularly major mergers.

After this limited retuning of parameters, the predicted LFs
agree with observations again, as shown in the bottom row of
Fig.~\ref{fig:field_lf}. The improvements in the predicted galaxy
sizes are largely retained.

The halo mass \-- stellar mass/total galaxy mass correlations predicted 
by the retuned model are very similar to those of the `Lacey16+cv' model 
(top row of Fig.~\ref{fig:mhalo_mgal_correlation}). This is not very 
surprising, because these two models are tuned to reproduce the $K$-band 
LF, which is tightly related to galaxy stellar masses. The stellar and 
total galaxy masses in haloes with $M_{\rm halo}\lsim 10^{12}\Msol$ in 
the retuned model are close to those in the `Lacey16+cv' model because 
the SN feedback strength in the retuned model is reduced and more mass 
can stay in galaxies. We checked that in haloes with $M_{\rm halo}\lsim 10^{12}\Msol$, 
the masses delivered to galaxies by cooling are still close to those 
in the model before retuning (i.e.\ `Lacey16+cv+new cool' model) and lower 
than in the `Lacey16+cv' model. This comfirms that when we include the 
reincorporated gas, the new cooling model predicts less cooling in low mass haloes.

For the retuned model as well as for the `Lacey16+cv' model, we also 
show the halo mass \-- stellar mass/total galaxy mass correlations 
separately for central galaxies and satellites. These correlations 
are shown in the bottom row of Fig.~\ref{fig:mhalo_mgal_correlation}. 
The differences between central and satellite galaxies are mainly in 
the halo mass range $M_{\rm halo}\lsim 10^{12}\Msol$. In this range, 
for a given halo mass, the satellites tend to have higher stellar 
mass, but lower galaxy mass, which implies they contain less cold 
gas than the central galaxies. Switching from the GFC1 model to 
the new cooling model does not change this difference between 
central and satellite galaxies.

%%%%%%%%%%%%%%%%%%%%%%%%%%%%%%%%%%%%%%%%%%%%%%%%%%%%%%%%%%%%%%%%%%%%%%%%%%%%%%

\section{summary} 
\label{sec:summary}
We have introduced a new model, better motivated 
and more self-consistent than previous models, for gas
cooling in haloes and accretion of gas onto galaxies for use in
semi-analytic models of galaxy formation. In this model we
explicitly calculate the contraction of the density profile of the
hot gas halo induced by cooling and by dark matter halo growth. This
contraction was not calculated explicitly in the previous \GALFORM
cooling models, nor in the \lgalaxy cooling model, while the
\MORGANA cooling model only considers the contraction of the hot gas
halo induced by cooling. We include the effect of the cooling
history of the hot gas on the current mass cooling rate by
estimating the total energy lost by cooling over the history of the
gas in the halo, using a new iterative scheme. We argue that our new
method for calculating mass accretion rates onto galaxies is more
accurate and more physically motivated than the other cooling models
mentioned above. In the new model we also follow the evolution of
the angular momentum distribution of the hot gas halo under the
effects of contraction of the hot gas distribution, which enables a
more detailed and self-consistent calculation of the angular
momentum of the gas accreted onto galaxies.

After setting out the methodology of the new cooling
model, we then compare its predictions to those of several previous
cooling models, including two older \GALFORM cooling models and the
cooling models in \lgalaxy and \MORGANA. The comparison is first
done for static dark matter haloes with masses in the range $10^{11}
- 10^{14}\Msol$. The comparison is then done for evolving dark
matter haloes with full merger histories, for haloes covering the same
mass range at $z=0$. Both of these comparisons include gas cooling
and accretion only, without any kind of feedback effects. Finally, we
investigate the effects of our new cooling model on a full galaxy
formation calculation, our starting point being the `Lacey16'
\citep{galform_lacey2015} \GALFORM model. Using the new cooling
model without any other adjustments results in too many bright
galaxies, and thus spoils the bright end of the predicted
galaxy luminosity functions. However, by slightly adjusting the values 
of three uncertain parameters relating to SN and AGN feedback and to the
timescale for dynamical friction, we are able to bring the \GALFORM
model predictions back into agreement with observations.

Compared to the cooling models previously used in \GALFORM, the
improved calculation of the cooling history and the detailed modeling
of the contraction of the hot gas halo significantly increase the mass
that cools in massive haloes. Some previous works
\citep[e.g.][]{de_lucia_2010_cmp,monaco_2014_comp} argued that the
\GALFORM cooling model tends to underestimate the gas mass that cools
in massive haloes, and proposed using a more centrally concentrated hot
gas density profile (e.g.\ a singular isothermal profile) to solve this
problem. However, in the new cooling model, the predicted cooled mass
becomes closer to the predictions of the \lgalaxy and \MORGANA cooling
models.

When comparing predictions between different cooling models for the
angular momentum of the cooled down gas, even larger differences are
seen than for the mass. The new cooling model tends to predict higher
specific angular momentum of the cooled down gas than the cooling
models previously used in \GALFORM. On the other hand, the predictions
of the new cooling model for the angular momentum are generally
smaller than those from the \lgalaxy and \MORGANA cooling models. This
is mainly because different models adopt different distributions for
the specific angular momentum of the hot gas, and different treatments
of the effects of cooling on these distributions.

In the full \GALFORM model with all other processes such as star
formation, supernova (SN) feedback and AGN feedback included, the new
cooling model tends to predict less gas cooling in lower mass haloes
($M_{\rm halo} \lsim 10^{12}\Msol$) than the cooling model previously
used in \GALFORM, because it models more correctly the effects of the
gas that is reincorporated into the hot gas halo after being ejected
by SN feedback. This effect improves the predicted size-luminosity
relation of both early-type and late-type galaxies relative to observations. However,
the improvement in the sizes of late-type galaxy is very small, which indicates that other
physical effects may be involved in explaining the discrepancy with
observations. For example, \GALFORM forces the stellar and gas disks to have the same scale radius, while in reality, the gas disk could be much more extended than the stellar disk. The inclusion of the new cooling model into \GALFORM 
and the retuning of a handful of uncertain parameters 
(see Table~\ref{table:retuned_parameter}) in the latest version of 
the model (`Lacey16') leads to an improved model, which supersedes 
previous versions of \GALFORM.

Having understood the behavior of the new cooling model, and having
compared the new cooling model to other cooling models, the next
step is to compare the predictions of the new cooling model with the
results from hydrodynamical simulations. We leave this comparison for
future work.

%%%%%%%%%%%%%%%%%%%%%%%%%%%%%%%%%%%%%%%%%%%%%%%%%%%%%%%%%%%%%%%%%%%%%%%%%%%%%%

\section*{Acknowledgements}
We thank Peter Mitchell for helpful comments on the paper draft.
This work was supported by the Science and Technology Facilities Council [ST/L00075X/1] 
and European Research Council [GA 267291]. This work used the DiRAC Data Centric system 
at Durham University, operated by the Institute for Computational Cosmology on behalf 
of the STFC DiRAC HPC Facility (www.dirac.ac.uk. This equipment was funded by a BIS 
National E-infrastructure capital grant ST/K00042X/1, STFC capital grant ST/K00087X/1, 
DiRAC Operations grant ST/K003267/1 and Durham University. DiRAC is part of the National 
E-Infrastructure.

\bibliographystyle{mn2e}

\bibliography{paper}

\appendix
\section{Approximate recursive equation for $E_{\rm cool}$}
\label{app:app_E_cool}
Here we consider the change of $E_{\rm cool}$ in a timestep $(t,t+\Delta t]$, 
and derive an approximate equation that relates $E_{\rm cool}(t)$ and 
$E_{\rm cool}(t+\Delta t)$. This equation can then be used to 
calculate $E_{\rm cool}$ at any given time recursively from the initial time $t_{\rm init}$. 

Within this timestep, the hot gas halo is treated as fixed, with its 
inner and outer boundaries respectively at $r_{\rm cool,pre}(t)$ and 
$r_{\rm vir}(t)$. By $t+\Delta t$, the gas between $r_{\rm cool,pre}(t)$ 
and $r_{\rm cool}(t+\Delta t)$ has cooled down.

From equation~(\ref{eq:new_cool_E_cool1}), one has
\begin{equation}
E_{\rm cool}(t+\Delta t)=4\pi\int_{t_{\rm init}}^{t+\Delta t} \int_{r'_{\rm p}(\tau)}^{r_{\rm vir}(\tau)}\tilde{\Lambda}\rho_{\rm hot}^2(r,\tau)r^2drd\tau,
\label{eq:E_cool_t+dt_a}
\end{equation}
where $\tilde{\Lambda}$ is the cooling function, $\rho_{\rm hot}(r,\tau)$ 
is the density of the hot gas at radius $r$ and time $\tau$, and 
$r'_{\rm p}(\tau)$ is the radius at $\tau$ of a shell that has radius 
$r_{\rm cool}$ at $t+\Delta t$. Note that here we use $r_{\rm cool}$ 
instead of $r_{\rm cool,pre}(t+\Delta t)$, because the hot halo is 
fixed here, and in the new cooling model, after the cooling calculation, 
the halo contraction would change $r_{\rm cool}(t+\Delta t)$ to $r_{\rm cool,pre}(t+\Delta t)$.

Then equation~(\ref{eq:E_cool_t+dt_a}) can be further expanded as
\begin{eqnarray}
E_{\rm cool}(t+\Delta t) & = & 4\pi\int_{t_{\rm init}}^{t+\Delta t} \int_{r_{\rm p}(\tau)}^{r_{\rm vir}(\tau)}\tilde{\Lambda}\rho_{\rm hot}^2(r,\tau)r^2drd\tau \nonumber \\
                         & - & 4\pi\int_{t_{\rm init}}^{t+\Delta t} \int_{r_{\rm p}(\tau)}^{r'_{\rm p}(\tau)}\tilde{\Lambda}\rho_{\rm hot}^2(r,\tau)r^2drd\tau \nonumber \\
                         & = & I_{1}-I_{2},
\label{eq:E_cool_t+dt_b}                       
\end{eqnarray}
where $I_{1}$ and $I_{2}$ represent respectively the two integrals 
in the above equation, and $r_{\rm p}(\tau)$ is the radius at $\tau$ 
of a shell that has radius $r_{\rm cool,pre}$ at $t+\Delta t$. Note 
that at $t+\Delta t$ the hot gas halo inner boundary is still at 
$r_{\rm cool,pre}(t)$ because the halo is assumed to be static over 
the interval $(t,t+\Delta t]$. Further
\begin{eqnarray}
I_{1} & = & 4\pi\int_{t_{\rm init}}^{t} \int_{r_{\rm p}(\tau)}^{r_{\rm vir}(\tau)}\tilde{\Lambda}\rho_{\rm hot}^2(r,\tau)r^2drd\tau \nonumber \\
      & + & 4\pi\int_{t}^{t+\Delta t} \int_{r_{\rm p}(\tau)}^{r_{\rm vir}(\tau)}\tilde{\Lambda}\rho_{\rm hot}^2(r,\tau)r^2drd\tau \nonumber \\
      & = & E_{\rm cool}(t)+\Delta t \times 4\pi\int_{r_{\rm cool,pre}}^{r_{\rm vir}}\tilde{\Lambda}\rho_{\rm hot}^2(r,t)r^2dr \nonumber \\
      & = & E_{\rm cool}(t)+L_{\rm cool}(t)\Delta t,
\label{eq:I_1_for_E_cool}
\end{eqnarray}
in which we have used equation~(\ref{eq:new_cool_E_cool1}) for the first 
integral in the above equation, while the second integral is simplified 
by the assumption that the hot gas halo is fixed within $(t,t+\Delta t]$, 
with the inner and outer boundaries $r_{\rm cool,pre}$ and $r_{\rm vir}$ 
respectively, and $L_{\rm cool}(t)$ is defined in equation~(\ref{eq:new_cool_L_cool1}). 

$I_{2}$ can be further written as
\begin{eqnarray}
I_{2} & = & 4\pi\int_{t}^{t+\Delta t} \int_{r_{\rm p}(\tau)}^{r'_{\rm p}(\tau)}\tilde{\Lambda}\rho_{\rm hot}^2(r,\tau)r^2drd\tau \nonumber \\
      & + & 4\pi\int_{t_{\rm init}}^{t} \int_{r_{\rm p}(\tau)}^{r'_{\rm p}(\tau)}\tilde{\Lambda}\rho_{\rm hot}^2(r,\tau)r^2drd\tau \nonumber \\
      & = & L'_{\rm cool}\Delta t + I_{3},
\label{eq:I_2_for_E_cool}
\end{eqnarray}
where $L'_{\rm cool}(t)$ is defined in equation~(\ref{eq:L_cool_p}), and the 
first integral in the above equation is simplified again because the hot 
gas halo is assumed fixed within $(t,t+\Delta t]$, while $I_{3}$ 
corresponds to the second integral above.

The integral $I_{3}$ represents the total energy radiated away by the 
gas within $r_{\rm cool,pre}\leq r\leq r_{\rm cool}$ from $t_{\rm init}$ 
to $t$, and it can be rewritten as the summation of $\delta E_{\rm cool}$ 
of each gas shell in this range, namely
\begin{eqnarray}
I_{3} & = & \int_{r_{\rm cool,pre}}^{r_{\rm cool}}\frac{\delta E_{\rm cool}}{\delta r}dr \nonumber \\
      & = & \int_{r_{\rm cool,pre}}^{r_{\rm cool}} \frac{\delta E_{\rm cool}}{\delta L_{\rm cool}}\frac{\partial L_{\rm cool}}{\partial r}dr \nonumber \\
      & = & \int_{r_{\rm cool,pre}}^{r_{\rm cool}} \tilde{t}_{\rm cool,avail}(r,t)\frac{\partial L_{\rm cool}}{\partial r}dr,
\label{eq:I_3_for_I_2}
\end{eqnarray}
in which we derive the third line from the second line by virture of 
the definition of $\tilde{t}_{\rm cool,avail}$ for an individual gas 
shell given in equation~(\ref{eq:t_cool_avail_def}).

Now consider that in the slow cooling regime, typically $r_{\rm cool}$ 
is close to $r_{\rm cool,pre}$, thus the radial dependence in 
$\tilde{t}_{\rm cool,avail}(r,t)$ can be ignored, while in the fast cooling 
regime, although $r_{\rm cool}$ could be much larger than $r_{\rm cool,pre}$, 
the cooling is so fast that halo growth and hot gas halo contraction only 
have weak effects on the cooling, and thus can only introduce weak dependence 
of $\tilde{t}_{\rm cool,avail}$ on $r$. In all, for $r_{\rm cool,pre}\leq r\leq r_{\rm cool}$, 
we can approximate $\tilde{t}_{\rm cool,avail}(r,t)\approx \tilde{t}_{\rm cool,avail}(r_{\rm cool},t)$, so
\begin{eqnarray}
I_{3} & \approx & \tilde{t}_{\rm cool,avail}(r_{\rm cool},t)\int_{r_{\rm cool,pre}}^{r_{\rm cool}} \frac{\partial L_{\rm cool}}{\partial r}dr \nonumber \\
      & = & \tilde{t}_{\rm cool,avail}(r_{\rm cool},t)L'_{\rm cool},
\label{eq:I_3_for_I_2_app}
\end{eqnarray}
where $L'_{\rm cool}(t)$ is defined in equation~(\ref{eq:L_cool_p}) and is the 
total cooling luminosity at $t$ of the gas between $r_{\rm cool,pre}\leq r\leq r_{\rm cool}$. 
Note that for hot gas halo that is fixed at all times, this approximation 
becomes exact. Based on this, one has
\begin{eqnarray}
I_{2} & \approx & [\Delta t + \tilde{t}_{\rm cool,avail}(r_{\rm cool},t)]L'_{\rm cool} \nonumber \\
      & = & \tilde{t}_{\rm cool,avail}(r_{\rm cool},t+\Delta t)L'_{\rm cool} \nonumber \\
      & \approx & t_{\rm cool,avail}(t+\Delta t)L'_{\rm cool} ,
\label{eq:I_2_for_E_cool_app}
\end{eqnarray}
in which we have used equations~(\ref{eq:t_cool_avail_a}) and (\ref{eq:t_cool_avail}). 

Substituting equations~(\ref{eq:I_1_for_E_cool}) and (\ref{eq:I_2_for_E_cool_app}) into 
equation~(\ref{eq:E_cool_t+dt_b}), one reaches the approximate recursive equation for 
$E_{\rm cool}$, i.e.\ equation~(\ref{eq:recursive_E_cool}).
%===========================================================================================================================================================
\section{Calculation of the change in the angular momentum distribution of the hot gas halo}
\label{app:app_j}
\subsection{Approximate calculation of $j_{\rm hot}[r(r')]$} 
\label{app:app_j_cal}
In the new cooling model, the hot gas halo evolves with the growth of the 
dark matter halo, and it also contracts in response to gas cooling, which 
removes pressure support from the central regions. These effects change 
the specific angular momentum distribution of the hot gas halo. We
assume that each spherical shell of hot gas conserves its specific
angular momentum, $j_{\rm hot}$, during this change, but the shell moves 
from $r$ to $r'$, and thus the angular momentum profile changes 
from $j_{\rm hot}(r)$ to $j'_{\rm hot}(r')=j_{\rm hot}[r(r')]$. As 
described in \S\ref{sec:j_calculation}, $r(r')$ can be determined through 
equation~(\ref{eq:m_conservation}), which is based on mass conservation 
for each shell. For the assumed form of the hot gas density profile, 
this equation does not have an explicit analytical solution, leading 
to no exact analytical expression for $j_{\rm hot}[r(r')]$. While $j_{\rm hot}[r(r')]$ 
can be derived numerically for each shell at every timestep, this is 
computationally expensive, and so here we present an approximate analytical 
expression that can be used instead. In Appendix~\ref{app:app_j_test}, we test 
the accuracy of this analytical approximation against a numerical solution 
of the same equations.

At the end of a timestep, the hot gas is distributed between $r_{\rm cool}$ 
and $r_{\rm vir}$, following a $\beta$-distribution with core radius, 
$r_{\rm core}$. Before the calculation of the next timestep, the effects 
of halo growth and hot gas halo contraction during the current timestep 
should be included. These effects redistribute the hot gas that was 
previously in this halo. The inner boundary of the hot gas moves from 
$r_{\rm cool}$ to $r_{\rm cool,pre}$, while the outer boundary moves 
from $r_{\rm vir}$ to $r'_{\rm vir}$. According to the assumptions in 
\S\ref{sec:new_cool_assumption}, this adjusted gas still follows a 
$\beta$-distribution, but with a new core radius $r'_{\rm core}$. 
As mentioned in \S\ref{sec:j_calculation}, this adjustment is only 
for the hot gas previously in this halo, while the newly added hot 
gas is assumed to mix with the hot gas halo after this adjustment. 
Therefore the total gas mass, $M_{\rm hot}$, before and after this 
adjustment is unchanged.

When considering the approximate calculation of $j_{\rm hot}[r(r')]$, it
is more convenient to work with the variables $x\equiv r/r_{\rm core}$ and 
$x'\equiv r'/r'_{\rm core}$ instead of $r$ and $r'$. Then the angular 
momentum profile after the adjustment of the hot gas halo can be written as 
$j_{\rm hot}[x(x')]$. The function $x(x')$ can be derived from 
equation~(\ref{eq:m_conservation}), which can be written as
\begin{equation}
M_{\rm hot}(<x)=M'_{\rm hot}(<x'), \label{eq:mass_equation}
\end{equation}
where $M_{\rm hot}(<x)$ is the mass of hot gas within radius $x$ according 
to the density profile before the adjustment induced by hot gas halo 
contraction and dark matter halo growth, while $M'_{\rm hot}(<x')$ is the 
mass of hot gas within $x'$ according to the density profile after this 
adjustment. As mentioned above, $x$ and $x'$ are respectively the radii of 
the same Lagrangian shell before and after the adjustment. Note that at 
the inner boundary the above equation satisfies the condition 
$M_{\rm hot}(<x_0)=M'_{\rm hot}(<x'_0)=0$, 
where $x_0=r_{\rm cool}/r_{\rm core}$ and $x'_0=r_{\rm cool,pre}/r'_{\rm core}$, 
while at the outer boundary it satisfies 
$M_{\rm hot}(<x_{\rm vir})=M'_{\rm hot}(<x'_{\rm vir})=M_{\rm hot}$, 
where $x_{\rm vir}=r_{\rm vir}/r_{\rm core}$ and 
$x'_{\rm vir}=r'_{\rm vir}/r'_{\rm core}$.

According to the assumed $\beta$-distribution, one has
\begin{equation}
M_{\rm hot}(<x)=\frac{M_{\rm hot}}{Y_{\rm vir}-Y_0}[x-\arctan(x)-Y_0], 
\label{eq:mass_before_adjuting}
\end{equation}
where $Y_{\rm vir}=x_{\rm vir}-\arctan(x_{\rm vir})$ and $Y_0=x_0-\arctan(x_0)$. 
Similarly, for the hot gas halo after the adjustment, one has
\begin{equation}
M'_{\rm hot}(<x')=\frac{M_{\rm hot}}{Y'_{\rm vir}-Y'_0}[x'-\arctan(x')-Y'_0], 
\label{eq:mass_after_ajusting}
\end{equation}
where $Y'_{\rm vir}=x'_{\rm vir}-\arctan(x'_{\rm vir})$ and $Y'_0=x'_0-\arctan(x'_0)$.

Substituting equations~(\ref{eq:mass_before_adjuting}) and (\ref{eq:mass_after_ajusting}) into 
equation~(\ref{eq:mass_equation}), one derives an implicit form for the function $x(x')$
\begin{eqnarray}
x-\arctan(x) & = & \frac{Y_{\rm vir}-Y_0}{Y'_{\rm vir}-Y'_0}[x'-\arctan(x')] \nonumber \\
             & + & \frac{Y'_{\rm vir}Y_0-Y_{\rm vir}Y'_0}{Y'_{\rm vir}-Y'_0} .
\label{eq:x_relation}
\end{eqnarray}

Equation~(\ref{eq:x_relation}) does not allow an explicit analytical
expression for $x(x')$. However, it is still possible to construct
simple analytical approximations for $x(x')$ in different ranges of
$x'$, and so to derive analytical approximations for $j_{\rm hot}[x(x')]$.

First note that typically $x'\leq x$, because the contraction moves
shells from large radii to small radii. When $x'$ is
large, both $x-\arctan(x)$ and $x'-\arctan(x')$ can be well
approximated by linear functions. These linear functions then lead to
a linear functional form for $j_{\rm hot}[x(x')]$. This linear functional
form can be kept during the recursion procedure, which is necessary
for deriving the specific angular momentum distribution from its
initial value, so for large enough $x'$, $j_{\rm hot}[x(x')]$ can
always be expressed as a linear function of $x'$.

On the other hand, when $x'$ is very close to $0$, a Taylor expansion gives 
$x'-\arctan(x')=x'^3/3-{x'}^5/5+O(x'^7)$. Note that this typically
happens in the slow cooling regime, in which the cooling is limited to
the central region of the halo and the induced contraction of the hot
gas halo is small in each timestep, so typically in this case $x$ is
also close to $0$, and the Taylor expansion is also a good
approximation for $x-\arctan(x)$, i.e.\ $x-\arctan(x)=x^3/3-{x}^5/5+O(x^7)$. 
These nonlinear terms in the Taylor expansions cause $j_{\rm hot}$ to 
gradually deviate from the assumed linear form before the starting of 
cooling. The nonlinear terms in the Taylor expansions are third and 
fifth order. This suggests the following expression for $j_{\rm hot}[x(x')]$
\begin{equation}
j_{\rm hot}[x(x')]=c_1x'^6+c_2x'^5+c_3x'^3+c_4x'+c_5 ,
\label{eq:j_app_small_x}
\end{equation}
where $c_1-c_5$ are coefficients and we include all terms with orders 
lower than $O(x'^7)$ that can be generated by the third and fifth order 
terms, while the linear term is added to include the initial linear 
form of the angular momentum profile.

When $x$ (and also $x'$) are either not very large or not close to
$0$, the function $x-\arctan(x)$ has a non-linear dependence, but not
so strong as in the case when $x$ is close to $0$. Thus, generally
speaking, $j_{\rm hot}[x(x')]$ in this regime can be expressed
approximately as a lower order polynomial, and here we choose a
second-order polynomial.

In summary, we adopt the following piecewise function as the
analytical approximation for $j_{\rm hot}[x(x')]$
\begin{equation}
j_{\rm hot}[x(x')]=\left\{
\begin{array}{lr}
a_1x'+a_2, & x'\geq 3.5 \\
a_3x'+a_4, & 2.0\leq x'< 3.5 \\
a_5x'^2+a_6x'+a_7, & 0.5 \leq x' < 2.0 \\
a_8x'^6+a_9x'^5+a_{10}x'^3 & \\
 +a_{11}x'+a_{12}, & 0.0\leq x'<0.5
\end{array}
\right. \label{eq:j_app}
\end{equation}
where $a_1-a_{12}$ are coefficients, with the coefficients in equation~(\ref{eq:j_app_small_x}) to be renamed as $a_8-a_{12}$.

The procedure is then as follows. At each timestep, several sample
points are taken over the whole range of $x'$, and then
equation~(\ref{eq:x_relation}) is solved numerically for these sample
points to find the corresponding $x$, with the specific angular
momentum distribution in the last timestep, $j_{\rm hot}[x(x')]$
being known for these sample points. Using these values,
equation~(\ref{eq:j_app}) then becomes a set of linear equations for the
coefficients $a_1-a_{12}$, which can be solved easily. Once these
coefficients are determined, then the approximate $j_{\rm hot}[r(r')]$
can be calculated for any value of $r'$ for the current timestep. 
Then the contribution from the newly added gas, $j_{\rm new}(r')$, can 
be added as described in \S\ref{sec:j_calculation}. Since it is assumed 
that $j_{\rm new}(r')\propto r'$, this further changes the coefficients 
of the first and zeroth order terms in equation~(\ref{eq:j_app}). After this, 
the angular momentum profile of this timestep is fully determined.

This approximation requires $9$ sample points for determining
$a_1-a_{12}$ (two adjacent $x'$ sections share one common sample
point), and so equation~(\ref{eq:x_relation}) needs to be solved for
$x(x')$ only $9$ times at each timestep. An alternative to this
approximate method would be to evaluate $j_{\rm hot}[r(r')]$ numerically
on a radius grid, which would require solving
equation~(\ref{eq:x_relation}) at each radius grid point, rather than at a
handful of sample points. The approximate method is
computationally much faster than the straightforward radius grid
method. Also, the approximate method only requires storing the $12$
coefficients, while the radius grid method requires storing the whole
radius grid and the numerical $j_{\rm hot}[r(r')]$ on it, and thus would
require much more computer memory.

%----------------------------------------------------------------------

\subsection{Comparison with direct calculation}
\label{app:app_j_test}
To assess the accuracy of the approximation introduced in the previous
section, we compared the angular momentum accretion rates for central
galaxies calculated using this approximation with those calculated
using a direct (but more computationally intensive)
calculation. This direct calculation evaluates $j_{\rm hot}(r)$
numerically on a radius grid at each timestep. The radius grid covers
the range between $r_{\rm cool,pre}$ and $r_{\rm vir}$ with $1000$
grid points.  $j_{\rm hot}(r)$ at a given timestep is calculated from
$j_{\rm hot}(r)$ at the previous timestep by solving
equation~(\ref{eq:x_relation}) for each grid point, and then using
equation~(\ref{eq:j_recursive}).

\begin{figure*}
 \centering
\includegraphics[width=1.0\textwidth]{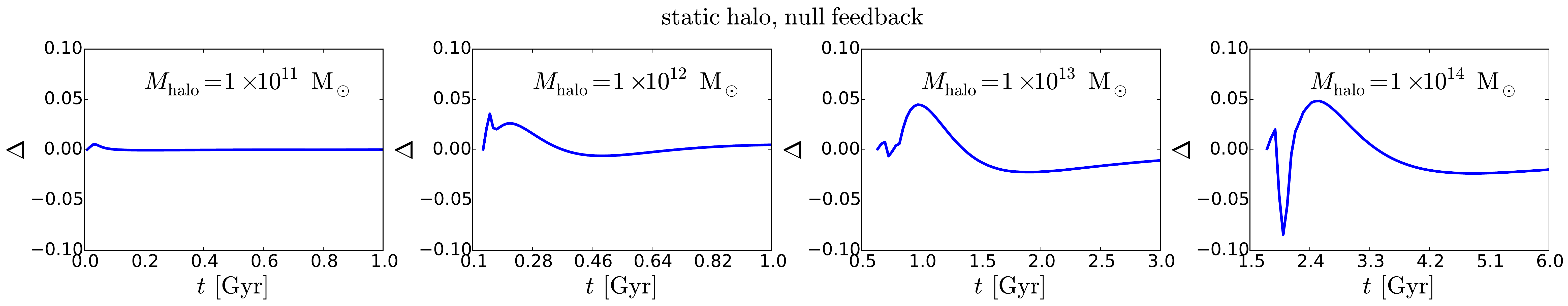}
\includegraphics[width=1.0\textwidth]{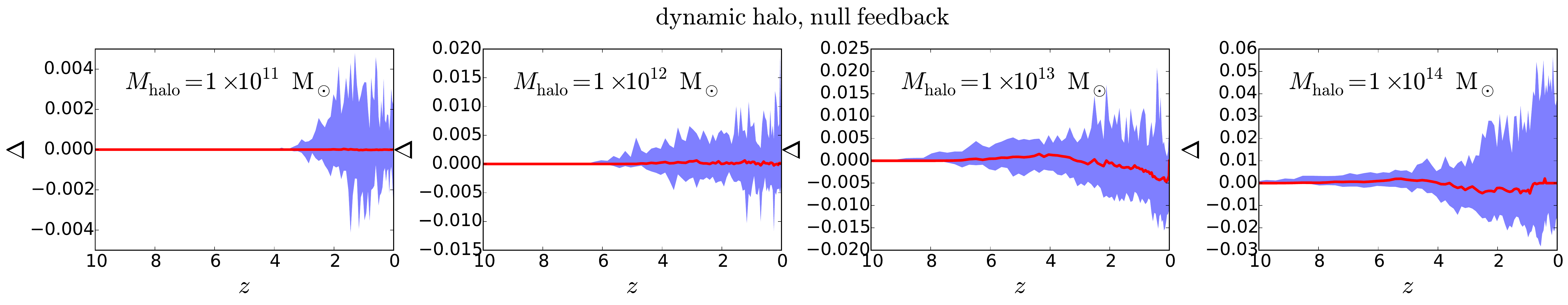}
\includegraphics[width=1.0\textwidth]{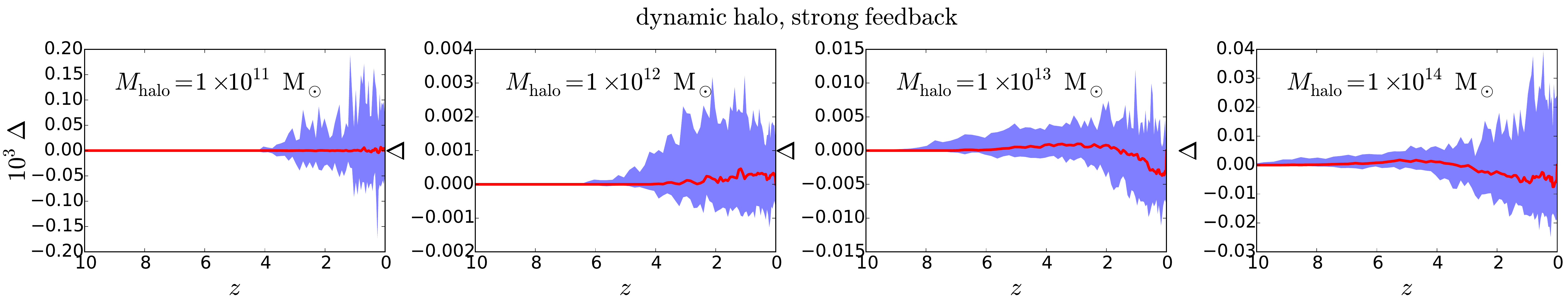}
\caption{The relative error, $\Delta$, in the angular momentum accretion rate
  calculated using the approximate method for evolving
  $j_{\rm hot}(r)$ compared with that obtained from the direct
  calculation. Results are shown for 3 cases (static halo without
  feedback, dynamically evolving halo without feedback, and
  dynamically evolving halo with strong supernova feedback) and 4
  different halo masses ($10^{11}$, $10^{12}$, $10^{13}$ and
  $10^{14}\,{\rm M}_{\odot}$). The $10^{11}\Msol$ static halo is at $z=3$, 
  and other static haloes are at $z=0$, while for dynamic haloes, these 
  masses are the halo masses at $z=0$. The dynamic halo cases use full halo
  merger histories, with $100$ Monte Carlo merger trees for each halo
  mass. For the dynamic halo cases, in each panel the solid line shows
  the median of the relative error, while the shaded region indicates
  the $5-95\%$ range. See text for more details. }

\label{fig:j_relative_error}
\end{figure*}

The comparison is done for three cases. The first one is for static
haloes, with no feedback. The second is for dynamically evolving haloes,
including full halo merger histories, but still without any feedback.
The third is also for dynamically evolving haloes, but with strong
supernova feedback. Here the supernova feedback is modeled as usual in
\GALFORM, with a mass ejection rate from the galaxy into the ejected
gas reservoir $\dot{M}_{\rm eject}=\beta \psi$, where $\psi$ is the
star formation rate and the mass-loading factor
$\beta=(V_{\rm c}/V_{\rm SN})^{-\gamma_{\rm SN}}$, with $V_{\rm c}$
being the circular velocity of the galaxy and $V_{\rm SN}$ and
$\gamma_{\rm SN}$ parameters. For the calculations here, we use
$V_{\rm SN}=320\kms$ and $\gamma_{\rm SN}=3.2$, which are close to the
values adopted in recent versions of \GALFORM.  The calculations are
done for four different halo masses, namely
$M_{\rm halo}=10^{11},\ 10^{12},\ 10^{13}$ and $10^{14}\Msol$, which
covers both the fast and slow cooling regimes. The $10^{11}\Msol$ static halo is at $z=3$, 
and other static haloes are at $z=0$, while for dynamic haloes, these 
masses are the halo masses at $z=0$. For the dynamically
evolving haloes, results are calculated for $100$ Monte Carlo merger
trees for each halo mass.

For each of these cases, the angular momentum accretion rate onto the
central galaxy due to the cooling flow, $\dot{J}_{\rm cool}$, is
calculated at each timestep, both for the approximate method in
Appendix~\ref{app:app_j_cal} ($\dot{J}_{\rm cool,app}$) and for the
direct calculation ($\dot{J}_{\rm cool,grid}$). The relative error, $\Delta$, is
then calculated as
$\Delta =(\dot{J}_{\rm cool,app}-\dot{J}_{\rm cool,grid})/\dot{J}_{\rm cool,grid}$. 
Fig.~\ref{fig:j_relative_error} shows this relative
error for the three cases and the four different halo masses. From
this figure, it can be seen that the relative error is generally less
than $10\%$, so the approximate method works well.

%===========================================================================================================================================================

\section{Random walk model for evolution of $\lambda_{\rm halo}$} 
\label{app:lambda_random_walk}
\subsection{Random walk model of halo spin evolution}
The evolution of halo spin results from the angular momentum and
mass brought into the halo by accretion and mergers. The angular momentum of the accreted material originates from the action of gravitational tidal torques at earlier times. This angular
momentum depends on the tangential component of the infall velocity. A
simple model for the halo growth is to assume that these
accretion/merger events are random, with random infall velocities. In
this case, the evolution of the halo spin accompanying the mass
accretion is a kind of random walk
\citep[e.g.][]{vitvitska_2002_spin}. For simplicity, we further assume
that this random walk for the halo spin is a Markov walk, meaning that
each step is statistically independent of previous steps.

In this picture, the spins of the descendant halo and its major
progenitors are related by a conditional spin distribution, which
gives the probability density for any given descendant spin value
given the spin and mass accretion history of the progenitor. We now
derive the form of this probability distribution for some plausible
assumptions. 

%----------------------------------------------------------------------
\subsection{Conditional distribution of descendant halo spin}
Mathematically, a random walk is described as a sequence of random
variables, $Y(x)$, where $x$ is the sequence index and $Y(x)$ is the
random variable at $x$, with its possible value $y$ and corresponding
probability distribution $P(y,x)$. For the random walk considered here,
we choose $x=\ln(M_{\rm halo}/M_{\rm i})$, where $M_{\rm halo}$ is the
mass of a given halo, and $M_{\rm i}$ is its initial mass. We choose
this form because it gives the same $\Delta x$ whenever the halo mass
has increased by a certain factor, and we expect that the change of
halo spin is more closely related to the fractional increase in halo
mass than to the absolute increase in mass.

N-body simulations of the formation of dark matter haloes by
hierarchical clustering show that the distribution of
$\lambda_{\rm halo}$ is well approximated by a lognormal, with median
$\lambda_{\rm med}$ and dispersion $\sigma_{\lambda}$ in
$\ln\lambda_{\rm halo}$ that are almost independent of the halo mass
and cosmological parameters
\citep[e.g.][]{spin_distribution_bett}. Motivated by this, we define
$Y=[\ln(\lambda_{\rm halo})-\ln(\lambda_{\rm med})]/\sigma_{\lambda}$.

For a Markov random walk, $P(y,x)$ is approximately described by the
Fokker-Planck equation:
%--------------------
\footnote{Strictly speaking, the Fokker-Planck equation is not valid for
  an arbitrarily sharp distribution like our initial condition
  $P(y,0)=\delta(y-y_0)$, but this distribution would be broadened
  quickly by diffusion. Thus the Fokker-Planck equation is expected still to
  be valid at times not too close to the initial time.}
%--------------------
\begin{equation}
\frac{\partial P}{\partial x}=-\frac{\partial}{\partial y}[a_1 P]
+\frac{\partial^2}{\partial y^2}[a_2 P] ,
 \label{eq:fp_equ}
\end{equation}
where $a_1$ and $a_2$ are two functions of $y$ and $x$. Given the
results for the spin distribution described above, we want
equation~(\ref{eq:fp_equ}) to have a steady-state asymptotic solution
$P(y,x) = 1/\sqrt{2\pi} \exp(-y^2/2)$, which corresponds to a
lognormal distribution for $\ln\lambda_{\rm halo}$ with parameters
that do not depend on $M_{\rm halo}$. For simplicity, we assume $a_2$
is a constant.  The requirement that
$P(y,x) = 1/\sqrt{2\pi} \exp(-y^2/2)$ be a steady-state solution then
leads to the relation $a_1=-a_2 y+c_0\exp (y^2/2)$, with $c_0$ a
constant. However, the term $c_0\exp(y^2/2)$ provides a drag towards
$y=+\infty$, which in terms of spin evolution is a trend for
$\lambda_{\rm halo}$ to become arbitrarily large, and this is
unphysical, so we set $c_0=0$, leading to $a_1=-a_2 y$.  In terms of
the random trajectories, $Y(x)$, the first term on the right hand side
of equation~(\ref{eq:fp_equ}) then represents a mean shift back towards
$Y=0$, while the second term represents a diffusion of $Y$.

With these choices for $a_1$ and $a_2$, the Fokker-Planck equation has
the following analytical solution (see e.g.\ \citet{fp_eqn_solution}
for details) for the initial condition $P(y,0)=\delta(y-y_0)$:
\begin{equation}
P(y,x | y_0,0)=\frac{1}{\sqrt{2\pi (1-e^{-2x/\tau})}}\exp\left[
  \frac{(y-y_0 e^{-x/\tau})^2}{2(1-e^{-2x/\tau})} \right] ,
\label{eq:conditional_p}
\end{equation}
where $\tau=1/a_2$ and $P(y,x | y_0,0)$ is the conditional
distribution of $y$ given $y=y_0$ at $x=0$.

Here $\tau$ serves as a relaxation scale for the variable $x$, with
the solution having roughly relaxed to the steady solution for
$x=\tau$. We choose $\tau=\ln 2$, so that the correlation between the
spin of a halo and its progenitor nearly disappears when it becomes
twice as massive as the progenitor. This value for $\tau$ was
originally chosen to approximately match the assumption made in
earlier \GALFORM models that a new spin is assigned randomly at every
halo formation event, defined as happening whenever the halo mass has
increased by a factor 2. However, we show below that this choice for
$\tau$ produces results for the spin evolution in quite good agreement
with N-body simulations. With the parameter $\tau$ fixed, and the
definitions of $Y$ and $x$, it is straightforward to derive the
corresponding conditional distribution for $\lambda_{\rm halo}$, with
which a halo's spin can be assigned given its progenitor spin and mass
growth history.

%----------------------------------------------------------------------
\subsection{Comparison with N-body simulations}
We test our simple random walk model for the evolution of
$\lambda_{\rm halo}$ by comparing its predictions with results from
\citet{vitvitska_2002_spin}, for haloes in cosmological N-body
simulations. Fig.~4 in \citeauthor{vitvitska_2002_spin} shows the
conditional probability distribution of $\lambda_{\rm halo}$ for
several ranges of initial spin and halo mass growth. Specifically,
they show three ranges for the initial spin, $\lambda_{\rm i}$, namely
$\lambda_{\rm i}<0.025$, $0.025<\lambda_{\rm i}<0.055$ and
$\lambda_{\rm i}>0.055$, and three ranges for the mass growth, which
are respectively $M_{\rm f}/M_{\rm i}<1.1$,
$1.1<M_{\rm f}/M_{\rm i}<1.25$ and $M_{\rm f}/M_{\rm i}>1.25$, with
$M_{\rm f}$ the halo mass after growth and $M_{\rm i}$ the mass before
growth.  In order to make a simple comparison between the results of
\citeauthor{vitvitska_2002_spin} and the predictions from our random walk
modeling, we estimate the typical value for each $\lambda_{\rm i}$ and
$M_{\rm f}/M_{\rm i}$ range, and then calculate the conditional
probability distribution using equation~(\ref{eq:conditional_p}).

We choose $\lambda_{\rm i}=0.019, 0.038, 0.08$ as typical values for
the three ranges $\lambda_{\rm i}<0.025$,
$0.025<\lambda_{\rm i}<0.055$ and $\lambda_{\rm i}>0.055$
respectively. These are the means over the corresponding ranges
according to the lognormal distribution of $\lambda_{\rm halo}$
measured from the same simulation.

For the mass ratio $M_{\rm f}/M_{\rm i}$, we set
$M_{\rm f}/M_{\rm i}=1$ as its lower limit, which means that the halo
mass is not allowed to decrease, while $M_{\rm f}/M_{\rm i}=2$ is set
as the upper limit. This is because \citeauthor{vitvitska_2002_spin}
always measure the change of halo spin between two adjacent N-body
snapshots, between which the physical time duration is relatively
short. Large values of $M_{\rm f}/M_{\rm i}$ would be caused by major
mergers instead of smooth accretion, and the number of major mergers
for a halo should be at most one in this short time duration. Thus, the
three ranges of $M_{\rm f}/M_{\rm i}$ in
\citeauthor{vitvitska_2002_spin} become $1<M_{\rm f}/M_{\rm i}<1.1$,
$1.1<M_{\rm f}/M_{\rm i}<1.25$ and $1.25<M_{\rm f}/M_{\rm i}<2$
respectively. We take the geometric mean of the range boundaries as
the typical value for the corresponding mass range, and this leads to
$M_{\rm f}/M_{\rm i}=1.049, 1.173, 1.581$ for the three ranges.

\begin{figure*}
\centering
\includegraphics[width=1.0\textwidth]{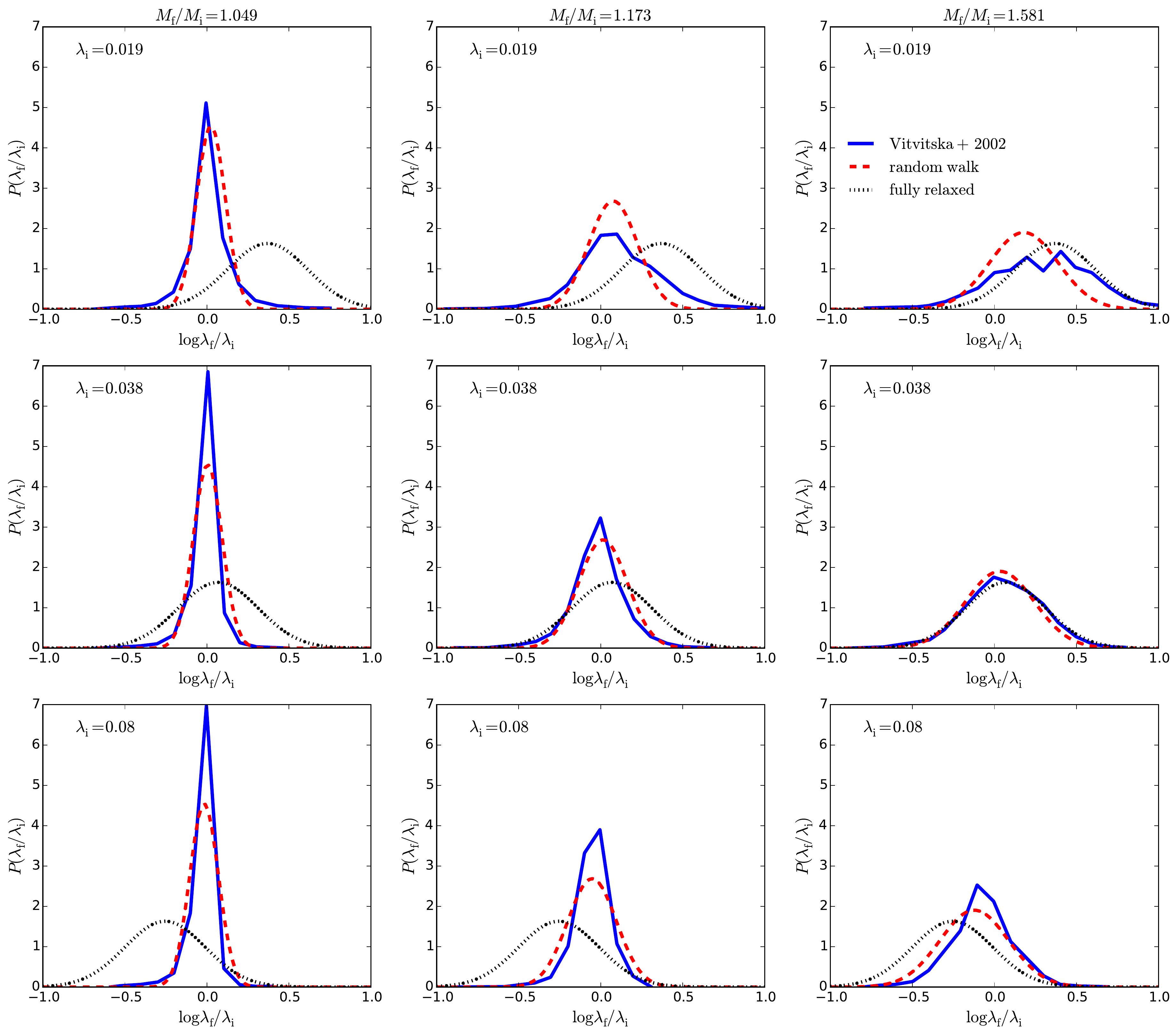}
\caption{Comparison of the conditional halo spin distributions
  predicted by our random walk model with measurements from N-body
  simulations in \citet{vitvitska_2002_spin}. The nine panels
  correspond to those in Fig.~4 of \citeauthor{vitvitska_2002_spin} Each
  row corresponds to a range of the initial spin, $\lambda_{\rm i}$,
  with our estimated typical $\lambda_{\rm i}$ for that range given in
  the upper left corner of each panel. Each column corresponds to a
  range of the ratio, $M_{\rm f}/M_{\rm i}$, with $M_{\rm f}$ and
  $M_{\rm i}$ being the halo masses at adjacent snapshots, and our
  estimated typical $M_{\rm f}/M_{\rm i}$ being shown at the top of
  the column. In each panel, the blue solid line is the conditional
  spin distribution from \citeauthor{vitvitska_2002_spin}, the red
  dashed line is the distribution calculated from
  equation~(\ref{eq:conditional_p}) based on our random walk model, and the
  black dotted line shows the fully relaxed distribution expected in
  the random walk model for reference.}

\label{fig:conditional_spin_p}
\end{figure*}

Using these estimated typical values, the corresponding conditional
distributions can be calculated for the random walk
model. Fig.~\ref{fig:conditional_spin_p} shows the comparison between
the predictions of our simple random walk model and the results
measured by \citeauthor{vitvitska_2002_spin} from their N-body
simulations. The agreement is acceptable for a simple comparison.

%===========================================================================================================================================================
\section{Simple AGN feedback model in GALFORM} 
\label{app:agn_feedback_model}

The AGN feedback model used in the `Lacey16' model was first introduced in 
\citet{galform_bower2006}. Specifically, it assumes that the AGN feedback 
is in the radio mode \citep{munich_model2}, in which a relativistic jet 
generated by supermassive black hole (SMBH) accretion heats the halo 
gas and thus suppresses cooling.

In \GALFORM there are two conditions for effective AGN feedback. Firstly, 
the halo gas should be close to the slow cooling regime, in which the 
cooling is slower than the gravitational infall and a quasi-hydrostatic 
hot gaseous halo exists. This is motivated by the idea that only the gas 
close to this regime can maintain its pressure and thus the jet can interact 
and heat the halo gas effectively. This condition is tested by comparing 
the cooling time scale, $t_{\rm cool}$, and the free-fall time scale, 
$t_{\rm ff}$, at the cooling radius, $r_{\rm cool}$. Specifically, AGN 
feedback is assumed to be effective only if
\begin{equation}
t_{\rm cool}(r_{\rm cool})/t_{\rm ff}(r_{\rm cool})>1/\alpha_{\rm cool} ,
\end{equation}
with $\alpha_{\rm cool}\sim 1$ an adjustable parameter. Consider that at 
earlier times the ratio $t_{\rm cool}(r_{\rm cool})/t_{\rm ff}(r_{\rm cool})$ 
is typically smaller, then increasing $\alpha_{\rm cool}$ causes AGN 
feedback to turn on earlier and thus enhances the suppression due to 
this feedback.

Secondly, the SMBH accretion rate should be significantly lower than 
the Eddington limit so that jets can be efficiently produced 
\citep{agn_jet_launch}, and the jet should be energetic enough to 
balance the cooling radiation. This motivates the following condition
\begin{equation}
f_{\rm Edd}L_{\rm Edd}(M_{\rm BH})>L_{\rm cool} ,
\end{equation}
where $f_{\rm Edd}\ll 1$ is a parameter, $L_{\rm Edd}(M_{\rm BH})$ is 
the Eddington luminosity of a black hole with mass $M_{\rm BH}$, and 
$L_{\rm cool}$ is the cooling luminosity of the hot gas halo. In the 
`Lacey16' model, $f_{\rm Edd}=0.01$.

Once the above two conditions are satisfied, the AGN feedback is assumed 
to be effective. In the GFC1 model, the increase of $r_{\rm cool}$ due 
to cooling is then set to zero, and then the associated mass and angular 
momentum cooling rates become zero. 

In the new cooling model, since a different procedure is used to calculate 
$t_{\rm cool,avail}$, some modifications are needed. Specifically, when 
AGN feedback turns on, the energy previously radiated away, $E_{\rm cool}$, 
is set to zero because the halo gas is heated up. This causes 
$t_{\rm cool,avail}$ to reduce to zero. With this, $r_{\rm cool}$ does not 
increase and the halo cold gas component stops growing immediately. If this 
component has nonzero mass, then it can still deliver cold gas to the central 
galaxy. When a halo is close to the slow cooling regime, the halo cold gas 
component typically is very small, so the cold gas accretion onto the central 
galaxy should stop shortly after AGN feedback turns on.

\end{document}